\documentclass[a4paper,11pt]{article}
\pdfoutput=1 

\usepackage{jcappub} 
\usepackage{hanging}

\usepackage{todonotes}

\usepackage[normalem]{ulem}
\usepackage{aas_macros}
\usepackage{graphicx}
\usepackage{graphics}
\usepackage{amsmath}
\usepackage{bm}
\usepackage{xcolor}
\usepackage{makecell}
\usepackage{hyperref}
\usepackage{verbatim}
\usepackage{comment}
\usepackage{lineno}
\usepackage{soul}
\usepackage{multicol,multirow}
\usepackage{fontawesome}
\usepackage{orcidlink}

\usepackage[draft,author=]{fixme}
\fxsetup{layout=footnote}

\newcommand{\zeff}{z_{\rm eff}}

\newcommand{\Gpc}{\ \text{Gpc}}

\newcommand{\hMpc}{\ h^{-1}\text{Mpc}}

\newcommand{\ihMpc}{\ h\,\text{Mpc}^{-1}}

\newcommand{\be}{\begin{equation}}
\newcommand{\ee}{\end{equation}}
\newcommand{\alphaBAO}{\alpha_{\rm BAO}}
\newcommand{\alphaAP}{\alpha_{\rm AP}}

\newcommand{\aiso}{\alpha_{\rm iso}}
\newcommand{\alap}{\alpha_{\rm AP}}

\newcommand{\aperp}{\alpha_\perp}
\newcommand{\apar}{\alpha_\parallel}
\newcommand{\hsc}[1]{{{{#1}}}}
\newcommand{\edited}[1]{{{{#1}}}}
\newcommand{\abacus}{{\tt Abacus}}
\newcommand{\abacussecond}{{\tt Abacus-2}}
\newcommand{\abacusfirst}{{\tt Abacus-1}}
\newcommand{\ezmock}{{\tt EZmock}}
\newcommand{\ezmocks}{{\tt EZmocks}}
\newcommand{\altmtl}{{\tt altmtl}}
\newcommand{\rascalc}{{\textsc {RascalC}}}
\newcommand{\alphaiso}{{\alpha_{\rm iso}}}
\newcommand{\alphaap}{{\alpha_{\rm AP}}}

\newcommand{\bgs}{{\tt BGS}}
\newcommand{\elgo}{{\tt ELG1}}
\newcommand{\elgt}{{\tt ELG2}}
\newcommand{\elgs}{{\tt ELG}s}
\newcommand{\lrgo}{{\tt LRG1}}
\newcommand{\lrgt}{{\tt LRG2}}
\newcommand{\lrgth}{{\tt LRG3}}
\newcommand{\lrgs}{{\tt LRG}s}
\newcommand{\lrgxelg}{{\tt LRG3$\times$ELG1}}
\newcommand{\lrgelg}{{\tt LRG3$+$ELG1}}
\newcommand{\qso}{{\tt QSO}}
\newcommand{\desidrone}{{DESI DR1}}
\newcommand{\desione}{{DESI DR1}}
\newcommand{\planck}{{\tt Planck} 2018-$\Lambda$CDM}

\newcommand{\desi}{{\tt DESI}}

\newcommand{\recsym}{{\bf RecSym}}
\newcommand{\reciso}{{\bf RecIso}}

\newcommand{\hinvMpccubed}{h^{-3} \, \text{Mpc}^{3}}
\newcommand{\hMpcinvcubed}{h^{3} \, \text{Mpc}^{-3}}
\newcommand{\hinvmpc}{h^{-1}\,{\rm Mpc}}
\newcommand{\hmpcinv}{h\,{\rm Mpc^{-1}}}
\newcommand{\rd}{r_{\mathrm{d}}}
\newcommand{\DMoverrd}{D_{\mathrm{M}}/r_{\mathrm{d}}}
\newcommand{\DHoverrd}{D_{\mathrm{H}}/r_{\mathrm{d}}}
\newcommand{\DHoverDM}{D_{\mathrm{H}}/D_{\mathrm{M}}}
\newcommand{\DVoverrd}{D_{\mathrm{V}}/r_{\mathrm{d}}}
\newcommand{\kmsMpcinv}{\,{\rm km\,s^{-1}Mpc^{-1}}}

\usepackage[nameinlink,noabbrev]{cleveref}
\crefname{equation}{Eq.}{Eqs.}
\crefname{section}{Section}{Sections}
\crefname{figure}{Figure}{Figures}
\crefname{table}{Table}{Tables}
\crefname{appendix}{Appendix}{Appendices}
\Crefname{figure}{Figure}{Figures}
\Crefname{equation}{Equation}{Equations}
\Crefname{section}{Section}{Sections}
\Crefname{table}{Table}{Tables}

\usepackage{xspace}

\title{\boldmath DESI 2024 III: Baryon Acoustic Oscillations from Galaxies and Quasars}

\author{{DESI Collaboration}:}
\emailAdd{spokespersons@desi.lbl.gov}
\affiliation{Affiliations are in Appendix \ref{sec:affiliations}}

\author[1]{{A.~G.~Adame},}
\author[2]{{J.~Aguilar},}
\author[3]{{S.~Ahlen}\orcidlink{0000-0001-6098-7247},}
\author[4]{{S.~Alam}\orcidlink{0000-0002-3757-6359},}
\author[5,6]{{D.~M.~Alexander}\orcidlink{0000-0002-5896-6313},}
\author[2]{{M.~Alvarez},}
\author[7]{{O.~Alves},}
\author[2]{{A.~Anand}\orcidlink{0000-0003-2923-1585},}
\author[8,7]{{U.~Andrade}\orcidlink{0000-0002-4118-8236},}
\author[9]{{E.~Armengaud}\orcidlink{0000-0001-7600-5148},}
% ...

\date{Accepted XXX. Received YYY; in original form ZZZ}

\abstract{
\edited{We present the DESI 2024 galaxy and quasar baryon acoustic oscillations (BAO) 
measurements using over 5.7 million unique galaxy and quasar redshifts in the range $0.1<z<2.1$. Divided by tracer type, we utilize 300,017 galaxies from the magnitude-limited Bright Galaxy Survey with $0.1<z<0.4$, 2,138,600 Luminous Red Galaxies with $0.4<z<1.1$, 2,432,022 Emission Line Galaxies with $0.8<z<1.6$, and 856,652 quasars with $0.8<z<2.1$, over a $\sim 7,500$ square degree footprint. 
The analysis was blinded at the catalog-level to avoid confirmation bias. All fiducial choices of the BAO fitting and reconstruction methodology, as well as the size of the systematic errors, were determined on the basis of the tests with mock catalogs and the blinded data catalogs. We present several improvements to the BAO analysis pipeline, including enhancing the BAO fitting and reconstruction methods in a more physically-motivated direction, and also present results using combinations of tracers. We employ a unified BAO analysis method across all tracers. We present a re-analysis of SDSS BOSS and eBOSS results applying the improved DESI methodology and find scatter consistent with the level of the quoted SDSS theoretical systematic uncertainties.
With the total effective survey volume of $\sim 18\Gpc^3$, the combined precision of the BAO measurements across the six different redshift bins is $\sim$0.52\%, marking a 1.2-fold improvement over the previous state-of-the-art results using only first-year data. We detect the BAO in all of these six redshift bins. The highest significance of BAO detection is $9.1\sigma$ at the effective redshift of 0.93, with a constraint of 0.86\% placed on the BAO scale.  We find our measurements are systematically larger than the prediction of the \planck\ at $z<0.8$. 
  We translate the results into transverse comoving distance and radial Hubble distance measurements, which are used to constrain cosmological models in our companion paper.
}
}
\begin{document}

\maketitle
\label{firstpage}

\section{Introduction}
\label{sec:intro}


Cosmology today is characterized by a well-established, simple phenomenological
model that explains observations over a broad range of scales and epochs. Of
particular relevance to this paper is the background expansion rate of the Universe
which is quantitatively described by the Hubble parameter $H(z)$. Assuming 
a homogeneous and isotropically expanding Universe, $H(z)$ 
is determined by the relative contributions of the various matter/energy components 
of the Universe, as well as its value at the present epoch, $H_0$ (the Hubble constant).
One of the great successes in cosmology is that
the parameters inferred from probes sensitive to the background expansion are (largely) consistent
with probes sensitive to the growth of cosmic structure.
Notwithstanding this concordance, 
modern cosmology faces
two outstanding problems. The first is to understand 
the nature of the energy that is causing the
accelerated expansion of the Universe. In particular, measurements of the
expansion history aim to constrain whether this accelerated expansion is
consistent with a cosmological constant, or if its source (``dark energy'')
evolves in time (most simply parametrized by an equation of state parameter $w \ne -1$) \cite{BAOreview2013}.
The second problem is the discrepancy in the measurement of the expansion rate
today, the Hubble constant $H_0$ \cite{CosmologyIntertwined}. 
Local distance ladder measurements using Cepheids
\citep{SHOES-2022} favor a higher value
($H_0 = 73.04 \pm 1.04\kmsMpcinv$), while measurements anchored at high
redshift \citep{SDSS-DR16-cosmology} favor a lower value ($H_0 = 68.18\pm0.79\kmsMpcinv$). The current and next generation of background expansion measurements, using a combination of standard candles and standard rulers, aim to address these problems by increasing the precision (and accuracy) of the expansion measurements over a wide range of redshifts. This paper presents the measurement of this background expansion using the standard ruler provided by the imprint of baryon acoustic oscillations (BAO) measured in the first year of data from the Dark Energy Spectroscopic Instrument (DESI) survey.

The baryon acoustic oscillation (BAO) method is one of the principal methods to
map the expansion history of the Universe. The physics of BAO has been discussed
extensively in a number of papers \citep{2005NewAR..49..360E,2010deot.book..246B,BAOreview2013}, 
including a companion to this
paper \cite{KP4s2-Chen}, and we refer the reader to these for a complete
discussion. Very simply, acoustic oscillations in the baryon-photon fluid
in the pre-recombination Universe imprint a characteristic scale in the
clustering of matter \cite{Peebles70,Sunyaev70}. This is manifested as a bump in the two-point correlation
function of matter, or equivalently as a series of oscillations in its Fourier
transform, the power spectrum. The comoving scale of this feature, $r_d \simeq 100 \hMpc$,
is given by the sound horizon at the end of baryon drag epoch and depends upon the photon and baryon content of the universe
\footnote{The relevant scale for BAO measurements is the horizon at the
drag-epoch, which is subtly different from the sound horizon $r_s$ that
determines the CMB acoustic peaks.} and is precisely constrained by cosmic
microwave background (CMB) measurements.
While the 3D matter distribution is not
directly measurable, the BAO feature is faithfully traced by \edited{galaxies, quasars}, and the
Lyman-$\alpha$ forest.
Measuring the apparent size of the BAO standard ruler perpendicular and parallel
to the line of sight constrains the angular diameter distance $D_A(z)$ and the
Hubble parameter $H(z)$. Using a fiducial power spectrum template as a ruler, these measurements are frequently expressed in terms of Alcock-Paczynski-like dilation parameters \cite{Alcock1979,Padmanabhan08}
\begin{equation}
    \qquad \qquad \alpha_{||} = \frac{H^{\mathrm{fid}}(z)r^{\mathrm{fid}}_{d}}{H(z)r_{d}}, \qquad \qquad \alpha_{\perp} = \frac{D_{A}(z)r^{\mathrm{fid}}_{d}}{D^{\mathrm{fid}}_{A}(z)r_{d}},
    \label{eqn:alpha_defs}
\end{equation}
where the ``fid'' denotes quantities measured in the fiducial cosmology. By simply comparing BAO measurements at different
redshifts, we can constrain the relative evolution of these quantities with
redshift. When combined with the CMB or BBN measurements that constrain the
physical scale of the BAO ruler, these relative measurements become absolute
measurements of $D_A(z)$ and $H(z)$\footnote{The relative measurements only require
the existence of a standard ruler but no knowledge of its physical size, while
physical distance measurements require the physical scale of the ruler.}. We
note that the measurements of $D_A(z)$ and $H(z)$ obtained are not
independent, but are correlated at $\sim 40\%$, where this correlation is
determined by the distribution of modes perpendicular/parallel to the line of
sight---motivated by this, \cref{eqn:alpha_defs} is often re-expressed
in terms of the geometric mean $\aiso$
and ratio $\alpha_\textrm{AP}$ 
of the dilation parameters:

\begin{equation}
    \qquad \qquad \aiso =  (\alpha_\parallel\alpha_\perp^2)^{1/3},
 \qquad \qquad \alap = \alpha_\parallel/\alpha_\perp. 
    \label{eqn:aisoap_defs}
\end{equation}
One of the features of the BAO method as a standard ruler is its 
relative insensitivity to astrophysical and observational systematics. At a 
theoretical level, this derives from the fact that the scale of the BAO 
feature ($\sim 100 \hMpc$) is much larger than the characteristic scales 
of nonlinear structure growth and galaxy formation ($\lesssim 10 \hMpc$). 
Furthermore, these effects can be significantly 
reduced by the ``reconstruction'' of the linear BAO signal. Reconstruction
effectively reverses the flows on large scales using the observed 
density field, undoing the effects
of gravitational evolution on the BAO feature
 \cite{Eisenstein07,Seo08,Padmanabhan09a,Seo2010}
The large scale of the BAO feature also makes it amenable to perturbative 
treatments, allowing for a very accurate theoretical understanding of any possible 
systematic effects. These effects have been studied in detail in the literature \cite{Padmanabhan09a,Noh09,White15,Hikage20};
\cite{KP4s2-Chen} provide a comprehensive review within the context of the 
precision of the DESI measurements.
On the observational side, the BAO method relies on a well-localized feature
in three dimensions, while most of the observational systematics are either 
two-dimensional (from the imaging surveys), or along the line of sight 
(from the selection function of galaxies). This allows for a robust extraction of the 
BAO signal even in the presence of significant observational systematics. This has been 
demonstrated previously in the literature \cite{Anderson12, Ross17,Ata2018,MerzImagingSysBAO}, and we explicitly show this for our 
data as well \cite{KP3s2-Rosado,KP3s4-Yu}.

The large scale and relatively low amplitude of the BAO feature required the 
advent of large volume galaxy surveys for it to be detected. 
The BAO feature was first detected in galaxy clustering by the SDSS \citep{BAO-discovery} and 2dFGRS \citep{BAO2dF} surveys.
The success of these measurements prompted the development of the next generation 
of spectroscopic surveys, most notably the 6dFGS \citep{SixdFBAO}, BOSS~\citep{SDSS-DR12-cosmology}, eBOSS~\citep{SDSS-DR16-cosmology} and WiggleZ \citep{Wigglez} surveys, 
that made distance measurements with increasing precision at redshifts $0 < z < 1$. 
The BOSS and eBOSS~\citep{LyABAO} surveys also demonstrated that the BAO method 
using the Lyman-$\alpha$ forest, both in the auto-correlation as well as 
cross-correlation with quasar samples, provide measurements at redshifts 
$2 < z < 4$. These BAO constraints connect the low-redshift SN distance 
measurements with the distance to the CMB last scattering surface in 
an inverse distance ladder, yielding very precise constraints on the 
expansion history of the Universe, and in particular the curvature of the 
Universe \cite{Planck2018,SDSS-DR16-cosmology}. On the other hand, these BAO measurements also provide a 
CMB-independent measurement of the Hubble constant using BBN to calibrate 
the BAO scale \cite{Addison13}. The Hubble constant measurements thus obtained are consistent 
with those from the CMB and provide an independent piece of evidence for 
the tension between the low and high redshift inferences of the Hubble constant.

BAO measurements at sub-percent precision are considered
the primary science targets of the Dark Energy Spectroscopic Instrument \citep[DESI;][]{DESICollaboration2016a},
along with novel constraints on theories of modified
gravity and inflation, and on neutrino masses. 
DESI, as a Stage-IV DE experiment,
aims to provide multiple sub-percent 
distance measurements 
over a broad $0 < z < 3.5$  
redshift range. 
DESI is in the process of a five-year survey over 14,000 deg$^2$, and will result
in a spectroscopic sample that will be an order of magnitude larger than 
previous surveys, both in the volume surveyed and in the number of galaxies 
measured. It achieves this with a combination of new instrumentation,
including a 5000-fiber multi-object spectrograph \citep{DESI2016b.Instr, DESI2022.KP1.Instr}
new imaging surveys and efficient target selection algorithms \citep{DESItarget}, 
and optimized data pipelines \citep{DESIspec}.
In addition, DESI builds in a number of internal systematics checks using
multiple tracer populations to probe common volumes. The early data from 
DESI was presented in \cite{DESI2023a.KP1.SV,DESI2023b.KP1.EDR}. These data were
used to make an initial BAO measurements in \citep{BAO.EDR.Moon.2023} and 
\citep{LyaBAO.EDR.Gordon.2023} which 
presented initial BAO measurements with the DESI galaxy and Lyman-$\alpha$ samples respectively.
These measurements were used to validate the DESI BAO pipeline, to demonstrate the 
statistical power of the DESI data, and set the stage for future analyses. This paper 
presents the BAO measurements from the galaxy samples in the first year of DESI data, 
\cite{DESI2024.IV.KP6} presents the BAO measurements from the Lyman-$\alpha$ forest, 
and \citep{DESI2024.VI.KP7A} presents the cosmological implications of these measurements.

\begin{table}
    \centering
    \small
    \resizebox{\columnwidth}{!}{%
        \begin{tabular}{|l|r|r|}
            \hline
             Ref.  &  Task & Section\\
            \hline
            \cite{BAO.EDR.Moon.2023} & First DESI BAO detection using early DESI data and BAO pipeline & --- \\
          \citep{2023MNRAS.524.3894R} &	Validation of semi-analytical/empirical covariance matrices for early DESI data  &	--- \\
             \cite{KP4s2-Chen} &	BAO Theory and Modelling Systematics & \cref{subsec:methods-fitting,subsec:sys-theory}\\
          \cite{KP4s3-Chen}&	Extensive comparison of reconstruction methods & \cref{subsec:methods-recon} \\
        {\cite{KP4s4-Paillas}} &	Optimal reconstruction of BAO for DESI 2024 &	\cref{subsec:methods-recon,sec:unblinding}\\
         & and the unblinding tests on the data & \\
           \cite{KP4s5-Valcin}	& Constructing the LRG and ELG combined tracers	& \cref{subsec:results-overlapping} \\
          \cite{KP4s6-Forero-Sanchez} &	Comparison between analytical and EZmock covariance matrices  & 	\cref{subsec:methods-cov}\\
            \cite{KP4s7-Rashkovetskyi}  &	 Analytical covariance matrices for correlation function for DESI 2024& \cref{subsec:methods-cov}\\
             \cite{KP4s8-Alves} &	Analytic covariance matrices of DESI 2024 power spectrum multipoles	& \cref{subsec:methods-cov}\\
             \cite{KP4s9-Perez-Fernandez} & Fiducial-cosmology-dependent systematics &\cref{subsec:sys-fiducialcosmo} \\
          \cite{KP4s10-Mena-Fernandez}  &	HOD-dependent systematics for LRGs  &	\cref{subsec:sys-hod} \\
            \cite{KP4s11-Garcia-Quintero}	& HOD-dependent systematics for ELGs &	\cref{subsec:sys-hod}\\
             \cite{KP3s2-Rosado} & The impact of the imaging systematics on BAO  & \cref{subsec:obssys}\\
            \cite{KP3s4-Yu} & The impact of the spectroscopic systematics on BAO & \cref{subsec:obssys}\\
            \cite{KP3s9-Andrade} & The tests of the catalog-level blinding method for DESI 2024 & \cref{subsec:blinding}\\\hline
             \end{tabular}
    }
    \caption{\label{tab:supportingpapers}
    The list of the papers supporting this paper and the corresponding sections 
    where their results are discussed.}
\end{table}

This paper is one of a series of Key Papers presenting measurements with the 
first year of data from the DESI survey, which is designated as \desidrone. 
While BAO is now a mature cosmological probe, the improved statistical precision
of the DESI project motivates reexamining all aspects of the pipeline. This
paper and its accompanying supporting papers (see \cref{tab:supportingpapers}) are the
result of this work. This paper serves both to present the actual DESI galaxy
BAO measurements and to summarize the key findings of the supporting
papers. 

Given the length of this paper, we present both an outline and an executive 
summary to guide the reader through this paper.
\cref{sec:catalog}  describes the data and the construction of the
large-scale structure (LSS) catalogs used. In particular, this includes discussing how these
data were blinded to the underlying cosmology, a first for a BAO analysis. 
\cref{sec:mocks} discusses the construction
of mock catalogs used to estimate the systematic and statistical errors and
to validate the pipelines. 
\cref{sec:methods} provides an overview of the entire BAO
measurement pipeline --- the measurement of the two-point functions, the
reconstruction pipeline, the definition of the model for the two-point
statistics, the fitting of the BAO scale, and the construction of the covariance
matrices. There are a number of improvements over previous analyses discussed here
--- these include adopting the \recsym\ reconstruction convention (see below),
revisiting the BAO fitting model including a new treatment of the broadband marginalization,
and a comprehensive treatment of estimating the covariance matrices using both 
analytical and mock-based methods.
\cref{sec:systematics} presents our systematic error budget. We approach this both with 
theoretical modeling as well as extensive tests on the data and mocks, clearly demonstrating 
that the systematic error level is significantly below the statistical errors of our sample.
Since this is the first time that the analysis of BAO measurements have been blinded at the 
catalog level, we present the process and tests used to determine when to unblind the data in 
\cref{sec:unblinding}.
\cref{sec:results} contains our results for the different samples,
including a comparison/reanalysis of previous BOSS/eBOSS data, while \cref{sec:interpretation}
combines these results, and presents the measured expansion history. 
\cref{sec:conclusion}
concludes with a discussion of these results and a look to future DESI results. \edited{Readers interested only in the results might focus on \cref{sec:results,sec:interpretation,sec:conclusion} and \cref{tab:finaldistance}.}

The analyses here use a fiducial cosmology to convert redshifts into distances; the
BAO distance measurements are relative to this fiducial cosmology. 
Our fiducial cosmology  matches the primary cosmology used in the AbacusSummit suite of simulations \citep{AbacusSummit}, 
which is a \planck\ cosmology \citep{Planck2018}.\footnote{We use the average cosmological parameter values from the 
\texttt{base\_plikHM\_TTTEEE\_lowl\_lowE\_lensing} chains.} We refer the reader to the 
above references\footnote{The cosmologies are also summarized here: 
\url{https://abacussummit.readthedocs.io/en/latest/cosmologies.html}.} for the complete 
specification of the cosmology, but the key parameters are 
$\omega_b = 0.02237$, $\omega_\mathrm{CDM} = 0.1200$, $h=0.6736$, 
$N_\mathrm{ur}=2.0328$ and one massive neutrino with $\omega_\nu = 0.00064420$.

\section{An Overview of the DESI samples and the LSS catalogs}
\label{sec:catalog}

\begin{figure}
\begin{center}
\includegraphics[width=0.6\textwidth]{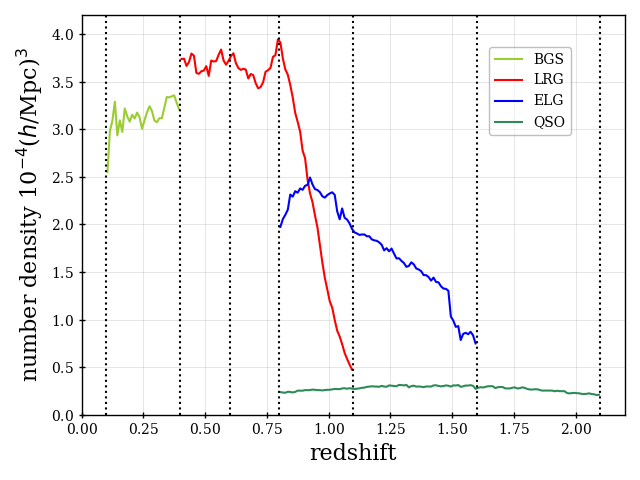}
\caption{The comoving number density as a function of redshift for the samples
used for the \desidrone\ \edited{galaxy/quasar} BAO measurements. Here, we show simply the observed number of redshifts, per comoving volume element.
The vertical lines represent the boundaries of the redshift bins used in this
analysis, except for the QSO sample which is analyzed as a single redshift bin.
}
\label{fig:nz}
\end{center}
\end{figure}

\subsection{\desidrone}
The DESI Data Release 1 (DR1; \cite{DESI2024.I.DR1}) dataset includes
observations using the DESI instrument \cite{DESI_instrument} on the Mayall
Telescope at Kitt Peak, Arizona during main survey operations starting from May
14, 2021, after a period of survey validation \cite{DESIsv}, through to June 14,
2022. \edited{DESI measures the spectra of 5,000 `targets' at once, using robotic positioners to place fibers in the focal plane at the celestial coordinates of the targets  \cite{DESIfocalplane,DESIfa}. The fibers are divided into ten `petals' and carry the light to a corresponding ten climate-controlled spectrographs. } The data was obtained via observations of `tiles', using an observing
strategy meant to prioritize completing observations in a given area of the sky \cite{DESIsops}.
Each tile represents a
specific sky position and set of associated targets \cite{DESItarget} assigned
to each robotic fiber positioner. DESI
dynamically divides its observing time into separate `bright' time and `dark'
time programs, depending on observing conditions. 
2744 tiles were observed in DR1
`dark' time and 2275 tiles were observed in `bright' time. 
Observations of the bright galaxy sample \cite{DESIbgstarget} happen in bright
time while the luminous red galaxies (LRGs \cite{DESIlrgtarget}), quasars (QSO
\cite{DESIqsotarget}), and emission line galaxies (ELGs \cite{DESIelgtarget})
are observed in dark time. 
These data were first processed by the DESI
spectroscopic pipeline \cite{DESIspec} the morning following observations for
immediate quality checks, and then in a homogeneous processing run (internally denoted as
`iron') to produce resulting redshift catalogs used in this paper and will be released in DR1.


\subsection{DR1 Large-scale structure catalogs}
\label{sec:lss}
The redshift and parent target catalogs were processed into large-scale
structure (LSS) catalogs and two-point function measurements as described in
\cite{desilss,DESI2024.II.KP3}. \cref{tab:Y1data} presents the basic
details on the tracer samples used in this paper; the catalogs used for the 
Ly-$\alpha$ BAO measurements are presented separately in \cite{DESI2024.IV.KP6}.
In total, over 5.7 million
unique redshifts are used for  \edited{galaxy and quasar} BAO measurements in DR1, a factor of $\sim3$ increase
compared to SDSS DR16 \cite{SDSS-DR16-cosmology}. 


A key component of the LSS catalogs is the matched random sample (`randoms'), which accounts for the survey geometry. \edited{The randoms were first produced to match the footprint of DESI target samples, as described in \cite{DESItarget}. These were then passed through the stage of the DESI \textsc{fiberassign} code that determines the `potential assignments' for each input random target, using all of the properties of the observed DESI DR1 tiles. The potential assignments do not require that the randoms are allocated a fiber in a full assignment run, they simply determine whether a fiber could reach their angular positions. Thus, this selection is one based on the individual position, and no fiber assignment effects are imprinted on the randoms. This procedure works to the angular scale at which the DESI \textsc{fiberassign} software can predict the focal plan position of targets to be observed, which is better than 1 arcsecond. It is thus far more accurate than trying to sample a pixellated angular mask. The process is detailed in \cite{desilss}. 

These potential assignments are cut to the same combination of `good' tiles and
fibers as the DR1 data samples. Subsequently, veto masks are
applied\footnote{These veto masks remove, e.g., regions influenced by bright
stars or nearby galaxies, area only assignable to higher priority targets, and
regions in  tail ends of the worst imaging conditions.} following the same
process as applied to DR1 data described in \cite{DESI2024.II.KP3}.} In DR1, the
randoms are normalized such that the ratio of weighted data and random counts is
the same in each of the distinct regions relevant to the photometry used to
target the sample. Similarly, the redshift distributions are matched between
data and randoms in each region separately. For all but the QSO sample, there
are two distinct photometric regions: data targeted from BASS/MzLS photometry
\cite{LS.Overview.Dey.2019} in the North Galactic Cap (NGC) and declination
greater than $32.375^\circ$ and those targeted from DECaLS photometry (in both
Galactic caps). For QSOs, the DECaLS sample is further divided into DES and
non-DES regions, as the target selection \cite{DESIqsotarget} is different in
those regions. 

Supporting studies help define and correct for variations in the selection
function due to the effects of imaging systematics on the input target samples
\cite{KP3s1-Zhou,KP3s2-Rosado,KP3s10-Chaussidon} and variations in the DESI
instrument's ability to successfully measure redshifts
\cite{KP3s4-Yu,KP3s3-Krolewski}. Our ability to simulate and correct for
incompleteness in target assignment is presented in
\cite{KP3s7-Lasker,KP3s6-Bianchi,KP3s5-Pinon}. The effects of all of those
issues are combined into a weight column in the data and random catalogs, meant
to be used for any subsequent calculations (i.e., both for reconstruction
and two-point statistics).  The number density of the DR1 DESI
samples varies strongly with both redshift and the number of overlapping tiles
(due to assignment completeness). Thus, `FKP'\footnote{`FKP' is for the three authors of the paper, Feldman, Kaiser, Peacock.} weights, inspired by \cite{fkp},
that are meant to maximize the signal to noise of clustering measurements at the
BAO scale with respect to such number density variations are also included in
all two-point calculations. Their calculation is fully detailed in \cite{DESI2024.II.KP3}.

 \begin{table*}
\centering
\begin{tabular}{|l|c|c|c|c|c|}
\hline
Tracer & redshift range & $N_{\rm tracer}$ & $z_{\rm eff}$   & $P_0(k=0.14)$ & $V_{\rm eff}$ (Gpc$^3$)\\  \hline
\bgs  & $0.1-0.4$ & 300,017 & 0.30  & $\sim 9.2\times10^3$ & 1.7  \\
\lrgo  & $0.4-0.6$ & 506,905 & 0.51  & $\sim 8.9\times10^3$ & 2.6 \\
\lrgt  & $0.6-0.8$ & 771,875 & 0.71  & $\sim 8.9\times10^3$ & 4.0  \\
\lrgth  & $0.8-1.1$ & 859,824 & 0.92 &  $\sim 8.4\times10^3$ & 5.0  \\
\elgo & $0.8-1.1$ & 1,016,340 & 0.95 & $\sim 2.6\times10^3$ & 2.0\\
\lrgelg & $0.8-1.1$ & 1,876,164 & 0.93 &  $\sim 5.9\times10^3$ & 6.5\\
\elgt  & $1.1-1.6$ & 1,415,687 & 1.32 & $\sim 2.9\times10^3$ &  2.7\\
\qso  & $0.8-2.1$ & 856,652 & 1.49 & $\sim 5.0\times10^3$ & 1.5 \\
\hline
\end{tabular}
\caption{\label{tab:Y1data}Statistics for each of the DESI tracer types used for the DESI Y1 BAO measurements presented in this paper. Redshift bins are non-overlapping, except for the shot-noise dominated QSO sample and the $0.8<z<1.1$ LRG and ELG. 
The effective volume calculation, $V_{\rm eff}$ provides a rough estimate for the relative amount of cosmological information in each redshift bin. See text for further details on the calculation of values in this table and \cite{DESI2024.II.KP3} for further details on the samples themselves. 
}
\end{table*}

In what follows,
we provide brief details on the properties of each sample, providing further
context for the statistics of each sample that are presented in  \cref{tab:Y1data}. The $z_{\rm eff}$
values are calculated weighting by the square of the weighted number density of randoms (with the weights---including FKP---described above), $n_{\rm ran}(z)$:
\begin{equation}
z_{\rm eff} =\frac{\int zr^2dr n^2_{\rm ran}(z)}{\int r^2dr n^2_{\rm ran}(z)},
\end{equation}
where $r$ is the comoving distance to the redshift $z$. These 
represent the redshift at which the BAO fit parameters, $\aiso$ and $\alap$, 
can be converted into physical distances (see \cref{sec:interpretation}).
The clustering amplitude 
is determined via a single parameter fit to the post-reconstruction power
spectra at wavenumbers $k<0.07\ihMpc$, assuming the linear matter power spectrum given
by our fiducial cosmology. The effective volume estimate is obtained in each
redshift bin via 
\begin{equation}
    V_{\rm eff} = 
    \int \left[\frac{\bar{n}_{\rm tracer}(z)P_0(k=0.14)}{1+\bar{n}_{\rm tracer}(z)P_0(k=0.14)}\right]^2 dV(z)
    \label{eq:veff}
\end{equation}
where for $P_0(k=0.14)$ we use the values listed in \cref{tab:Y1data}, which are rounded numbers taken from the DR1 $P_0(k)$ measurements.\footnote{Specifically, the ones that remove the effect of angular separations less than 0.05 degrees \cite{KP3s5-Pinon}.}
The choice of $k=0.14 \ihMpc$ was chosen to be most representative for BAO measurements \cite{FontRibera2014}. The comoving number 
density as a function of redshift for each sample is shown in \cref{fig:nz}.






\paragraph{The Bright Galaxy Sample (BGS):} 

The final BGS sample used for our DR1 BAO measurements comprises of $300, 017$ 
redshifts in the interval $0.1 < z < 0.4$ over $7,473 \deg^{2}$, with an assignment completeness of $61.6\%$ (see \cite{DESI2024.II.KP3} for full details). The completed DESI BGS sample is expected to have $\sim80\%$ assignment completeness over 14,000 deg$^2$, implying the DR1 sample contains just over 2/5 of the final DESI BGS sample. The minimum bound of $z=0.1$ is chosen to minimize any effects of bright limits (\cite{BGS.TS.Hahn.2023}) on the target sample. The maximum bound is chosen to separate the sample from the LRG sample (described next).

The nominal BGS sample is flux-limited \cite{DESIbgstarget} and thus has strong variation of its number density with redshift. In order to produce a more homogeneous sample a cut on the DR1 BGS sample was engineered to produce a sample of roughly constant number density of $5 \times 10^{-4} \;\hMpcinvcubed$, for a  sample with 100\% fiber assignment completeness. Thus, the DR1 number density is roughly constant at $3 \times 10^{-4} \;\hMpcinvcubed$, averaged over the DR1 footprint, as can be seen in the light green curve in \cref{fig:nz}. The sample was defined using $k$-corrected $r$-band absolute magnitudes from \cite{fastspecfit,fastspecfit_code} and a correction for evolution that is linear in redshift (matching that used for \cite{DESI2023b.KP1.EDR}). The cuts make the sample a close match to both the LRG number density (for a complete sample) at $z=0.4$ and the clustering amplitude, as can be seen by comparing to the red curve in \cref{fig:nz} and the $P_0$ entries in \cref{tab:Y1data}. The density is high enough to make shot-noise a minor contribution to the BGS statistical uncertainty in the DR1 two-point measurements at BAO scales. See \cite{DESI2024.II.KP3} for more details.

\paragraph{The Luminous Red Galaxy Sample (LRG):} 

The DESI DR1 LRG sample used for BAO measurements consists of $2, 138, 600$ good redshifts over $5,840 \deg^{2}$ in the redshift interval $0.4 < z < 1.1$. The DR1 assignment completeness is $69.2\%$ in the Y1 sample; this is expected to increase to $90\%$  over $14,000 \deg^{2}$ in the complete survey, so the DR1 LRG sample is thus approximately 30\% of the full DESI LRG sample. 

The lower redshift bound was chosen to separate the sample from BGS, as most low-redshift LRG targets are also BGS targets and we are able to match the densities of the two samples at $z=0.4$ (when accounting for assignment completeness). The complete LRG sample has a nearly constant number density over $0.4 < z < 0.8$ of $5 \times 10^{-4} \;\hMpcinvcubed$, which becomes just over $3.5 \times 10^{-4} \;\hinvMpccubed$ when averaged over the DR1 area, as shown in \cref{fig:nz}. The upper bound of $z<1.1$ was chosen to align with the ELG redshift bins. As for previous LRG samples, the clustering is significantly biased compared to the matter distribution, which enhances the BAO signal. The clustering amplitude is approximately constant over the entire redshift range, though there is some evolution in both the clustering amplitude and composition of the sample for $z>0.95$, as documented in \cite{EDR_HOD_LRGQSO2023}, which causes the slight decrease in the $b\sigma_8$ clustering amplitude for the $0.8<z<1.1$ bin.

This sample is further split for the clustering analysis into 3 disjoint redshift ranges, $0.4 < z < 0.6$ (\lrgo), $0.6 < z < 0.8$ (\lrgt), and $0.8 < z < 1.1$ (\lrgth), \edited{with} this higher redshift sample overlapping the low-redshift ELG sample. The redshift binning was chosen in part to align with the SDSS BAO measurements used in \cite{SDSS-DR16-cosmology}. The redshift bin $0.4<z<0.6$ matches one of the SDSS bins. Despite SDSS using $0.6<z<1.0$ for one of their LRG bins, the SDSS sample is dominated by galaxies with $z<0.8$ and their effective redshift is slightly lower than that of our $0.6<z<0.8$ LRG sample ($z_\textrm{eff}=0.7$ compared to $z_\textrm{eff}=0.71$). \edited{This allows us to present a direct comparison between the \desidrone\ measurements and the SDSS measurements (\cref{sec:comparison}).} The $0.8<z<1.1$ redshift bin is aligned with the lower redshift bin chosen for ELGs (described next), which allows a combined LRG+ELG sample to be created and used in the $0.8<z<1.1$ redshift range.

\paragraph{The Emission Line Galaxy Sample (ELG):} 

The DR1 ELG sample defined for clustering analysis in \cite{DESI2024.II.KP3} comprises $2,432,022$ good redshifts in the interval $0.8 < z < 1.6$ over a footprint\footnote{The footprint sizes differ for different dark time tracers due to the differences in the veto masks applied to each; full details are in \cite{DESI2024.II.KP3}.} of $5,914 \deg^{2}$. DESI ELGs are assigned fibers at the lowest priority of any DESI target class, and in DR1, the fiber assignment completeness is only $35.3\%$. This should increase to over 60\% in the final dataset, with the footprint growing to 14,000 deg$^2$. The DR1 sample is thus less than 1/4 of the final DESI ELG sample. The clustering amplitude is the lowest of any of the DESI target classes, consistent with the general knowledge that actively star-forming galaxies are generally less clustered than passive galaxies. These lower clustering amplitudes explain why the effective volume for the $0.8<z<1.1$ ELG sample is significantly lower than the LRG sample in the same redshift bin, despite having more galaxies. One can see from \cref{eq:veff} that there is a leading factor of $P_0^2$. As the DESI survey progresses, the ELG completeness will increase and thus $N_{\rm tracer}/V_{\rm tot}$ will grow larger and make the clustering amplitude less relevant to the effective volume calculation.

The sample is split into 2 disjoint redshift ranges for measuring two-point clustering: $0.8 < z < 1.1$ (\elgo) and $1.1 < z < 1.6$ (\elgt). The split at $z=1.1$ is meant to align with the maximum redshift of the LRG sample. At $z<0.8$, the ELG number density drops below that of the LRG sample, even after accounting for fiber assignment incompleteness (see \cite{DESI2024.II.KP3}).  Further trends with the imaging depth become severe, as shown in \cite{KP3s2-Rosado}. The upper limit of $z<1.6$ is applied as the OII emission line doublet used to secure redshifts for the ELG sample is at a longer wavelength than covered by the DESI spectrographs for $z>1.6$.

\paragraph{Combined LRG and ELG (\lrgelg):}

A combined LRG and ELG sample is used for $0.8<z<1.1$. Ignoring any possibility of sample variance cancellation, the optimal BAO information extracted from data in this redshift range should come from combining the two samples into one, with an appropriate weighting given the difference in clustering amplitude. The construction of this combined sample and its validation are presented in \cite{KP4s5-Valcin} and results from it are denoted `\lrgelg'. Further details, including its motivation, are discussed in \cref{subsec:results-overlapping}.

\paragraph{The Quasar Sample (QSO):} 
The DR1 QSO sample used for BAO measurements consists of $856,652$ good redshifts with $0.8 < z < 2.1$. The sample covers an area of 7,249 deg$^2$, within which the assignment completeness is $87.6\%$. The QSO assignment completeness in the completed DESI survey, which spans 14,000 deg$^2$, is expected to be greater than $99\%$. Thus, the \desidrone\ QSO sample contains just under half of the number expected for the completed survey.

The upper bound of $z<2.1$ is chosen as $z>2.1$ quasars are used as sight-lines along which Lyman-$\alpha$ forest absorption is measured and then used
to obtain BAO measurements, as described in \cite{DESI2024.IV.KP6}. The $z>0.8$ choice is somewhat arbitrary, but aligns both with the lower bound chosen by SDSS and with the redshift bin edge adopted for LRG and ELG. The clustering amplitude of the sample is intermediate between that of the LRG and ELG samples, which makes it the most highly biased sample, given the low $\sigma_8(z_{\rm eff})$ at its effective redshift.










\subsection{Blinding}\label{subsec:blinding}
 In order to protect against confirmation bias in our \desidrone\ \edited{galaxy} BAO measurements, the location of the BAO feature in our measured two-point functions for galaxies and quasars was intentionally shifted from its true values by an unknown amount by applying a blinding shift at the catalog level following the methods proposed in \cite{brieden2020}. Full details of the specific methodology applied to \desidrone\ are presented in \cite{KP3s9-Andrade}.\footnote{The blinding applied to Lyman-$\alpha$ forest BAO studies was different and is described in \cite{DESI2024.IV.KP6}.} We summarize some basic details here. First, $w_0,w_a,f_{\rm NL}$ were randomly chosen from a pre-defined list of values,\footnote{The values were recorded in a combination of files, designed to prevent accidental unblinding, which were then read from in order to keep the $w_0,w_a,f_{\rm NL}$ values the same for all catalog versions. 
 } which were bounded to keep $\alpha_{\rm iso}$ within 3\% of its fiducial value within the redshift range $0.4<z<2.1$. The primordial non-Gaussianity parameter was allowed to vary within $-15<f_{\rm NL}<15$.  The choice of these bounds represented one of the final necessary choices by the DR1 cosmological analysis team prior to obtaining the first (blinded) \desidrone\ BAO measurements. The DESI redshifts were then coherently shifted based on the expected difference in redshift 
between the chosen $w_0,w_a$ and $w_0=1,w_a=0$, at the comoving coordinates initially determined from the true redshift and the fiducial cosmology. Redshifts were also shifted based on estimates of the local density in order to blind structure growth methods. Rather than being drawn randomly, a shift in the growth rate, $f$, was chosen so that it would compensate for the expected change in the monopole of the redshift-space clustering, up to 10\% of the expected change using a linear redshift-space distortions model. The $f_{\rm NL}$ shift was implemented via weights and demonstrated to have no effect on BAO measurements, beyond small random fluctuations from the effective re-weighting. Further description of the \desidrone\ blinding methodology and validation is in \cite{KP3s9-Andrade}.


The cosmology of the shift was kept constant with every catalog update and never revealed. When the DR1 analyses matured to the point where all choices were frozen (see \cref{sec:unblinding} for the tests that were first performed on the shifted data), the LSS catalogs without any shifts applied were then used for the first time, to produce two-point clustering measurements \hsc{(i.e., `unblinded')}. \edited{After the unblinded results were revealed, an error was found in the LSS catalogs related to the completeness weights that are assigned to random points.\footnote{Small regions with no observations were erroneously assigned weights of 0.} The main effect of this on BAO measurements is to re-weight the contributions from different areas of the footprint. The greatest change to the BAO fits was a $-0.7\sigma$ shift in the $\alap$ result for LRGs with $0.4<z<0.6$. The magnitude of all other shifts were less than $0.5\sigma$, and  most were less than 0.2$\sigma$.
Only two pieces of the BAO analysis were updated \hsc{after unblinding}: 1) the final choice on the covariance matrix and 2) the use of the LRG+ELG combined sample in the $0.8<z<1.1$ redshift bin}\hsc{; however, both of these updates were decided/planned before unblinding}.

\section{Mocks}
\label{sec:mocks}

\begin{table}
\centering
\small
    \resizebox{\columnwidth}{!}{%
    \begin{tabular}{|l|r|r|}
    \hline
        Ref   & Task/Paper &  Mocks  \\
    \hline
    \cite{KP4s2-Chen} &  BAO Theory and Modelling Systematics &  \abacusfirst\ cubic, CV	\\
     \cite{KP4s3-Chen}&	Extensive comparison of reconstruction methods &\abacusfirst\ cubic, Y5\\
     \cite{KP4s4-Paillas}  &	Optimal reconstruction of BAO for DESI 2024  & \abacussecond/\ezmock\ DR1 \\
      & and the summary of the unblinding test & \\
     \cite{KP4s5-Valcin}	& Constructing the LRG and ELG combined tracers	& \abacussecond\ DR1 \\
    \cite{KP4s6-Forero-Sanchez} &	Comparison between analytical and EZmock covariance matrices  & \ezmock\ DR1 \\
     \cite{KP4s7-Rashkovetskyi}  &	Analytical covariance matrices for correlation function for DESI 2024 &\abacussecond/\ezmock\ DR1 \\
     \cite{KP4s8-Alves} &	Analytic covariance matrices of DESI 2024 power spectrum multipoles &\abacussecond/\ezmock\ DR1  \\
     \cite{KP4s9-Perez-Fernandez} & Fiducial-cosmology-dependent systematics & \abacussecond\ DR1   \\
    \cite{KP4s10-Mena-Fernandez}  &	HOD-dependent systematics for LRGs & \abacusfirst\ cubic, CV \\
     \cite{KP4s11-Garcia-Quintero}	& HOD-dependent systematics for ELGs &  \abacusfirst\ cubic, CV \\
    \hline
    \end{tabular}
    }
\caption{
The table summarizes the principal set of simulations used for each supporting task/paper and the sections of this paper that make use of them. 
\abacussecond\ and \ezmock\ DR1 duplicate the footprint of the \desidrone, the variation of completeness with target number density, and an approximate fiber assignment effect. The tasks that need to simulate the \desidrone\ survey realism concluded their tests with the DR1 mocks. \abacusfirst~Y5 approximately traces the expected DESI~Y5 footprint. 
 The tasks that needed to test the theoretical systematics with the least level of noise utilized the cubic simulations; in this case, we used \abacusfirst\ cubic instead of \abacussecond, as the former was completed first and was sufficient for our analysis.  CV denotes the control variate technique that was applied to the \abacusfirst\ cubic \citep{Hadzhiyska23}. The CV technique allows to reduce sample variance noise from the two-point statistics by utilizing a highly correlated surrogate based on the Zeldovich approximation.}\label{tab:tasksims}
\end{table}

Realistic and accurate mock simulations form the backbone of our analysis. They allow us to test the limitations of our theoretical models in the presence of non-linear evolution and galaxy-halo physics. In addition, they allow us to assess our ability to mitigate imperfections in our survey caused by effects such as atmospheric conditions, foreground astrophysical systems, and instrument limitations. Building a single set of mock simulations that can be used for all these tests is not feasible because the combination of volume, resolution and number of realizations needed is beyond our current computational resources. Therefore, we build a series of DESI mock simulations focusing on the different aspects of theoretical and observational systematics. A specific set of simulations used for each task is listed in \cref{tab:tasksims}.

We built two kinds of mock simulations. The first called \abacus\ used the high-resolution \textsc{AbacusSummit} simulation suite and hence produced highly accurate non-linear structure. The second called \ezmock\ were computationally very cheap to produce over large volumes and resulted in highly accurate linear scales but non-linear scales that were not that well controlled.

All mocks are produced at the Planck 2018 $\Lambda$CDM cosmology, specifically the mean estimates of the Planck TT,TE,EE+lowE+lensing likelihood chains: $\Omega_c h^2 = 0.1200$, $\Omega_b h^2 = 0.02237$, $\sigma_8 = 0.811355$, $n_s = 0.9649$, $h = 0.6736$, $w_0 = -1$, and $w_a = 0$ \citep{Planck2018}.

\subsection{\abacus }

\textsc{AbacusSummit} is a large suite of high-resolution gravity-only N-body
simulations using the \textsc{Abacus} N-body code \citep{AbacusSummit,
abacusnbody}. These simulations provide us with realizations of the density
field and dark matter halos in cubic boxes with a range of cosmologies. The
entire suite consists of over 150 simulations at 97 different cosmologies. This study makes use of the
25 ``base'' boxes of the \planck\ cosmology, each of which contains
$6912^3$ particles within a $(2h^{-1}$Gpc$)^3$ volume corresponding to a particle
mass of $2.1 \times 10^9$~M$_\odot/h$.\footnote{For more details, see
\url{https://abacussummit.readthedocs.io/en/latest/abacussummit.html}}

The dark matter halos are identified with the {\sc CompaSO} halo finder \citep{compaso}. We also run a post-processing ``cleaning'' procedure to remove over-deblended halos in the spherical overdensity finder, and to intentionally merge physically-associated halos that have merged and then physically separated \citep{2021Bose}. 

The dark matter halo catalogs are then populated with galaxies using an extended halo occupation distribution (HOD) model with the \textsc{AbacusHOD} code \citep{abacushod}.
These \abacus\ mocks were produced
in two generations called \abacusfirst\ and \abacussecond.  The main difference between the two comes from the fact that they were produced at different times to be able to make \edited{early} progress on \edited{testing the analysis pipeline} while we collect more data and improve our model of the \edited{survey and} instrument. The \abacusfirst\ used very early version of the DESI early data release (DESI-EDR)\cite{DESI2023b.KP1.EDR} to find the best fit halo occupation distribution model whereas \abacussecond\ used the final DESI-EDR after correcting for all the systematics and including a detailed model for DESI focal plane effects.

\paragraph{\abacusfirst }
These mocks were produced by fitting the galaxy two-point correlation function 
averaged in angular bins
at small scales using \abacus\ halos and a flexible halo occupation 
distribution model (HOD) \citep{2022AbacusHOD} to populate these halos with galaxies.
We found best-fit HOD parameters at each available snapshot between redshift of 0 to 2 in the \textsc{AbacusSummit} suite. The details of HOD models used are described in \citep{AlamMulti}. We note that satellite galaxies were distributed using NFW profiles fit to the density profile of each halo in the simulation. For the QSO mocks we also included an additional velocity dispersion to account for the significant QSO redshift errors. Each tracer at each redshift is populated over all 25 base boxes, giving a total volume of 200$h^{-3}$Gpc$^{3}$. 

\paragraph{\abacussecond }
The HOD parameters are tuned to the final DESI EDR redshift-space two-point correlation function measurements. We refer the readers to \cite{EDR_HOD_LRGQSO2023, EDR_HOD_ELG2023, EDR_BGS_ABACUS} for the exact HOD models and calibration. The final cubic mocks consists of BGS samples at $z = 0.1, 0.2, 0.4$, LRG samples at $z = 0.5, 0.8, 1.1$, ELG samples at $z = 0.95, 1.1, 1.325$, and QSO samples at $z = 1.1, 1.4, 1.7$. Each tracer at each redshift is populated over all 25 base boxes, for a total volume of 200$h^{-3}$Gpc$^{3}$. We also provide Zeldovich control variates (ZCV) mock simulations with suppressed sample variance for all \textsc{AbacusSummit} realizations (see \cite{Hadzhiyska23} for description of the technique). 

\subsection{\ezmock\ }
In order to generate large simulation volumes for covariance matrices and
pipeline validations, we use the \ezmock\ code \citep{Chuang:2014vfa} which can be calibrated to
accurately reproduce the two- and three-pt clustering on the scales relevant for this
analysis without the cost of a full N-body simulation. 
It has been widely used in eBOSS \citep{Zhao2021} and DESI
\citep{Zarrouk:2020hha}.  

The method comprises two steps: constructing a
dark matter density field and populating galaxy catalogs. The dark matter
density field is based on the 
the Zel'dovich approximation \citep{Zeldovich:1969sb}.
To populate the resulting density field with galaxies, \ezmock\ uses an effective bias model to
account for non-linear evolution and galaxy bias. The latest description of the
effective bias model can be found in \cite{Zhao2021}. We produced two
generations of DESI \ezmock\ by fitting the two-point clustering of \abacusfirst\
and \abacussecond\  in order to give equivalent covariance matrices. We produced
1000 realizations of each generation of \ezmock\ with a box side of 2$h^{-1}$Gpc. These provide the covariance matrix for equivalent 2$h^{-1}$Gpc
\abacus\ mocks. We also produced 1000 realizations of each generation of
\ezmock\ with a box side of 6$h^{-1}$Gpc in order to fit the volume occupied by
the \desidrone\ data without any repetition of structure to validate our 
covariance matrices for the full survey volume.

\subsection{Simulations of DR1}

Both the \abacussecond ~ and 6$h^{-1}$Gpc \ezmock ~have been used to simulate the \desidrone\ LSS dataset \cite{KP3s8-Zhao}. 
A brief outline of the steps to create such mocks is as follows: For both, the first step is to transform the box coordinates to angular sky coordinates and redshifts. Then, the data are sub-sampled as a function of redshift such that the total projected density matches that of the given target sample and the $n(z)$ (after accounting for redshift failures) matches that of the observed DR1 sample \cite{DESI2024.II.KP3}. This provides a simulated DESI target sample. The simulated target sample is then cut so that it covers the same sky area as the DESI target samples. Then, matching the process applied to randoms described in Section \ref{sec:lss}, it is passed through the stage of the DESI \textsc{fiberassign} code that determines the `potential assignments' for each simulated target, using all of the properties of the observed DESI DR1 tiles. These potential assignments are cut to the same combination of `good' tiles and fibers as the DR1 data samples. Subsequently, veto masks are applied following the same process as applied to DR1 data described in \cite{KP3}.

The process described above reproduces the small-scale structure of the DESI DR1 footprint, but does not impart any incompleteness within it. For the \abacussecond, mock LSS catalogs (hereafter mocks) were produced with three variations in the fiber assignment completeness. These are:
\begin{itemize}
    \item The `{\bf complete}' mocks that have no assignment incompleteness added and thus can be used as a baseline comparison for understanding the effect of the incompleteness. 
    \item The `{\bf altmtl}' mocks that represent our most realistic simulations of the DR1 data. They apply the process described in \cite{KP3s7-Lasker} to apply the DESI \textsc{fiberassign} code to tiles in the same ordering and cadence in a feedback loop to the target list as occurred for the observed data. The process was demonstrated to perfectly reproduce DESI fiber assignment on real DESI targets, with no approximations. 
    \item The `{\bf fast-fiberassign}' mocks that emulate the fiber assignment process by sampling from the average targeting probability of the galaxies multiple times, learned from the data as a function of number of overlapping tiles and local (angular) clustering. The final sample is obtained by recombining the multiple realisations in such a way that deliberately creates a small-scale exclusion effect, which approximately reproduces the fiber-collisions pairwise incompleteness. 
    The process is much faster than the \altmtl\ and is described and validated in \cite{KP3s6-Bianchi}. 
    
\end{itemize}
The computation time required for the \altmtl\ mocks prohibits it from being run on all 1000 \ezmocks. Thus, we apply only the fast-fiberassign process to the \ezmocks.

All flavors of mocks go through the process of assigning redshifts and weights to randoms in the same way as for the real data samples and are normalized within the same regions, etc (e.g., all integral constraints effects, e.g., described in \cite{deMattia19IC}, are the same between data and mock LSS catalogs) following the prescription in \cite{DESI2024.II.KP3}.
More details about the creation and validation of the different mock flavours can be found in \cite{KP3s8-Zhao}.

\section{Methods}
\label{sec:methods}


This section summarizes the various methods used in the BAO analyses that follow. We refer the reader
to the referenced supporting papers for more detail and validations. 

\subsection{Two-point function codes}
\label{sec:2pt}
The BAO measurements derive from the two-point clustering statistics of the data, the correlation function in configuration-space and the power spectrum in
Fourier space. The techniques for computing these are now well established. We
use the Landy-Szalay estimator \cite{Landy1993} for the correlation functions
(modified as in \cite{Padmanabhan12} for the reconstructed data) and an FKP
based estimator (\cite{yamamoto2006,Hand2017:1712.05834v1}) for the power
spectrum. A more detailed discussion of our particular implementations can be
found in \cite{DESI2024.II.KP3}. The specific codes are
\textsc{pycorr}\footnote{\url{https://github.com/cosmodesi/pycorr}}  for
correlation functions and
\textsc{pypower}\footnote{\url{https://github.com/cosmodesi/pypower}} for power
spectra. We compress the angular dependence (to the line of
sight) into Legendre multipoles; our analyses rely on the $\ell=0$ (monopole)
and $\ell=2$ (quadrupole) components. 
The galaxies are weighted by terms to account for the selection 
function and to optimally measure two-point statistics (FKP weights), 
both summarized in \cref{sec:catalog} and fully defined in
\cite{KP3}, unless otherwise noted.

Since the clustering measurements are consistent for both Galactic caps, we combine 
these measurements when constructing our data vectors.
In configuration-space, the combination is performed by
summing the pair counts computed in each region independently. Similarly, the
power spectrum estimates are obtained for each Galactic cap and the measurements
are then combined by averaging the two power spectra, weighting by their respective normalizations \cite{DESI2024.II.KP3}.
The number
of randoms used is more than 50$\times$ the size of the data for all correlation
functions measurements and more than 100$\times$ for all power spectra\footnote{The large numbers of randoms make the statistical fluctuations from the random catalog negligible.}. 

The DESI fiber assignment imprints structure into the two-point statistics,
especially on small scales. This can be mitigated very effectively by removing
small-angle/small-separation pairs in the two-point statistics
\cite{KP3s5-Pinon}. While such an approach is necessary for analyses that use
the full shape of the two-point function, we expect these fiber assignment issues to have no measurable
impact on our extraction of the BAO distance scales. This is both a result of
the fact that the BAO feature is at large scales, and that we marginalize out
the overall shape of the two-point function in our analysis. We validate this with mocks with and without fiber assignment and find no impact at greater than 2$\sigma$ significance (\cref{subsec:obssys}). Note that this does
not include accounting for the overall completeness, which we do in our galaxy
weights. 
Given this insensitivity,
we do not include any additional corrections for fiber assignment in our analysis.

\subsection{Density-field reconstruction}\label{subsec:methods-recon}

Density-field reconstruction \cite{Eisenstein07} is now a well-established element of galaxy BAO analyses, as it robustly eliminates biases due to the nonlinear evolution of the density field and improves the statistical precision of the BAO method. Beyond the standard reconstruction method proposed in \cite{Eisenstein07}, which has been widely applied to observational datasets \cite{Padmanabhan12, Anderson12, Tojeiro2014:1401.1768, Kazin2014:1401.0358, Alam17, Gil-Marin2020}, there are several improved reconstruction algorithms that have been proposed in the literature \cite{Seo2010:0910.5005, Schmittfull2017:1704.06634, Hada2018:1804.04738}. Although these methods have significant promise for reconstructing the linear density field at small scales \hsc{at the very low shot noise regime,} the improvements in the BAO distance measurements are marginal \hsc{at the galaxy number densities of \desidrone}. Considering this and the robustness and simplicity of the original method, we restrict ourselves to using it for this work, with the modifications described below.

\begin{itemize}
    \item One of the main differences from previous applications of
    reconstruction in SDSS is the use of the \recsym\ convention \cite{White15}.
    \recsym\ shifts the tracers and randoms from the LSS catalogs in the same
    way using the redshift-space displacement, preserving RSD in the
    post-reconstruction clustering. The \reciso\ convention
    \cite{Padmanabhan12} used in BOSS and eBOSS approximately removes 
    RSD, resulting in more isotropic clustering post-reconstruction.
    We adopt the \recsym\ convention as our baseline since it is the choice
    that fully removes the nonlinear damping and shift of the BAO due to
    large-scale modes \cite{KP4s2-Chen} and avoids artefacts in the correlation function 
    on small scales.
    However, \cite{KP4s4-Paillas} shows that the DESI BAO constraints
    in practice are rather insensitive to this choice (a similar conclusion was
    found in \cite{SeoBAOmodel2016}).
    \item We tested the sensitivity of the reconstruction method to the choice of scale that is used to smooth the density field. Using the blinded DESI data and mocks that match the expected clustering properties of \desidrone, we determined the optimal smoothing scale to be used when reconstructing each DESI target sample, as described in \cite{KP4s4-Paillas} and presented in \cref{tab:reconstruction_parameters}.
    
\end{itemize}

\hsc{Our reconstruction uses \textsc{pyrecon},\footnote{\url{https://github.com/cosmodesi/pyrecon}} a \textsc{Python} package developed by the DESI collaboration. This comprehensive toolkit offers a diverse range of reconstruction algorithms and accommodates various conventions, and provides the flexibility to process periodic-box simulations or survey data with non-uniform geometries. } 
In terms of the numerical implementation of the reconstruction method, we adopt an efficient algorithm based on iterative Fast Fourier Transforms introduced in \cite{Burden2015:1504.02591v2} as our baseline, which we find highly consistent with the output from \textsc{MultiGrid}.\footnote{An algorithm that follows the multigrid relaxation technique with a V-cycle based on damped Jacobi iteration.} 
\hsc{The iterative Fast Fourier Transform we adopt is different from the method in eBOSS. In eBOSS \cite{Bautista2021,SDSS-DR16-cosmology}, the method iteratively updates the locations of the galaxy and  random particles, which we will denote as \textsc{IFFTP}, while the method we adopt for \desione\ iteratively updates the density of given meshes (hereafter, simply \textsc{IFFT}). This is the first time that the IFFT method (the latter) has been implemented in data analysis. An extensive comparison of reconstruction algorithms in the context of DESI is presented in \cite{KP4s3-Chen}.}

\begin{table}
    \centering
    \vspace{0.5em}
    \begin{tabular}{|c|c|c|c|c|}
    \hline
    Tracer & Redshift range & Linear bias $b$ & Growth rate $f$ & Smoothing scale $\Sigma_{\rm sm}$\\
    \hline
    \bgs & 0.1--0.4 & 1.5 & 0.68 & $15\, h^{-1}{\rm Mpc}$\\
    \lrgs & 0.4--1.1 & 2.0 & 0.83 & $15\, h^{-1}{\rm Mpc}$ \\
    \elgs & 0.8--1.6 & 1.2 & 0.90 & $15\, h^{-1}{\rm Mpc}$ \\
    \qso & 0.8--2.1 & 2.1 & 0.93 & $30\, h^{-1}{\rm Mpc}$ \\
    \hline
    \end{tabular}
    \caption{Redshift range, tracer linear bias, growth rate of structure, and smoothing scale assumed when reconstructing each DESI target sample. The smoothing scale was chosen after testing different values for each tracer, as detailed in \cite{KP4s4-Paillas}.
    }
    \label{tab:reconstruction_parameters}
\end{table}

\cref{tab:reconstruction_parameters} summarizes the different hyperparameters that we calibrated when reconstructing the \desidrone\ samples. The catalogs were reconstructed across the entire redshift range of each tracer simultaneously, assuming a value of the growth rate of structure determined by our fiducial cosmology and the effective redshift of each sample. 


\subsection{\edited{Defining the clustering model}}\label{subsec:methods-fitting}

We next describe the BAO fitting method used in the galaxy DR1 analysis. We design this method to fully isolate the BAO feature within the broader two-point clustering measurements by combining a physically motivated theory model from quasi-linear theory and a parameterised model to marginalise over non-linearities that may otherwise affect our measurements of the BAO scale. The design of our method is based on detailed past investigations, starting from the original works of \cite{Eisenstein07} with improvements through the eras of BOSS and eBOSS \cite{Anderson12,Anderson14,VargasMagana2014,VargasMagana2016:1610.03506v2,Ross17,Beutler17,Gil-Marin2020,Bautista2021}. Although these references have demonstrated the BAO fitting methodology to be robust at a level beyond that required for these previous surveys, there remained some inconsistencies in the modelling of the power spectrum and correlation function, and some arbitrariness in the choice of free parameters dependent upon the specific configuration and signal to noise of the measurements. As such, the modelling choices motivated or adopted in previous works have all been revisited in \cite{KP4s2-Chen} to ensure a robust fit for the \desione\ results, using a method based on the allowed physical degrees of freedom that can also be consistently used for future surveys without significant modification. We have taken particular care to develop a more consistent modelling in Fourier and configuration-space, and to better motivate (or remove the need for) different modelling choices. Where such choices remain, we have quantified their systematic differences in the BAO constraints (see Section~\ref{sec:systematics}). In this subsection, we provide an overview of the official DESI galaxy BAO modelling prescription and summarise those changes.

\subsubsection{Fourier-space fitting framework}

Our generic model for the galaxy power spectrum as a function of scale $k$ and (cosine) angle $\mu$ in the ``true'' cosmology can be written as
\begin{equation} \label{eq:generic_model}
    P(k, \mu)= \mathcal{B}(k, \mu) P_{\rm nw}(k) + \mathcal{C}(k, \mu)P_{\rm w}(k) + \mathcal{D}(k)\,,
\end{equation}
where $P_{\rm nw}(k)$ and $P_{\rm w}(k)$ denote the smooth (no-wiggle) and BAO (wiggle) components of the linear power spectrum, respectively, which are obtained using the \textit{peak average} method from \cite{Brieden2022}. The linear matter power spectrum template is predicted from \textsc{class}\footnote{\url{https://github.com/lesgourg/class_public}} using our fiducial cosmology (see \cref{sec:intro}). Generally, the term $C(k,\mu) P_{\rm w}(k)$ encompasses the BAO component we are interested in, damped by non-linear galaxy motions \cite{Eisenstein07}; while $\mathcal{B}(k, \mu) P_{\rm nw}(k)$ uses quasi-linear theory to model the smooth component of the galaxy clustering, and $\mathcal{D}(k)$ is our parametric model to account for additional non-linearities and observational effects. This model generally matches that used in previous BAO analyses \cite{Anderson12,Anderson14,Kazin2014:1401.0358,Alam17}, but the exact forms of each component differ and so will be discussed later in this section.

In order to make contact with the apparent size of the BAO seen in the fiducial cosmology, $C(k,\mu) P_{\rm w}(k)$ is evaluated at \cref{eqn:alpha_defs},
\begin{align}
    k^\prime  =  \frac{\alpha_\text{AP}^{1/3}}{\alpha_\text{iso}}\left[ 1+\mu^2_{\rm obs} \left( \frac{1}{\alpha_\text{AP}^2}-1 \right) \right]^{1/2} k_{\rm obs}
    \label{eq:dilationk}
\end{align}
and
\begin{align}
    \mu^\prime  = \frac{\mu_{\rm obs}}{\alpha_\text{AP}}\left[ 1+\mu_{\rm obs}^2 \left( \frac{1}{\alpha_\text{AP}^2}-1 \right) \right]^{-1/2},
\label{eq:dilationmu}
\end{align}
where the subscript `obs' is used to distinguish between \textit{observed} coordinates and measurements (assuming a fiducial cosmology) and those in the (unknown, but to be constrained) \textit{true} cosmology. Note that this transformation is not just a coordinate transformation between the true and fiducial cosmologies but also includes a rescaling of the BAO template from the template cosmology, reflecting that what is measured is the \textit{apparent} size of the BAO relative to the template sound horizon. The true, fiducial and template cosmologies need not be the same (\cite{KP4s9-Perez-Fernandez}, see also \cref{subsec:sys-fiducialcosmo}), but the latter two are usually equated for simplicity. The rescaling in \cref{eq:dilationk,eq:dilationmu} in principle should not apply to the non-BAO parts of the power spectrum. In order to prevent accidentally using broadband information in the smooth component in the \desione\ analysis, we hence evaluate the smooth components directly at the observed coordinates, without dilation. The model for the power spectrum multipoles is thus
\begin{align}
    P_{\ell, \rm obs}(k_{\rm obs}) = \frac{2\ell+1}{2} \int_{-1}^1 &d\mu_{\rm obs}\ \mathcal{L}_\ell(\mu_{\rm obs}) \big[\mathcal{B}(k_{\rm obs}, \mu_{\rm obs}) P_{\rm nw, obs}(k_{\rm obs}) \nonumber \\
    &+ \mathcal{C}(k^\prime(k_{\rm obs},\mu_{\rm obs}), \mu^\prime(k_{\rm obs},\mu_{\rm obs}))P_{\rm w}(k^\prime(k_{\rm obs},\mu_{\rm obs}))\big] + \mathcal{D}_\ell(k_{\rm obs}).
    \label{eq:pow_spec_multipoles}
\end{align}
This is similar to that used in \cite{Beutler17}, but with the key difference that the term $\mathcal{B}(k, \mu) P_{\rm nw}(k)$ is not dilated. As a final technicality, we note that our rescaling is applied to the entire BAO component; strictly speaking the prefactor $C(k,\mu)$ is not subject to the exact same rescaling as $P_{\rm w}$, but the differences will be degenerate with the free parameters within $C(k,\mu)$ itself. We will now describe each remaining model component in turn.

Following previous BAO studies \cite{Seo2016, Beutler17}, we adopt the following parametric form for $\mathcal{B}(k, \mu)$: 
\begin{equation}
    \mathcal{B}(k,\mu) = \left(b_{1}+f\mu^{2}[1 - s(k)]\right)^{2} F_{\rm fog}\,,
    \label{eqn:Bkmu}
\end{equation}
where $F_{\rm fog} = \left(1 + \frac{1}{2} k^2\mu^{2} \Sigma_s^2\right)^{-2}$ accounts for the `Fingers of God' effect due to halo virialization \citep{Jackson72,Park94} via a single free smoothing scale $\Sigma_{s}$. The term $\left(b_{1}+f\mu^{2}[1 - s(k)]\right)^{2}$ is a generalised form of the Kaiser factor \citep{Kaiser1987} that also accounts for impact of reconstruction; here $b_{1}$ is the linear galaxy bias for the particular sample we are fitting, while $f$ is the linear growth rate of large scale structure, both of which are free parameters in our model. For pre-reconstruction and the \textbf{RecSym} convention, $s(k) = 0$, while for \textbf{RecIso}, $s(k) = \exp\left[-(k \Sigma_{\rm sm})^2/2 \right]$ and $\Sigma_{\rm sm}$ is the smoothing scale we applied to the density field during the reconstruction process.

The function $\mathcal{C}(k, \mu)$ captures the anisotropic non-linear damping of the BAO feature on top of linear theory. Similarly to the smooth component above we have that it takes the form (\cite{KP4s2-Chen}, see also ref.~\cite{Sugiyama24})
\begin{equation}
    \mathcal{C}(k,\mu) = \left(b_{1}+f\mu^{2}[1 - s(k)]\right)^{2}\exp\left[-\frac{1}{2}k^2\biggl(\mu^{2}\Sigma_{||}^2 + (1-\mu^{2})\Sigma^{2}_{\perp}\biggl)\right]
    \label{eq:baodamping}
\end{equation}
where $\Sigma_{||}$ and $\Sigma_{\perp}$ model the damping for modes along and perpendicular to the line of sight. The FoG factor in Equation~\ref{eqn:Bkmu} is dropped here due to its high correlation in fits with the damping parameters \cite{KP4s2-Chen}. In \textbf{RecIso} two caveats apply: (a) the smoothing kernel $s(k)$ is always evaluated in the observed coordinates $k_{\rm obs}$, since it is defined in the fiducial cosmology and (b) the simple exponential form here is approximate and only holds on small scales where the contribution from the randoms is negligible. At intermediate scales the damping due to long-wavelength modes takes on a more complex form since the randoms and galaxies are displaced by different amounts, which is one of the reasons we choose \textbf{RecSym} as our default convention, as it is the unique choice that removes the nonlinear damping and shift due to long-wavelength modes. However, we note that past BAO measurements have often empirically employed the exponential form along with the prefactor in \cref{eqn:Bkmu} for \textbf{RecIso}, and we will continue to do so here in tests involving this scheme.

Finally, the $\mathcal{D}(k)$ factor captures any deviation from linear theory in the broadband shape of the power spectrum multipoles. Past analyses have used a polynomial form for this \cite{Anderson12,Anderson14,Beutler17,Ross17,Gil-Marin2020,Bautista2021}, although a single exact equation cannot be provided here for comparison due to different studies using polynomial equations with different numbers of terms. To improve on this, in \desidrone\ we instead parameterize it using a spline basis with bases separated by a single user defined scale $\Delta$,
\begin{equation} \label{eq:spline_basis}
    \mathcal{D}_\ell(k) = \sum_{n=-1}^{n_{\mathrm{max}} }a_{\ell,n} W_3\left(\frac{k}{\Delta} - n\right)\,
\end{equation}
where $W_3$ is a piecewise cubic spline kernel \citep{Chaniotis2004, Sefusatti2016} and $n_{\mathrm{max}}$ sets the number of broadband terms we consider.
A suitable $\Delta$ is chosen under the premise that the spline basis is able to match the broadband shape of the power spectrum without reproducing the BAO wiggles themselves. This sets a limit on the choice of $\Delta$ to be larger than half the BAO wavelength ($\pi/r_{\mathrm{d}}$ where $r_{\mathrm{d}}$ is the sound-horizon at the baryon-drag epoch). We hence use twice this minimum value, $\Delta = 2 \pi / r_{\mathrm{d}} \simeq 0.06\hmpcinv$. The number of broadband terms $n_{\mathrm{max}}=7$ for our default fitting procedure then arises from considering how many spline terms of width $\Delta$ are required to fully span the $0.02\hmpcinv < k < 0.30\hmpcinv$ range of our power spectrum measurements, i.e., there is no need to specify a choice for $n_{\mathrm{max}}$. Unlike previous analyses, our new method hence provides a more physically motivated and less arbitrary broadband model.

Finally, the model multipoles need to be convolved by the data window function. This can be accomplished via matrix multiplication $\tilde{P}_{\tilde{\ell}}(\tilde{k}_i) = \sum_j W_{\tilde{\ell}\ell}(\tilde{k}_i,k_j)P_{\ell}(k_j)$ which can be compared directly to the data vector. This follows previous approaches \cite{Ross2013, Beutler21}. To ensure accuracy in this convolution, the unconvolved model is evaluated at $k$-points within $0.001 \hmpcinv < k < 0.35 \hmpcinv$ and separated by $\Delta k=0.001\hmpcinv$. The computational binning is thus five times finer than the $k$-bins of the data vector $\tilde{k}$, while covering a larger $k$-range. For the theory input to $P_\ell(k_j)$ we include angular dependence up to the hexadecapole. This is because power from these higher order moments can still `leak' into the observed multipoles (monopole and quadrupole, or monopole only in case of 1D fits) due to the convolution with the window function $W_{\tilde{\ell}\ell}(\tilde{k},k)$ --- in principle higher order multipoles can also enter but their contributions are negligible on the scales we fit. 

In summary, our model for the BAO can be evaluated for comparison to data using Eq.~\ref{eq:pow_spec_multipoles}, with the functional forms and free model parameters described in Eqs.~\ref{eqn:Bkmu}-~\ref{eq:spline_basis}. Fitting for these free parameters proceeds as detailed in Section~\ref{sec:model_fitting}.


\subsubsection{Configuration-space framework}

Our approach for modelling the correlation function very closely follows that for the power spectrum, more so than in previous works \cite{Anderson14,Ross17}. We start with the power spectrum multipoles in fiducial coordinates obtained from \cref{eq:pow_spec_multipoles}, before the window function convolution, and \textit{without} the $\mathcal{D}_\ell(k)$ terms. These multipoles are Hankel-transformed to configuration-space to yield the correlation function multipoles in the same coordinates
\begin{equation}
    \xi_{\ell, \rm obs}(s) = i^\ell  \int_0^\infty \frac{dk\ k^2}{2\pi^2} j_\ell (ks) P_{\ell, \rm obs} (k) \,,
\end{equation}
where $j_\ell$ are the spherical Bessel functions. As a direct transform of the power spectrum, the correlation function model hence also contains the same BAO dilation parameters, BAO and Fingers of God damping parameters, linear galaxy bias and growth of structure, with the same physical interpretation. We evaluate our theory model in narrow bins of $\Delta s=1\hinvmpc$. We match our wider $\Delta s=4\hinvmpc$ measurement binning by averaging the theory, weighted by the number of random - random pair counts in each fine $\Delta s=1\hinvmpc$ bin.

For the remaining broadband modelling, we Hankel-transform the same spline basis functions $\mathcal{D_{\ell}}(k)$ as used for the power spectrum.  However, all of these except for the $n=[0,1]$ terms of ~\cref{eq:spline_basis} in the quadrupole quickly go to zero on large scales, and so do not need to be included given the choice of fitting scales we use in configuration-space model. These two terms are explicitly evaluated as
\begin{equation}
    \Delta^3 B_{2,n}(s\Delta) = i^2 \int \frac{dk\ k^2}{2\pi^2} \ \,W_3\left(\frac{k}{\Delta} - n\right)\ j_2(k s)
    \label{eqn:corrspline}
\end{equation}
for $n=0$ and 1.
However, we also introduce two additional nuisance terms for each multipole to account for the potential impact of uncontrolled large-scale data systematics. Such effects can be confined to $k < k_{\rm min}$ in Fourier space and removed by truncating the range of scales used in our fit to the power spectrum data. However, in the configuration-space, these are modelled using
\begin{equation}
    \tilde{\mathcal{D}}_\ell(s) = {b}_{\ell,0} + {b}_{\ell,2} \left( \frac{s k_{\rm min}}{2\pi}\right)^{2}.
\end{equation}
with $k_{\mathrm{min}}=0.02\ihMpc$. In summary, the configuration-space broadband terms comprise of:
\begin{align}
    \tilde{\mathcal{D}}_{0}(s) &= b_{0,0} + b_{0,2} \left( \frac{s k_{\rm min}}{2\pi}\right)^{2}\, , \\
    \tilde{\mathcal{D}}_{2}(s) &= b_{2,0} + b_{2,2} \left( \frac{s k_{\rm min}}{2\pi}\right)^{2} + \Delta^3 \left( a_{2,0}B_{2,0}(s\Delta) + a_{2,1} B_{2,1}(s\Delta) \right)
\end{align}
with $B_{2,0}(s\Delta)$ and $B_{2,1}(s\Delta)$ given by \cref{eqn:corrspline}.



\subsubsection{Polynomial-based broadband modeling}
\label{subsubsec:polymodel}
For select comparisons with previous BOSS/eBOSS literature, we will occasionally compare our DESI results using the default spline-based broadband model with a more traditional polynomial-based model. In Fourier space, this latter choice consists of writing $\mathcal{D}_{\ell}(k)=\sum_{n=-1}^{3}a_{\ell,n}k^{n}$. In configuration-space, this is $\tilde{\mathcal{D}}_{\ell}(s)=\sum_{n=0}^{2}\tilde{a}_{\ell,n}s^{-n}$. Note that even when using this polynomial option, we still include all the other improvements made in DESI 2024, except for the different broadband functions. 

\subsubsection{Main differences between DESI 2024 and previous BAO modelling}\label{subsec:sdssvsdesifitting}

Here we provide a summary of the most important differences in our DESI 2024 BAO modeling compared to the previous methods used in BOSS and eBOSS \cite{Anderson12,Anderson14,Ross17,Beutler17,Gil-Marin2020,Bautista2021}. This covers only those differences we deem most relevant to the interpretation of our fitting results, or that contribute to our systematic error budget. A more comprehensive list including more minor changes can be found in \cite{KP4s2-Chen}. Overall, we find that the sum total of the differences between our new methodology and that used in BOSS/eBOSS result in only a small systematic error, far below \desidrone precision (See \cref{subsec:sys-theory}).

\begin{enumerate}
    \item{As described in \cref{subsec:methods-recon}, our BAO fitting model is calibrated for \textbf{RecSym} as a fiducial reconstruction procedure. BOSS/eBOSS used the \textbf{RecIso} convention, with caveats as described around Eq.~\ref{eq:baodamping}. The \textbf{RecSym} form of reconstruction is more physically motivated, and leads to a much simpler form for the BAO damping. We note however that \cite{KP4s4-Paillas} demonstrate consistent constraints on \desidrone\ blinded data even when using \textbf{RecIso}.}
    \item{BOSS/eBOSS most frequently used a polynomial-based broadband (\cref{subsubsec:polymodel}, although with varying numbers of free-parameters between different studies). We prefer to use a spline-based model to marginalise over non-linear physics in the broadband as this leads to a less arbitrary choice in the number of free parameters, and greater consistency between the power spectrum and correlation function. Though we believe our new approach to be better, we detect small differences between the two approaches for $\alphaap$ which we adopt into our systematic error budget.}
      \item{When including FoG damping, BOSS/eBOSS analyses usually applied this equally to both the wiggle and no-wiggle components (i.e., \cite{Ross17,Beutler17}). Ref.~\cite{KP4s2-Chen} (figure 12 therein) demonstrated that this introduces a degeneracy between the FoG and BAO damping parameters. We hence include FoG damping \textit{only} in the no-wiggle component, which removes this degeneracy. The freedom we give to the two BAO damping parameters is sufficient to account for non-linearities that can move them away from their theoretical values, and we find no systematic differences between the two methods.}
     \item{We treat the correlation function model purely as the Hankel transform of the power spectrum, applying all our modelling choices and BAO dilation to the power spectrum first. The exception is the broadband terms, which are applied separately to the power spectrum and correlation function; nonetheless, the form of these for the correlation function is still based purely on what one obtains from Hankel transforming the spline-based broadband model for the power spectrum. This is an improvement in the consistency of the modelling compared to \cite{Anderson14, Ross17}. 
     Nonetheless, we tested both methods on our mocks and data and conservatively adopt the small difference in $\alphaap$ into our systematic error budget.}
\end{enumerate}

\subsection{\edited{Fitting the clustering data}}
\label{sec:model_fitting}

The \edited{galaxy} DESI 2024 BAO results in this work are obtained using \textsc{desilike},\footnote{\url{https://github.com/cosmodesi/desilike}} a \textsc{python} package that provides a common framework for writing DESI likelihoods. \edited{The BAO theory and likelihood is implemented in JAX~\citep{jax2018github}\footnote{\url{https://jax.readthedocs.io/en/latest/}}. Even though gradient-based sampling methods were implemented, we found that with analytic marginalization over broadband parameters that leaves a few sampled parameters, and using Jax just-in-time compilation and parallelization capabilities, the ensemble sampler \textsc{emcee}~\citep{emcee}\footnote{\url{https://github.com/dfm/emcee}} provided well-sampled posterior estimates in a just a few minutes. In addition to MCMC sampling, we also perform posterior profiling using \textsc{desilike}'s wrapping of Minuit \citep{minuit}.}


During the course of this work, we also used/developed a fully independent galaxy BAO fitting pipeline (\textsc{Barry}\footnote{\url{https://github.com/Samreay/Barry}} \cite{Hinton2020}), for some of the supporting papers and with which we tested the consistency of our results prior to unblinding. This latter code was also used to demonstrate that our results are independent of the choice of MCMC/Nested sampling algorithm used \cite{KP4s2-Chen}. We adopt \textsc{desilike} as our official pipeline, owing to the greater computational speed offered by its JAX implementation, and its better integration within the wider set of DESI pipelines used for producing the clustering measurements (\cref{sec:2pt} \cite{DESI2024.II.KP3}), and for the cosmological interpretation of our BAO constraints \cite{DESI2024.VI.KP7A}.

For our default fitting, we adopt Gaussian priors on the BAO and Finger-of-God damping parameters and flat priors for all other model parameters. We parameterize the linear RSD though a parameter $d\beta = f / f_\mathrm{fid}$ where $f_\mathrm{fid}$ is the fiducial value for the growth rate at the effective redshift of the sample. Furthermore, $a_{0, n}, a_{2, n}$ are scaled by the amplitude of the fiducial no-wiggle power spectrum at the pivot points $k_{p} = n \Delta$ before injecting them into \cref{eq:spline_basis}. All of these are simple rescalings of the parameters described in \cref{sec:model_fitting} to improve the convergence of the fitting. 

All our priors are listed in \cref{tab:priors}. The choices for the Gaussian priors, particularly their central values, were informed by a combination of theoretical calculations, measurements of the cross-correlation between the initial and post-reconstruction density fields in \edited{\abacussecond\ DR1} simulations, and by running fits to mock catalogs for each tracer. By fitting the average over many mocks, we ensured that the signal-to-noise ratio is large enough to let the damping parameters vary freely during the fit. The resulting central values for the priors are given in \cref{tab:priors2}. In \cref{sec:results}, we show that the recovered BAO parameters from fits to \desidrone\ are largely insensitive to this choice of priors.

We perform two-dimensional fits to monopole and quadrupole data for \lrgs\ and \elgt. For the \bgs, \elgo, and \qso\ samples, we perform only one-dimensional fits to the monopole due to their relatively noisier clustering measurements \edited{based on the unblinding tests detailed in \cref{sec:unblinding}}. In these 1-D cases, $\alphaap$ and $d\beta$ are set to unity.

\begin{table}
    \centering
    \resizebox{\columnwidth}{!}{%
        \begin{tabular}{|l|c|c|c|}
            \hline
           Parameter & $P(k)$ prior & $\xi(r)$ prior & Description \\ \hline
           $\alphaiso$ &[0.8, 1.2]  &[0.8, 1.2]  & Isotropic BAO dilation. \\
          $\alphaap^*$ & [0.8, 1.2] & [0.8, 1.2] &  Anisotropic (AP) BAO dilation. \\
           $\Sigma{\perp}$ & $\mathcal{N}(\Sigma^{\mathrm{fid}}_{\perp},1.0)$ &$\mathcal{N}(\Sigma^{\mathrm{fid}}_{\perp}, 1.0)$ & Transverse BAO damping [$\hinvmpc$]\\
            $\Sigma{\parallel}$ & $\mathcal{N}(\Sigma^{\mathrm{fid}}_{\parallel},2.0)$ &$\mathcal{N}(\Sigma^{\mathrm{fid}}_{\parallel}, 2.0)$ & Line-of-sight BAO damping [$\hinvmpc$]\\
            $\Sigma_s$ & $\mathcal{N}(2.0,2.0)$ & $\mathcal{N}(2.0,2.0)$ & Finger of God damping [$\hinvmpc$] \\
            $b_1$ & $[0.2, 4]$ & $[0.2, 4]$ & Linear galaxy bias \\
            $d\beta^*$ & $[0.7, 1.3]$ & $[0.7, 1.3]$ & Linear RSD parameter\\
            $a_{0, n}$ & $\mathcal{N}(0, 10^4)$ & N/A  & Spline parameters for the monopole\\        
            $a_{2,n}^*$ & $\mathcal{N}(0, 10^4)$ & $\mathcal{N}(0, 10^4)$ & Spline parameters for the quadrupole\\
            $b_{0,n}$ & N/A & [$-\infty, \infty$]  & Unknown large scale systematics\\
            $b_{2,n}^*$ & N/A & [$-\infty, \infty$]  & Unknown large scale systematics\\
            Fitting range &[0.02, 0.3]$\ihMpc$ & [48, 152]$\hMpc$ & Measurement bin edges \\
            Data binning & $0.005\ihMpc$ & $4\hMpc$ &Measurement bin width \\
        \hline
        \end{tabular}
    }
    \caption{The free parameters and their priors for Fourier-space (FS) and configuration-space analyses. $\mathcal{N}(\mu, \sigma)$ refers to a normal distribution of mean $\mu$ and standard deviation $\sigma$, $[x_{1}, x_{2}]$ to a flat distribution between $x_{1}$ and $x_{2}$ inclusive. For faster fitting and convergence, we re-parameterise our linear RSD parameter using $d\beta = f / f_\mathrm{fid}$, where $f_\mathrm{fid}$ is fiducial growth rate. $a_{0, n}, a_{2, n}$ are also scaled by the amplitude of the fiducial no-wiggle power spectrum at the pivot points $k_{p} = n \Delta$ before injecting them into \cref{eq:spline_basis}. Parameters with superscript `$*$' are fixed to the following values when only a 1D fit is performed: $\alphaap=1$, $d\beta=1$, $a_{2,n}=0$, $b_{2,n}=0$. The central values of the Gaussian BAO damping parameters, denoted by superscript `fid' vary between tracers and so are provided separately in \cref{tab:priors2}. \edited{$\Sigma{\perp}$, $\Sigma_{\parallel}$ and $\Sigma_s$ are additionally bounded between $[0, 20]$.} The total number of free parameters for our 1D and 2D fits are 15 and 25 respectively for $P(k)$; and 7 and 13 respectively for $\xi(r)$. The corresponding numbers of degrees-of-freedom for our 1D and 2D fits are 41 and 87 for $P(k)$; and 19 and 39 for $\xi(r)$.
    \label{tab:priors}
    }  
\end{table}
\begin{table}
    \centering
    \begin{tabular}{|l|l|c|c|c|c|}
        \hline
        \multirow{2}{*}{Tracer} & \multirowcell{2}{Redshift\\range} & \multicolumn{2}{c|}{$\Sigma^{\mathrm{fid}}_{\perp} [\hinvmpc]$} & \multicolumn{2}{c|}{$\Sigma^{\mathrm{fid}}_{\parallel} [\hinvmpc]$}\\
         & & Pre-recon & Post-recon & Pre-recon & Post-recon \\ \hline
        BGS & 0.1 - 0.4 & 6.5 & 3.0 & 10.0 & 8.0 \\
        LRG & 0.4 - 0.6 & 4.5 & 3.0 & 9.0 & 6.0 \\
        LRG & 0.6 - 0.8 & 4.5 & 3.0 & 9.0 & 6.0 \\
        LRG & 0.8 - 1.1 & 4.5 & 3.0 & 9.0 & 6.0 \\
        ELG & 0.8 - 1.1 & 4.5 & 3.0 & 8.5 & 6.0 \\
        ELG & 1.1 - 1.6 & 4.5 & 3.0 & 8.5 & 6.0 \\
        QSO & 0.8 - 2.1 & 3.5 & 3.0 & 9.0 & 6.0 \\
        \hline
    \end{tabular}
    \caption{The central values of the BAO damping parameters used for our default Gaussian priors (see \cref{tab:priors}) when fitting each of the DESI DR1 tracers. These are obtained from a combination of theoretical calculations, measurements of the cross-correlation between the initial and post-reconstruction density fields in \edited{\abacussecond\ DR1} mocks, and by running fits to data vectors averaged over many realizations of these mocks for each tracer.
    \label{tab:priors2}}
\end{table}

For Fourier-space fits, the theory model always includes all three multipoles for the product with the window matrix, but the broadband terms for the multipoles that are not fitted (e.g. quadrupole and hexadecapole for 1-D fits) are set to $0$ as they are mostly unconstrained. 


In the rest of this work we mainly report on the BAO scaling parameters $\alpha_{\text{iso}}$ and $\alpha_{\text{AP}}$, but in doing so are fully marginalizing over the other parameters of our model, including the correlation function/power spectrum broadband parameters, the galaxy bias, and the damping parameters.



\subsection{Covariance Matrices}
\label{subsec:methods-cov}

\begin{figure}
\begin{center}
\includegraphics[width=0.99\textwidth]{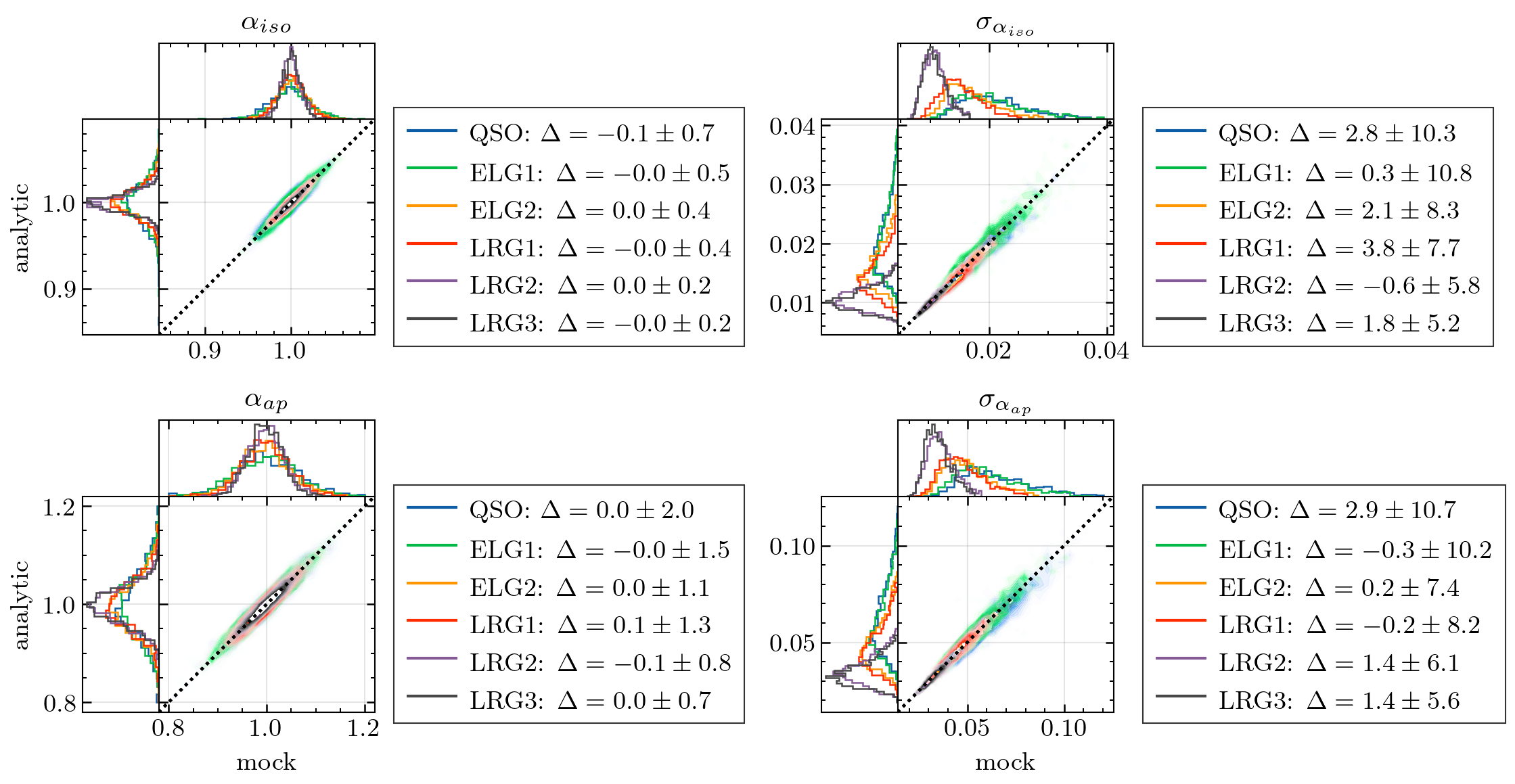}
\caption{Comparison of fits using the {\sc RascalC} semi-analytical and
mock-based covariance matrices. The figure demonstrates the consistency of the inferred
distance scales and errors using our two approaches for computing the covariance matrices.
\edited{The panels on the left show the BAO distance scales 
$\alpha_{\rm iso}$ and $\alpha_{\rm AP}$, while the panels on the right show the
errors on these parameters. The $x$-axis show results using the mock-based
covariances in the fits, while the $y$-axis show results using the {\sc RascalC} semi-analytical covariances.}
The {\sc RascalC} matrices have been tuned to the ensemble of mocks. Further tests
were performed using matrices tuned to a single mock which yield consistent
results. 
The $\Delta$ values in the legends on the left show the fractional mean and standard deviation
(in percent, relative to the analytic value)
of the difference between the $\alpha$ values, while the values on the right 
correspond to the difference in the errors. 
}
\label{fig:covariance_comparison}
\end{center}
\end{figure}

Our analysis here has used both analytic and mock-based approaches to computing the covariance matrices for the two-point functions. We summarize the construction and validation of these covariance matrices below.

The analytic covariances assume Gaussian covariances based on the observed clustering of the galaxies and account for the effects of the survey geometry and selection function. 
Note that while we largely limit to the \edited{disconnected} terms, these are derived from the observed nonlinear clustering of the data or mocks.

The covariance matrices for the correlation functions are generated with the {\sc RascalC} code\cite{rascal,rascal-jackknife,RascalC,RascalC-legendre-3},\footnote{\url{https://github.com/oliverphilcox/RascalC}} a variant of which was used for the BAO analysis of the early DESI data \cite{RascalC-DA02}. In addition to the \edited{disconnected} term, the code also includes an adjustable shot-noise parameter that serves as a proxy for the missing connected 4-point terms in the covariance matrix. This parameter can be calibrated with data jackknife resampling (or mocks).
A complete discussion of this method is in \cite{KP4s7-Rashkovetskyi}, along with the validation of the data pipeline on mocks demonstrating high consistency with their sample covariance, in particular for errorbars of BAO scale parameters.
The code used to generate the covariance matrices for this analysis is publicly available.\footnote{\url{https://github.com/misharash/RascalC-scripts/tree/DESI2024/DESI/Y1}}

The covariances for the power spectrum are based on the formalism by \cite{Wadekar2019}. Our implementation is described in detail in \cite{KP4s8-Alves}.\footnote{The code is available at \url{https://github.com/cosmodesi/thecov}} For the power spectrum we limit to the Gaussian terms and do not include any higher-order corrections. 

In addition to the analytic covariances, we also use the \ezmocks\ described
previously to generate sample covariance matrices. The advantage of \edited{the \ezmocks\ covariances} is
that we can \edited{directly} assess the impact of observational systematics like fiber assignment.  These mocks are used to test the approximations
made in the generation of the analytic covariance matrices. A detailed
comparison of the two approaches for generating covariance matrices is in \cite{KP4s6-Forero-Sanchez}.
In particular, we quantify how the fitting of the BAO scales and the corresponding
errors depends on the choice of the covariance matrix used. \cref{fig:covariance_comparison} compares the
$\alpha$ values and their errors for the full sample of mocks obtained from fits in configuration space. We find that the
analytic and mock-based covariances yield answers that are consistent to (much)
better than 0.1\% for $\alpha_\mathrm{iso/AP}$ and generally $< 2\%$ for the errors. Both these values sit comfortably within the respective standard deviations for all the galaxy samples. \edited{Note that, in addition to the consistent average BAO scales and average errors between the two types of covariance matrices, the individual measurements demonstrate high correlation, typically ranging between 0.965-0.980, as depicted by the distribution thickness in \cref{fig:covariance_comparison}.} We stress that these tests are done using analytic covariances tuned on \ezmocks, thus making the sample covariance baseline a fair estimate of the true covariance. \edited{While \cref{fig:covariance_comparison} shows results for a matrix built using the mean of \ezmock\ clustering and the shot-noise rescaled for an optimal match to the sample covariance, \cite{KP4s6-Forero-Sanchez} shows that such a matrix is consistent with matrices built from a single \ezmock\ realisation and the shot-noise rescaling based on a jackknife covariance estimate as is done using observational data.} 

\edited{A caveat is that we have not attained the desired level of agreement between the analytic and the \ezmock\ covariance matrices in Fourier space as of the time of writing this paper, as will be presented in \cite{KP4s6-Forero-Sanchez}. Hence, we consider our default results in configuration space the most robust measurements.}

Given this level of agreement\edited{, particularly in configuration space}, we adopt analytic covariance matrices \edited{built based on the unblinded \desidrone\ data and calibrated using jackknife resampling of this same data} for our analyses in this paper. This choice allows us to tune our covariances to match the observed clustering of the data after we unblind the data, and to avoid the small discrepancies between the clustering seen in the data and the mocks. We note that we are focusing on the BAO observables here, although preliminary work seems to suggest that the analytic covariances may also work well for other observables, at least on large scales.

\section{Systematic Error Budget}
\label{sec:systematics}


\edited{This section provides an overview of the systematic errors on the BAO scales stemming from various components of our analysis pipeline, as determined through the tests performed in our supporting papers (\cref{tab:supportingpapers}). Although individual supporting papers may use a more stringent limit for investigating systematics, the overarching rule we use is to count systematics when we detect an effect that exceeds $3\sigma$, compared to the statistical error associated with the mock test. We will conclude this section by presenting the combined systematic errors on the BAO scales.}
\begin{table}
    \centering
    \begin{tabular}{|ccc|}
    \hline
    Name/Description &  $\sigma_{\alpha_{\mathrm{iso}}}$ & $\sigma_{\alpha_{\mathrm{AP}}}$ \\\hline
    Non-linear mode-coupling & $<0.1\%$ &
    $<0.1\%$ \\
    Relative velocity effects  & $<0.05\%$ &
    $<0.05\%$ \\
    Broadband modelling & $<0.02\%$ &
    $0.11\%$ \\
    BAO wiggle extraction & $<0.02\%$ &
    $<0.09\%$ \\
    Dilating smooth vs. wiggle & $<0.02\%$ &
    $<0.09\%$ \\
    Modelling $\xi(s)$ from $P(k)$ &  $<0.01\%$ &
    $0.12\%$ \\
    \hline
    \textbf{Combined} &  $ 0.1\%$ & $0.2\%$\\
    \hline
    \end{tabular}
    \caption{Different contributions to the DESI galaxy BAO theory and modelling systematic error budget for $\alpha_{\mathrm{iso}}$ and $\alpha_{\mathrm{AP}}$ considered in this work, taken from \cite{KP4s2-Chen}. The first two rows denote theory cases, where the exact contributions depend on the nature of the tracers (galaxy bias, redshift, etc.), we have estimated a conservative upper value considering all DESI tracers/redshift bins. The other rows are obtained from fitting the \abacusfirst\  cubic, control-variate (CV) measurements while varying our modelling assumptions/choices. \edited{Only two of the tests show that our results can potentially be subject to a small nonzero bias.} The remainder return no difference within the statistical precision available from the simulations and so are reported as upper bounds. The combined theory and modelling systematic error is obtained by summing the two solid detections in quadrature and rounding up to account for the remaining upper limits.}
    \label{tab:theory}
\end{table}

\subsection{Theoretical Systematics}\label{subsec:sys-theory}

Ref.~\cite{KP4s2-Chen} investigates the systematic error induced by various approximations made when modelling the BAO. These can potentially arise from either physical effects that are dropped from the model used in BAO fits, e.g., the \edited{shrinkage} of the BAO feature due to non-linear clustering/evolution in the galaxy distribution \cite{ESW2007,Crocce2008,SeononlinearBAO,Padmanabhan09b,Sherwin12} or the imprint of relative baryon-dark matter perturbations.  There can also be residual errors due to (small) imperfections in the numerical choices made in the broadband model and extraction of the BAO template. 

The physical processes affecting the BAO feature are comprehensively computed in \cite{KP4s2-Chen} within perturbation theory. The impact of the numerical choices made in the modeling is measured by fitting to our high-precision \abacusfirst\ cubic, control-variate (CV) measurements while changing the numerical prescriptions used in the modelling. The robustness of the BAO means that even with the large volume and sample-variance cancellation of the \abacusfirst\ cubic mocks most of the modelling choices we test give consistent constraints within the statistical uncertainties, and we report only upper bounds on the potential systematic error. We obtain positive detections of only two (small) systematic modelling errors. The first arises from comparing our preferred spline-based broadband model with the historically used polynomial-based model. The second arises from comparing our new approach of modelling the correlation function as the Hankel transform of the dilated power spectrum multipoles with free growth rate parameter, to the BOSS approach \cite{Ross17} of transforming the undilated multipoles and applying the dilation \textit{after} the transformation, then varying two different bias parameters for the monopole and quadrupole rather than the growth rate parameter. In both cases, we argue that our new method is better physically motivated and preferred, but as these are nonetheless modelling choices that also passed the systematic checks used in previous studies, we absorb these differences into our systematic error budget. 

Using these two detections, upper bounds on the other modelling considerations, and folding in the expected theoretical upper limits from nonlinearities, we obtain a (conservative) systematic error on $\alpha_{\rm iso, ap}$ of $0.1$ and $0.2\%$, respectively. The different contributions to this total are in \cref{tab:theory}.

\begin{table}
    \centering
    {\renewcommand{\arraystretch}{1.3}
    \begin{tabular}{|c|c|c|c|c|c|}
        \hline
         & Parameter & $\sigma_\text{HOD,BGS}$ &  $\sigma_\text{HOD,LRG}$ &  $\sigma_\text{HOD,ELG}$ &  $\sigma_\text{HOD,QSO}$ \\ \hline
        \multirow{2}{*}{$P(k)$}
        & $\aiso$ & 0.19\% &  0.066\% &0.047\% & 0.12\%  \\
        & $\alap$ & ---  &  0.094\% & 0.089\% & --- \\ \hline
        \multirow{2}{*}{$\xi(r)$} 
        & $\aiso$ & 0.20\% &   0.074\% & \bf{0.17\%} & 0.07\%  \\
        & $\alap$  & ---  & \bf{0.187\%} & 0.11\% & ---  \\
    \hline
    \end{tabular}
    }  
     \caption{\label{tab:hodsystematics_summary}Summary table for the estimation of HOD-dependent systematics for individual tracers. The table is split into Fourier and configuration space results, and we include the systematics for both $\aiso$ and $\alap$. 
    The results for LRGs and ELGs are presented in detail in \cite{KP4s10-Mena-Fernandez} and \cite{KP4s11-Garcia-Quintero}. The results for BGS and QSO are derived consistently to LRGs and ELGs. \edited{The bold-faced numbers indicate a detection of systematics; otherwise, the numbers reflect the statistical precision associated with no detection. As conservative upper limits, we adopt systematic error on $\alpha_{\rm iso, ap}$ of $0.1$ and $0.2\%$ for the HOD-dependent systematics.}
    }    
\end{table}

\subsection{HOD-dependent systematics}\label{subsec:sys-hod}
Systematic errors can be introduced when we infer the underlying BAO scale in the matter density field from the BAO measurement using a specific choice of LSS tracer, which is typically a biased tracer. Studies have shown that such systematics are subdominant compared to the shift in the BAO scale due to structure growth and redshift-space distortions; moreover, they effectively vanish after reconstruction \cite{Rossi:2020wxx}. Given the precision of the state-of-the-art DESI DR1 data and that the new major tracer, the ELGs, believed to exhibit distinct halo occupation features compared to LRGs, this test is revisited. Our supporting papers, \cite{KP4s10-Mena-Fernandez} and \cite{KP4s11-Garcia-Quintero}, extensively test the systematics for the different assumptions of the halo occupation distribution (HOD) for LRGs and ELGs, respectively, using DESI mock catalogs constructed from Abacus simulations. \cite{KP4s10-Mena-Fernandez} studied nine different variations of HODs that are within $3\sigma$ of the best-fit HOD parameters to the One-Percent Survey \cite{EDR_HOD_LRGQSO2023}. \cite{KP4s11-Garcia-Quintero} studied the impact of the ELG HODs using 22 different HOD models, \edited{including the standard models used for \abacusfirst\ and \abacussecond}  as well as extended models that include galactic conformity, assembly bias, and modified satellite profile adopted from \cite{EDR_HOD_ELG2023}. 

The amplitude of HOD systematics is estimated using the same methodology in both of our supporting papers, \cite{KP4s10-Mena-Fernandez} and \cite{KP4s11-Garcia-Quintero}. \edited{We compute the differences in $\aiso$ and $\alap$ between every pair of HOD models, averaging over the 25 realizations. We compute the significance of this difference to assess the systematic detection level. If we do not detect systematics at the level of 3$\sigma$ in terms of the typical dispersion of the average differences from the mocks, we use the distribution of the differences between all pairs of HOD models and quote the 68\% region of the distribution as the systematic error. On the other hand, if any of the paired HODs shows a non-zero difference at 3$\sigma$ or above, we take the shift between that pair of HODs as the systematic error. If there is more than one pair of HODs with a non-zero difference above 3$\sigma$, we quote the maximum of these shifts.}
\cref{tab:hodsystematics_summary} shows the summary of the systematics for each tracer. The systematics for BGS and QSO are derived consistently.

Note that the quoted systematic error, in the case of no detection, often merely reflects the limited sample variance cancellation due to the different subsampling noise. Even for the detection cases, an analysis like this depends on how extreme the HOD models we decide to compare are; we compare the models that span the $\pm 3\sigma$ contours around the best-fit HODs, which is a reasonable choice to incorporate viable HOD parameter values allowed by the data. In addition, this test includes the systematics in the process of reconstruction as well, as we fix the galaxy bias input to reconstruction, while the actual HOD model (and, therefore, the effective bias) is being varied, with variations with respect to the input bias scaling up to 10\%.
The theoretical systematics reported by \cite{KP4s2-Chen} already include a contribution due to the reference HOD model assumed in the simulation they utilized, and therefore the HOD systematics reported in this section should be considered as additional systematics on the BAO scales that can be introduced by assuming different HODs than the reference HOD model tested in \cite{KP4s2-Chen}. 


Using the results presented in \cref{tab:hodsystematics_summary}, we adopt a conservative approach and consider a common HOD-dependent systematic error for all tracers of 0.2\% for both $\aiso$ and $\alap$.



\subsection{Observational Systematics}\label{subsec:obssys}

Similar to results found in SDSS (e.g., \cite{Ross2017}), we find that observational systematics have a negligible impact on our BAO measurements. We therefore do not add any additional observational systematic uncertainty. We briefly outline the studies justifying this conclusion.

\begin{table}[]
    \centering
    \small
    \begin{tabular}{|l|c|c|c|r|r|}
    \hline
     Tracer       & Recon   & $\langle \alpha^{\rm complete}_{\rm iso} \rangle$   & $\langle \alpha^{\rm complete}_{\rm AP}\rangle$   & $\langle \alpha^{\rm complete}_{\rm iso} - \alpha^{\rm altmtl}_{\rm iso}\rangle$   & $\langle\alpha^{\rm complete}_{\rm AP} - \alpha^{\rm altmtl}_{\rm AP}\rangle$ \\
    \hline
     ${\tt BGS}$  & Post    & $0.9998 \pm 0.0038$                 & ---                                & $0.00288 \pm 0.00229$                                                 & ---                                                                   \\
     ${\tt LRG1}$ & Post    & $0.9996 \pm 0.0025$                 & $0.9934 \pm 0.0091$                & $-0.00182 \pm 0.00163$                                                & $0.00374 \pm 0.00455$                                               \\
     ${\tt LRG2}$ & Post    & $0.9967 \pm 0.0022$                 & $1.0091 \pm 0.0075$                & $0.00207 \pm 0.00144$                                                 & $0.00096 \pm 0.00392$                                              \\
     ${\tt LRG3}$ & Post    & $0.9958 \pm 0.0019$                 & $0.9854 \pm 0.0063$                & $0.00240 \pm 0.00161$                                                 & $-0.00407 \pm 0.00512$                                              \\
     ${\tt ELG1}$ & Post    & $1.0015 \pm 0.0035$                 & ---                                & $0.00118 \pm 0.00422$                                                 & ---                                                               \\
     ${\tt ELG2}$ & Post    & $0.9988 \pm 0.0030$                 & $1.0025 \pm 0.0100$                & $0.00387 \pm 0.00270$                                                 & $0.00387 \pm 0.00676$                                              \\
     ${\tt QSO}$  & Post    & $0.9968 \pm 0.0033$                 & ---                                & $0.00207 \pm 0.00361$                                                 & ---                                                               \\
    \hline
    \end{tabular}
    \caption{\edited{The mean and the error associated with the mean of the BAO measurements from the 25 \abacussecond\ DR1 mocks using the \desidrone\ baseline BAO method, as presented in \cite{KP4s4-Paillas}.  Results indicate that fits to the `complete' mocks, i.e., before applying fiber assignment effects, align with unbiased BAO measurements within statistical error. Furthermore, the mean difference between the \ altmtl\ mocks and the ` complete' mocks demonstrates the recovery of unbiased BAO estimates despite the presence of the fiber assignment effect. \lrgelg\ is omitted since it is the combination of \lrgth\ and \elgo.}}
     \label{tab:BAOaltmtlmocks}
\end{table}

Observational systematics in DESI can be broadly divided into three classes: fiber assignment incompleteness, spectroscopic systematics, and imaging systematics. Any effects of fiber assignment incompleteness can be tested by comparing results on our DR1 mocks (described in \cref{sec:mocks}) that have no fiber assignment incompleteness (`complete') and those that have had fiber assignment run in a way that mimics the DESI DR1 observing history (`\altmtl'). Each have been run through the DESI DR1 BAO measurement pipeline and consistent results were found \cite{KP4s4-Paillas}, \edited{as summarized in \cref{tab:BAOaltmtlmocks}. The table first shows that the BAO fits from \abacussecond\ DR1 `complete' mocks are consistent with no bias. The \altmtl\ mocks are consistent with the `complete' mocks within 1.5$\sigma$, demonstrating that we can recover unbiased BAO measurements in the presence of the fiber assignment. The differences in $\aiso$ appear systematic. The weighted mean (assuming no correlation) is $\langle \alpha^{\rm complete}_{\rm iso} - \alpha^{\rm altmtl}_{\rm iso}\rangle_{\rm all} = 0.00149\pm0.00077$ and thus just under 2$\sigma$. This is not significant enough for us to add as a systematic uncertainty, but does suggest a closer look is warranted in future DESI studies.}

\begin{figure*}
\centering
\includegraphics[width=0.49\linewidth]{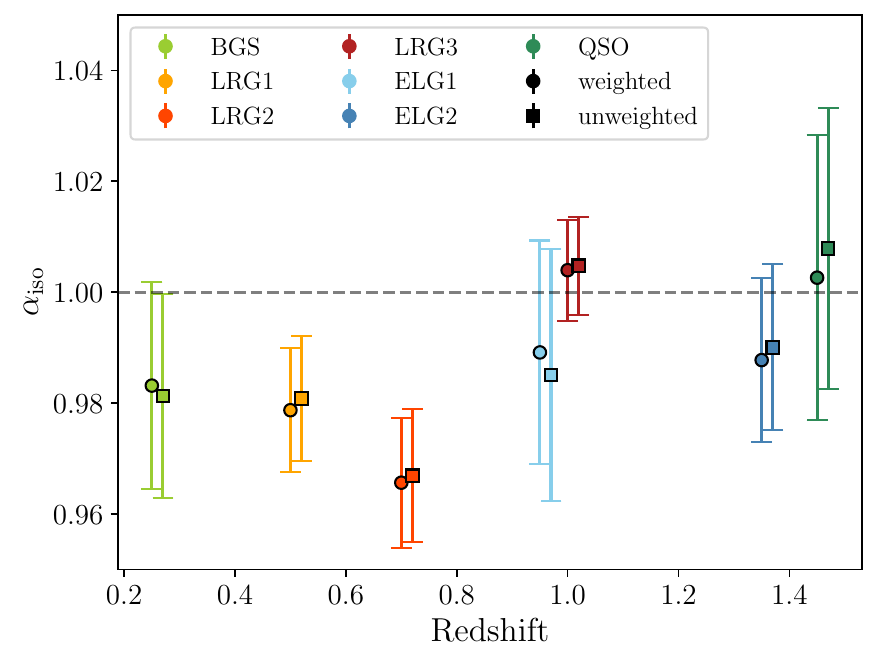}
\includegraphics[width=0.49\linewidth]{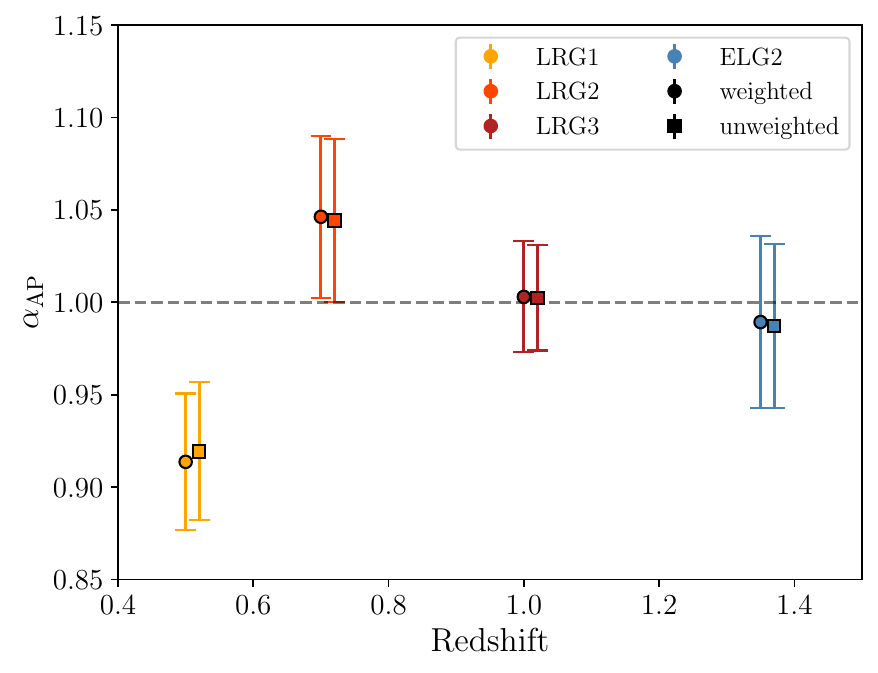}
\caption{The impact of imaging systematics on the BAO scales. In detail, we show the difference in the derived BAO scales when we use the mitigation method for imaging systematics and when we use the raw data (`No weight'). The two measurements for each tracer are slightly displaced horizontally for a better visualization. The BAO scales are very stable even when there was no mitigating and even after density field reconstruction using such raw data. Again, the BAO is robust against the imaging systematics.}\label{fig:imagingsysBAO}
\end{figure*}

Significant trends between the observed DESI galaxy density and the observational properties of the imaging data used to define the target samples (e.g., the imaging depth) are found for all DESI tracers \cite{KP3}. These trends are removed from the DR1 LSS catalogs via weights that are added to the sample based on regression analysis against maps of the observational properties. The overall impact of the weights on the DR1 clustering measurements is quite large in significance; the $\chi^2$ difference between weighted and unweighted clustering for ELGs is greater than 1000 \cite{KP3s2-Rosado}. However, the impact on BAO measurements is negligible, as shown in \cref{fig:imagingsysBAO}. When comparing the post-reconstruction BAO measurements with and without weights to correct for imaging systematics, the variation is found to be at most $0.27 \sigma$ (for the QSO $\aiso$) and is typically less than $0.2\sigma$ on any of the measured BAO parameters ($\aiso$ and $\alap$), without any coherent direction across tracers. The weights modulate the density field obtained from the LSS catalogs and thus there is expected to be some fluctuation in any measured parameter that is proportional to the original uncertainty, purely from the degree to which the weighting de-correlates the weighted and unweighted data. Thus, shifts as a fraction of the uncertainty on the measured BAO parameters are the most relevant thing to quote here. The 0.2$\sigma$ level of variation is equivalent to a correlation of 0.99 between the weighted and unweighted results. This is thus similar to that one would expect from randomly applying weights with the same variance as the imaging systematic weights to the sample. We stress that the weighted clustering results are expected to be much closer to the `true' clustering than those without the imaging systematic weights. Thus, even if these shifts were not consistent with random variation, they would represent an extreme scenario.

Spectroscopic systematics refer to both the effect of errors in the DESI redshift estimation and unphysical galaxy/quasar variations in the fraction of successful redshifts as a function of parameters relating to DESI hardware and observing conditions. Similar to the corrections for imaging systematics, trends with spectroscopic properties can be regressed against, and corrected with, weights. Such weights, $w_{\rm zfail}$, are determined as a function of effective observing time for all DESI tracers, as described in \cite{KP3s4-Yu,KP3s3-Krolewski}. Despite the significance of the observed trends, \cite{KP3s4-Yu} find that difference in clustering between results including $w_{\rm zfail}$ or not yields a maximum $\chi^2$ difference of just 0.09 (for ELG samples at both redshift range) across all of the pre-reconstruction $\xi_{0,2}$ within the BAO fit range. \edited{Such a small $\chi^2$ difference implies at most a 0.3$\sigma$ shift in any potential parameter; the differences are not localized and any effect on BAO measurements is thus expected to be negligible.}
Further corrections, such as extra weights to remove the remaining dependence of the success rate on the focal plane location, were also studied in \cite{KP3s4-Yu} and found to have similarly small effects on the measured clustering. The size of the effect of redshift success trends on the DESI DR1 two-point functions is thus negligible for BAO measurements.

Finally, redshift errors have been characterized
again in DR1. These are divided into two components: a typical uncertainty, characterized by a narrow Gaussian or Lorentizian profile, and a rate of catastrophic failure. The potential impact of the typical uncertainty on clustering measurements have been studied with galaxy mocks generated with the UNIT-SHAM method \citep{EDR_SHAM_UNIT2024} and AbacusSummit-HOD models \citep{EDR_HOD_LRGQSO2023,EDR_HOD_ELG2023}. In \cite{EDR_SHAM_UNIT2024} the impact on ELGs was negligible at all scales and the effect for LRGs was $\sim0.05\sigma$. In DR1, the ELG redshift uncertainty is even smaller \cite{KP3s4-Yu}, and LRGs show the same level of uncertainties \cite{KP3s3-Krolewski}. The impact on measured QSO clustering was substantially greater, due to the larger uncertainty on their redshifts. \edited{However, the effect of redshift uncertainties as an additive component to the velocity dispersion is included for both LRG and QSO in the DESI Y1 mocks, where no detectable bias in BAO parameters is found.}

In studying catastrophic failures in the redshift measurements of DESI ELGs (using, e.g., repeated observations of the same ELG), \cite{KP3s4-Yu} find that the most prominent feature is confusion at $z\sim1.32$ between residuals from sky spectra and [O\,\textsc{ii}] emission. The total failure rate from repeat spectra was found to be 0.27\%, much smaller than the worst case of 1\% \cite{ELG.TS.Raichoor.2023}. However, the observed failure patterns with 0.27\% rate show similar clustering impacts to the assumed 1\% failures for ELG mocks at $1.1<z<1.6$ due to the prominent confusion at $z\sim1.32$. Nevertheless, their impact on the measured configuration space clustering is $<0.01\sigma$ at scales $s>60\hMpc$ and thus should not impact BAO measurements.
 
\begin{figure}
    \centering
    \includegraphics[width=0.7\textwidth]{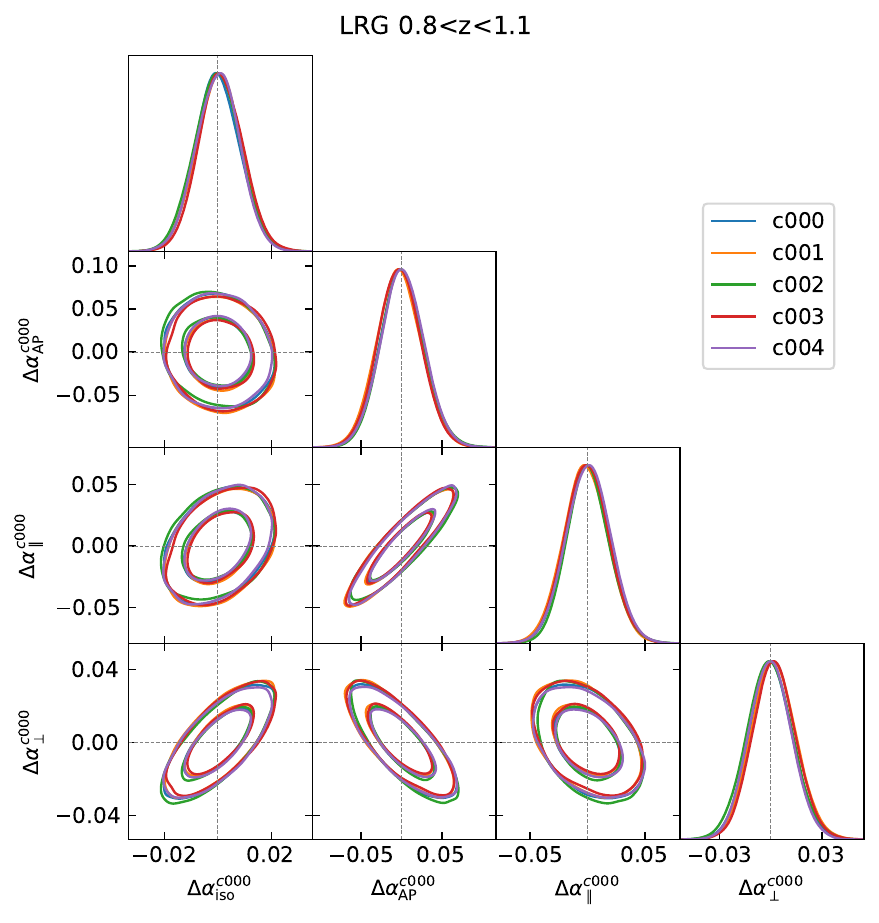}
    \caption{Contour plots of the differences of the (rescaled) alpha values measured with the different \textsc{AbacusSummit} cosmologies using \desidrone\ data. As an example, the LRG high redshift bin is shown. \edited{The quantities $\Delta \alpha^{c000}$ are defined as follows. For a given cosmology c00x, the derived alpha values are rescaled to c000 units and the values measured directly with c000 are subtracted.}}
    \label{fig:fidcosmo_contour}
\end{figure}

\subsection{Systematics due to the assumption of the fiducial cosmology} \label{subsec:sys-fiducialcosmo}


In this subsection, we summarize the impact of using a wrong fiducial cosmology in the BAO analysis, which were extensively studied in \cite{KP4s9-Perez-Fernandez}. 
 
The choice of reference cosmology comes into play at three different stages. First, we assume a set of cosmological parameters when converting redshift measurements into distances; we term this the {\bf grid cosmology}. The difference between the grid and true cosmologies causes a distortion of the BAO scale along and across the line of sight \citep{Alcock1979}, which is quantified by the parameters $q_{\parallel} = H_\textrm{grid}(z)/H_\textrm{true}(z)$ and $q_{\perp}=D_\textrm{true}(z)/D_\textrm{grid}(z)$ respectively. \edited{Without the three-dimensional standard ruler like BAO}, this effect is somewhat degenerate with the redshift-space distortions, but with a sufficiently large data set, such as we have with DESI, \edited{and with the BAO feature} it is possible to distinguish between the two \citep{Ballinger96,SeoBAOFisher}. Second, a {\bf template cosmology} is chosen in order to compute the linear power spectrum, which is then used to create the model power spectrum for the fitting \edited{($P_{\rm nw}$ and $P_w$ in \cref{eq:generic_model}).} The effect of fixing the template is interpreted as an additional isotropic rescaling of the distances by a factor of $r_d^{\rm tem}/r_d^{\rm true}$. Lastly, the values for the linear bias $b_1$ and the growth rate $f(z)$ input into the reconstruction algorithm are cosmology dependent\edited{, affecting the estimation of the displacement field}. The separate effect of both the template and the values assumed for reconstruction have been comprehensibly studied in the past \citep{Thepsuriya_2015, Carter2020, Bernal20, Pan2023}, as well as the joint effect of consistently changing the reference cosmology in the whole pipeline \citep{VargasMagana2016:1610.03506v2},  while the separate effect of the grid is less explored \citep{Sanzwuhl_2024}. Potential systematic shifts of the order of a few tenths of a percent in the alpha values have been reported in the most extreme scenarios. 

We studied the effect of analysing mocks with \desidrone\ geometry with different fiducial cosmologies. Namely, the four secondary \textsc{AbacusSummit} \citep{Maksimova2021} cosmologies (\cref{tab:cosmologies}), which \edited{deviate from the Planck 2018 cosmology in various aspects. For example, from \cref{tab:cosmologies}, c002 introduces an offset on the distance to redshift relation at the level of $\sim4.7-7.5$\% and an offset in the sound-horizon scale at the level of 6.8\%. Up to 2.9\% of the anisotropic distortion (AP) is also tested with c002.}

Our baseline test consists of simultaneously changing the cosmology for grid, template and reconstruction (the separate contributions of grid and template are explored in the companion paper). The fiducial DESI cosmology c000 is used to create the \abacussecond\ DR1 mock catalogs, but the different grid cosmologies c001-c004 are used to convert redshifts into distances to compute the pair counts and power spectrum.  The change in template is implemented by simply changing the linear power spectrum computed with \textsc{class} and then following the fitting procedure with \textsc{desilike} as described in \cref{subsec:methods-fitting}. The values for the linear bias and growth rate are \edited{derived assuming each fiducial cosmology and introduced in reconstruction.}

We compare the expected BAO scales (the last two columns in \cref{tab:cosmologies}) with the derived BAO scales for all tracers and for each of the wrong fiducial cosmologies using 25 \abacussecond\ DR1 mocks,
and regard a shift detection for values above a 3$\sigma$ \edited{based on the dispersion of the 25 mocks}. \cref{tab:fidsystematics_summary} summarises our results in \cite{KP4s9-Perez-Fernandez}. For cosmology c003 we measure a systematic shift of the order of 0.2\%
for $\alpha_{\rm iso}$ that can be neatly explained by the phase shift in the BAO due to the difference in $N_{\rm eff}$ (see for example \citep{Baumann_2017}). Since this effect comes from the template and it is a matter of how the re-scaling by the sound horizon is interpreted in the BAO measurements, we opt to not include it as part of the systematic error budget. 
\edited{A detection at the level of 0.1\% is reported for c002 from \cite{KP4s9-Perez-Fernandez}; otherwise we do not see a detection of bias.} Thus we adopt a conservative value of 0.1\% for both $\alpha_{\rm iso}$ and $\alpha_{\rm AP}$. Furthermore, we tested the consistency of the BAO fits against the choice of fiducial cosmology for the blinded Y1 catalogs, as part of the unblinding tests described in \cref{sec:unblinding}. Likewise, we conduct an analogous comparison with the unblinded data, for which consistent results are observed for the different cosmologies (\cref{fig:fidcosmo_contour}).


\begin{table}
   \small
    \centering
    \resizebox{\columnwidth}{!}{%
        \begin{tabular}{|c|c|c|c|c|c|}
            \hline
            Cosmology & Description &  $\frac{r_{d, \rm tem}}{r_{d,\rm true}}$  & $\frac{\alpha_{\rm iso, fid}}{\alpha_{\rm iso, true}}$ & $\frac{\alpha_{\rm iso, fid}}{\alpha_{\rm iso, true}}$  & $\frac{\alpha_{\rm AP, fid}}{\alpha_{\rm AP, true}}$ \\
             &  & (template) & (grid) & (template + grid) & (grid) \\
            \hline
            c000 & Planck 2018 $\Lambda$CDM & 1.000& 1.000 -- 1.000 & 1.000 -- 1.000 & 1.000 -- 1.000 \\
            
            c001 & Low $\omega_c$ $\Lambda$CDM & 1.056& 0.963 -- 0.989 & 1.017 -- 1.044 & 0.976 -- 0.991 \\
            
            c002 & Thawing dark energy & 0.932& 1.047 -- 1.075 & 0.976 -- 1.002 & 1.011 -- 1.029 \\
            
            c003 & $N_{\rm eff} = 3.70 $ & 1.022& 0.983 -- 0.995 & 1.005 -- 1.017 & 0.989 -- 0.996 \\
            
            c004 & $\sigma_8 = 0.75$ $\Lambda$CDM & 1.000& 1.000 -- 1.000 & 1.000 -- 1.000 & 1.000 -- 1.000 \\
            \hline
        \end{tabular}
    }
     \caption{\edited{The level of scale dilation and anisotropy on the $\alpha$ parameters we introduce in the test of the fiducial cosmology, by adopting four fiducial cosmologies (c001-c004) when analyzing the mock catalogs generated using the true cosmology (c000). Each of c000-c004 is based on \texttt{AbacusSummit}\citep{Maksimova2021}. The effect of the grid cosmology on the scale dilation depends on redshift and we therefore list the ranges for $0.1<z<2.1$. Note that, with c002, we introduce an offset on the distance to redshift relation at the level of $\sim 4.7-7.5$\% and an offset in the sound-horizon scale at the level of 6.8\%. Up to 2.9\% of the anisotropic distortion (AP) was also tested with c002. The last two columns are the expected BAO scales we should recover. A significant (3$\sigma$) offset of the derived BAO scale from the last two columns is considered as a bias.} \label{tab:cosmologies}}
\end{table}
\begin{table}
    \centering
    {\renewcommand{\arraystretch}{1.3}
        \begin{tabular}{|c|c|c|c|}
            \hline
            Space & Parameter & c000-c00X & c000-c003\\ \hline
            \multirow{2}{*}{Fourier}
            & $\aiso$ & \edited{0.1\%} &  0.2\% \\
            & $\alap$ & None (0.1\%) &  None (0.1\%) \\ \hline
            \multirow{2}{*}{Config.} 
            & $\aiso$ & None (0.1\%) & 0.2-0.3\%\\
            & $\alap$  & None (0.1\%) & None  (0.1\%)\\
            \hline
        \end{tabular}
    }  
    \caption{Estimation of systematics due to the assumption of an incorrect fiducial cosmology, based on the test by \cite{KP4s9-Perez-Fernandez}.  We show the residual bias in the BAO scales after accounting for the expected shift due to using incorrect grid and template cosmologies (c00X, where $X=1$, 2, 4).  Unless the residual bias in the BAO is detected at a significance greater than $3\sigma$ (in terms of the standard deviation of the mean of the mocks), we consider there to be no net bias in the BAO scale due to the fiducial cosmology and put an ad-hoc value of 0.1\%. We detected a significant bias in $\aiso$ for c003, which assumes an incorrect $N_{\rm eff}$ in the BAO template. Systematics from an incorrect $N_{\rm eff}$ are not included in our systematic budget, as this concerns the interpretation of the measured BAO scale.}
    \label{tab:fidsystematics_summary}
\end{table}

\subsection{Combining all systematics}\label{sec:combinesys}
\begin{table}[!h]
    \centering
    \begin{tabular}{|l|c|l|}
        \hline
        Systematic &  Error (in percent) & Comments \\
        \hline
        Theoretical & 0.1 ($\alpha_{\rm iso}$), 0.2 ($\alpha_{AP}$) & Includes fitting methodology and choices, \\ 
        & & as well as expected impacts from\\
        & & galaxy bias and cosmology misestimation \\
        \hline
        Observational &  &  \\
        a. Imaging & Not detected  & Tested on the data, with the \\
    & &    largest change being 0.3\% seen for the \\
        & & \elgo, and the rest being 0.1\%. \\
        
     b. Spectroscopic & Not detected  & \edited{Tested with the mocks on the clustering level.} \\
        c. Fiber assignment & Not detected & \edited{The  test was finalized after unblinding.}\\\hline
        HOD & 0.2 & Only one detected statistically significant pair \\
        & & Limited by statistical precision of mocks \\
        & & Note : some of this error is already included in \\
        & & the theory budget \\\hline

        $N_{\rm eff}$ & 0.2 ($\alpha_{\rm iso}$) & Bias for $N_{\rm eff}=3.7$\\
        
        Fiducial $D_A(z)$ & $< 0.1$ & May require iteration post-unblinding  \\
        & & if best-fit is far from fiducial \\
        & & Upper limit based on statistical precision \\\hline

        Reconstruction & Not detected & No significant effects from different \\\hline
        & & algorithms etc. \\\hline

        Covariances & Not detected & Based on comparisons between analytic \\
        & & and mock covariances \\
        
        \hline
    \end{tabular}
    \caption{Summary of our individual systematic errors. In some cases, we do not have any statistically significant
    detection of a systematic bias. }
    \label{tab:systematicbudget}
\end{table}

\begin{table}
    \centering
    {\renewcommand{\arraystretch}{1.3}
    \begin{tabular}{|c|c|c|c|c|c|c|}
        \hline
        & Tracer & {$\sigma_\text{BGS}$}  & \multicolumn{2}{c|}{$\sigma_\text{\lrgs,\elgs}$} & {$\sigma_\text{QSO}$} \\ \hline
      {Space}  & {Source}     & $\aiso$ (\%)   & $\aiso$ (\%)  & $\alap$ (\%)   & $\aiso$ (\%)   \\ \hline
      $\xi(r)$ & Theory (\cref{tab:theory}) & 0.1&  0.1 & 0.2 & 0.1 \\
      $\xi(r)$ & HOD   (\cref{tab:hodsystematics_summary}) & 0.2&  0.2  & 0.2 & 0.2  \\
      $\xi(r)$ & Fiducial (\cref{tab:fidsystematics_summary}) & 0.1&  0.1  & 0.1 & 0.1 \\ \hline
      $\xi(r)$ &  Total  & 0.245 &  0.245 & 0.3 & 0.245 \\ \hline
      $P(k)$ & Theory (\cref{tab:theory}) & 0.1& 0.1 & 0.2 & 0.1 \\
      $P(k)$ & HOD   (\cref{tab:hodsystematics_summary}) & 0.2&  0.1 & 0.1 & 0.12  \\     
      $P(k)$ & Fiducial (\cref{tab:fidsystematics_summary}) & 0.1&0.1  & 0.1 & 0.1 \\ \hline
            $P(k)$ &  Total  & 0.245 &   0.18 & 0.245 & 0.19  \\ 
    \hline
    \end{tabular}
    }  
    \caption{The combined non-zero systematics. The total error in each space is obtained by summing contributions in quadrature and rounding to two significant digits.  For the fiducial cosmology systematics, we do not include the bias that can be introduced due to an incorrect assumption of $N_{\rm eff}$ in the fitting model. Although the tracer-by-tracer estimation of the systematics, such as the HODs, return smaller systematics for some tracers and in $P(k)$, here we will adopt the worst case of the systematics across all tracers, $0.245\%$ for $\aiso$ and 0.3\% for $\alap$ for both $\xi(r)$ and $P(k)$.   }
    \label{tab:finalsystematics}
\end{table}

\edited{In this subsection, we combine all the systematics in the basis of $\aiso$ and $\alap$, and then project these to the systematics in $\alpha_\perp$ and $\alpha_\parallel$. It is imperative to include all pertinent systematics, while setting the upper limit of systematics conservative enough to avoid adding systematics after unblinding. }

The summary of the systematic tests is in \cref{tab:systematicbudget}. From \cite{KP4s3-Chen,KP4s4-Paillas}, we conclude that the systematics due to the reconstruction choice, within a few comparably optimal options we narrowed down, is negligible. Moreover, the systematics due to incorrect estimation of bias and incorrect fiducial cosmology during reconstruction are already accounted for by the HOD systematics (\cref{tab:hodsystematics_summary}) and the fiducial cosmology systematics budgets (\cref{subsec:sys-fiducialcosmo}). We also conclude that the observational effects (\cref{subsec:obssys}) and the covariance choice (\cref{subsec:methods-cov} and \cite{KP4s6-Forero-Sanchez}) are negligible.

\cref{tab:finalsystematics} shows the combination of all systematic budgets based on \cref{tab:systematicbudget}.
We opt for the most conservative approach, namely, directly adding in quadrature individual systematic budgets. 
\edited{This approach} only marginally increases the final error by \edited{$\sim$5\% even for the best $\aiso$ measurement}.

Assuming the negligible correlation between the systematics in $\aiso$ and $\alap$, which was the reason we have been evaluating the systematics using this basis, we derive

\begin{equation}
\mathcal{C}^{\rm Sys}_{\rm \aiso,\alap}  = 
\begin{pmatrix}
0.245^2 & 0\\
0 & 0.3^2 
\end{pmatrix}\label{eq:sysisoap},
\end{equation}
where the matrix elements are the fractional (\%) covariances.

We projecting this to the systematics in $\alpha_\perp$ and $\alpha_\parallel$
following

\begin{align}
{\rm C}^{-1}_{q_iq_j} =  \frac{d\ln p_a}{d\ln q_i} \mathcal{C}^{-1}_{p_a,p_b} \frac{d\ln p_b}{d\ln q_j} \label{eq:alphatransform}
\end{align}
where $q_i,q_j=\alpha_\perp, \alpha_\parallel$ and $p_a,p_b=\aiso, \alap$. With $\mathcal{C}_{p_a,p_b} = \mathcal{C}^{\rm Sys}_{\rm \aiso,\alap}$ and
\begin{equation}
\frac{d\ln p_a}{d\ln q_i}  = 
\begin{pmatrix}
2/3 & 1/3\\
-1 & 1 
\end{pmatrix},
\end{equation}

we find

\begin{align}
{\rm C}^{\rm Sys}_{\alpha_\perp,\alpha_\parallel} =
\begin{pmatrix}
0.265^2& 0.478(0.265\times 0.316)\\
0.478(0.265\times 0.316)& 0.316^2
\end{pmatrix}\label{eq:sysperppar}.
\end{align}
This again is the fractional (\%) covariances.
\edited{For the best BAO measurement, \lrgelg, this increases an error on $\apar$ by 1.5\% and $\aperp$ by 2.2\%.}

Note that, unlike the statistical constraints in \cref{tab:Y1unblinded}, this synthetic matrix for systematics (\cref{eq:sysperppar}) shows a positive cross-correlation, which is mainly because the variances in $\aiso$ and $\alap$ are more comparable, unlike the statistical constraints that are subject to a particular combination of contribution of transverse and parallel wavemode. We add $\mathcal{C}^{\rm Sys}_{\rm \aiso,\alap}$ and ${\rm C}^{\rm Sys}_{\alpha_\perp,\alpha_\parallel}$ to the corresponding statistical BAO covariance to construct the final posteriors. 


\section{The Unblinding Tests}
\begin{table*}
    \centering
    \resizebox{\columnwidth}{!}{
        \begin{tabular}{|l|c|}
        \hline
        {\bf Test} & {\bf Results} \\\hline
        $\chi^2$ acceptable? & Yes  \\\hline
        $\alpha$ consistent between pre and post? & Yes \\\hline
        $\alpha$ consistent between NGC+SGC and NGC? & Yes \\\hline
        Consistency between North and South? & \edited{Tested at the level of the clustering (\cref{sec:catalog})} \\\hline
        $\alpha$ consistent across different fiducial cosmologies? & Yes \\\hline
        $\alpha$ consistent between $\xi(r) $ and $P(k)$? & Yes \\\hline
        Are the error bars reasonable? & \edited{Consistent with the mocks}  \\\hline
        $\alpha$ consistent between LRG and ELG $0.8<z<1.1$ & Yes \\\hline
        \end{tabular}
    }
    \caption{\label{tab:Y1checklist} Unblinding checklists.  For each entry in the 
    checklist, if the difference observed in the blinded data is within the full range covered by the 25 \abacussecond\ DR1 mocks, we consider that the test was passed. \edited{For \bgs, \elgo, and \qso, 1-D BAO fits pass these tests while the other tracers pass these tests with 2-D as well as 1-D fits.}}
    
\end{table*}

\label{sec:unblinding}

\edited{One manifestation of confirmation bias is the tendency to attribute a higher level of systematic uncertainty to outcomes that are unexpected or perceived as outliers. To prevent this from affecting the DESI DR1 BAO measurements, we determined the systematic error budget detailed in \cref{tab:systematicbudget} prior to unblinding.}

Before unblinding, we also required the following set of the conditions to be met.

\begin{enumerate}
\item  The official \desione\ galaxy BAO analysis pipeline must be determined, including the density field reconstruction setup and the BAO fitting setup. We determined these configurations based on the tests on the blinded data and the mocks. These mocks were calibrated against the \desi\ One Percent sample rather than \desione. While this was a limitation potentially affecting optimization for the unblinded data, it was a necessary trade-off due to the blinding process. 
\item  The consistency tests in the checklist (\cref{tab:Y1checklist}) were to have been completed on the blinded data to the fullest extent possible. For each entry in the checklist, if the difference observed in the blinded data is within the entire range covered by the 25 \abacussecond\ DR1 mocks, we consider that the test was passed. 
\end{enumerate}

\vspace{0.5cm}
\cref{tab:Y1checklist} presents the list of the tests we performed before unblinding. The tests are extensively described in \cite{KP4s4-Paillas}. Based on the tests, we decided the following.

\begin{enumerate}
\item We decided the optimal reconstruction scheme and the smoothing scales for each tracer \cite{KP4s4-Paillas}. In particular, we decided to take the reconstructed QSO BAO fit as our fiducial choice regardless of its actual reconstruction efficiency when applied to the unblinded \desione\ data.
\item We decided the prior ranges of the BAO damping scales, as detailed in \cite{KP4s2-Chen}.
\item We decided to perform 1-D BAO fits for the lower signal-to-noise samples, i.e., \bgs, \elgo, and \qso. 2-D BAO fits on these samples showed either frequent outliers (i.e., failed to locate the BAO) in the mocks or failed some of the consistency tests in \cref{tab:Y1checklist} with the blinded data. 
\item \hsc{We decided to utilize the combined tracer of \elgo\ and \lrgth\  for $0.8<z<1.1$ bin to carry out the cosmology analysis. The information from the lowest signal-to-noise tracer, \elgo, is propagated to cosmological constraints only through the combined tracer analysis. } 
\end{enumerate}

\vspace{0.5cm}
We also decided on the aspects of the analysis that could be re-done after unblinding.

\begin{enumerate}
    \item  \edited{The calibration of covariance matrices could be finalized after unblinding, based on \ezmock\ and analytic covariance matrices. Consequently, the final decision on the covariance matrices between the numerical and analytic covariances could also be made at this stage. This was partly due to the unavailability of \ezmocks\ for DR1 during the unblinding tests, and also because we could not determine beforehand how well the clustering in the \ezmocks\ and \abacussecond\ would align with that of the unblinded data.} 
    \item Given the bias expected with an incorrect assumption of $N_{\rm eff}$ in the template (\cref{subsec:sys-fiducialcosmo}), we allowed a potential iteration of the BAO fits using a different assumption of $N_{\rm eff}$ \edited{if we were to find the best fit $N_{\rm eff}$ from the \desidrone\ cosmology analysis different to the value we assumed in the fitting}.
    \item An update on the catalogs could be made if an error were to be found in checks unrelated to the BAO fitting. The error would then be fixed before assessing the impact of the correction on the BAO fits. \edited{As detailed in \cref{subsec:blinding}, an error was indeed identified in the LSS catalogs after unblinding and promptly rectified. The most significant shift on the BAO measurement observed was 0.7$\sigma$ shift in $\alap$ for \lrgo, while other shifts are typically less than 0.2$\sigma$.  }
    \item HOD systematics for QSO and BGS would be determined after unblinding, \edited{as a result of prioritizing the highest signal-to-noise tracers such as \lrgs\ and \elgs.}
\end{enumerate}

\edited{After unblinding, we discovered a minor discrepancy between the fitting model outlined in \cite{KP4s2-Chen} (and \cref{sec:model_fitting} in this paper) and its implementation in the code. 
The correction of the code resulted in BAO measurement changes of less than a maximum of $0.3 \sigma$, with most shifts under $0.1 \sigma$, and no systematic correlation in the shifts across redshift bins.} 

\section{Results}
\label{sec:results}

In this section, we present the results of the BAO analysis using the unblinded \desidrone\ catalogs. 

\begin{figure*}
\centering
\includegraphics[width=0.38\linewidth]{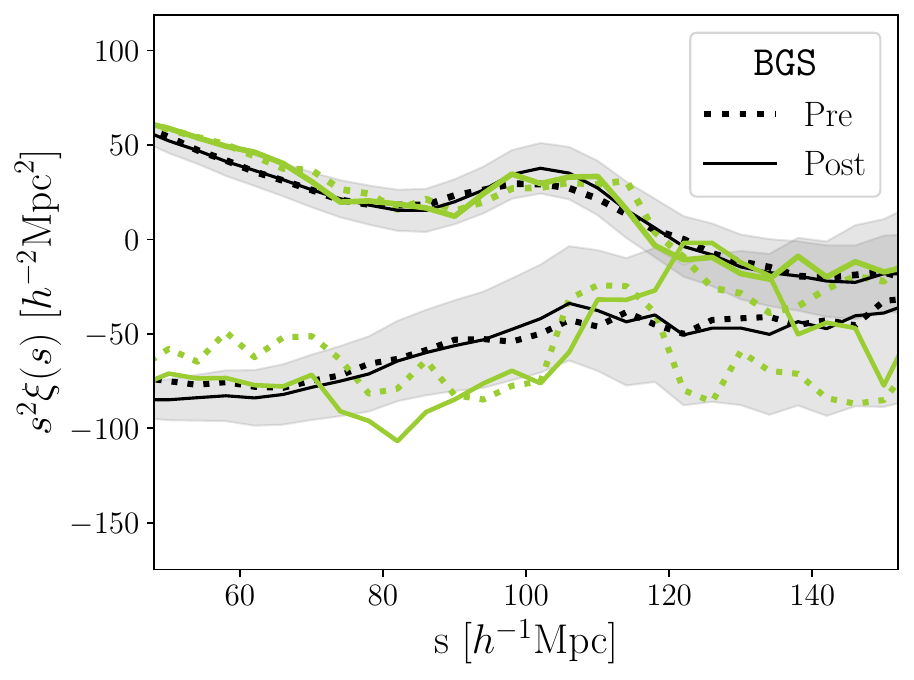}
\includegraphics[width=0.38\linewidth]{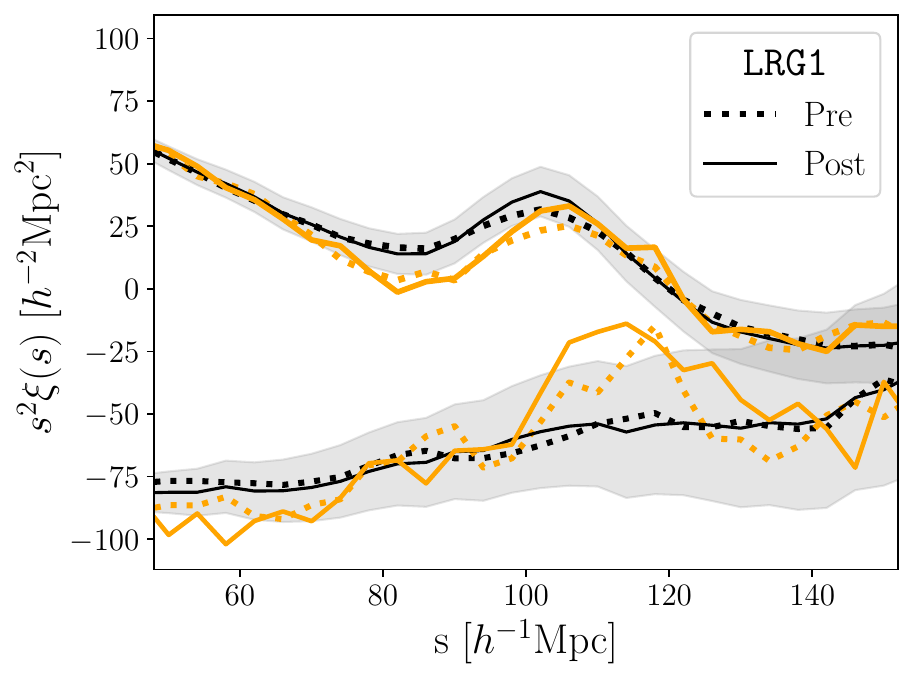}
\includegraphics[width=0.38\linewidth]{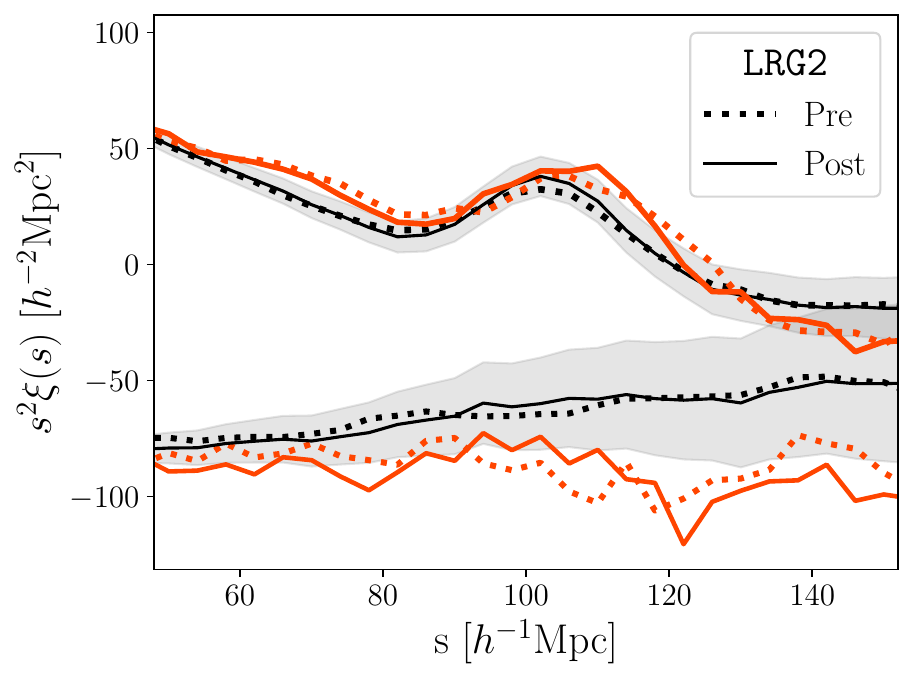}
\includegraphics[width=0.38\linewidth]{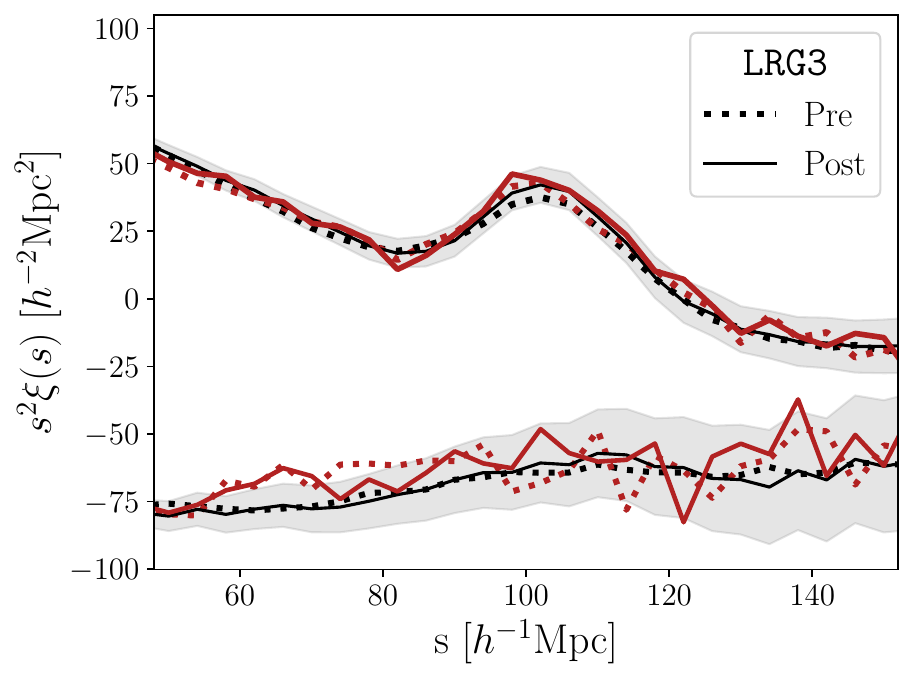}
\includegraphics[width=0.38\linewidth]{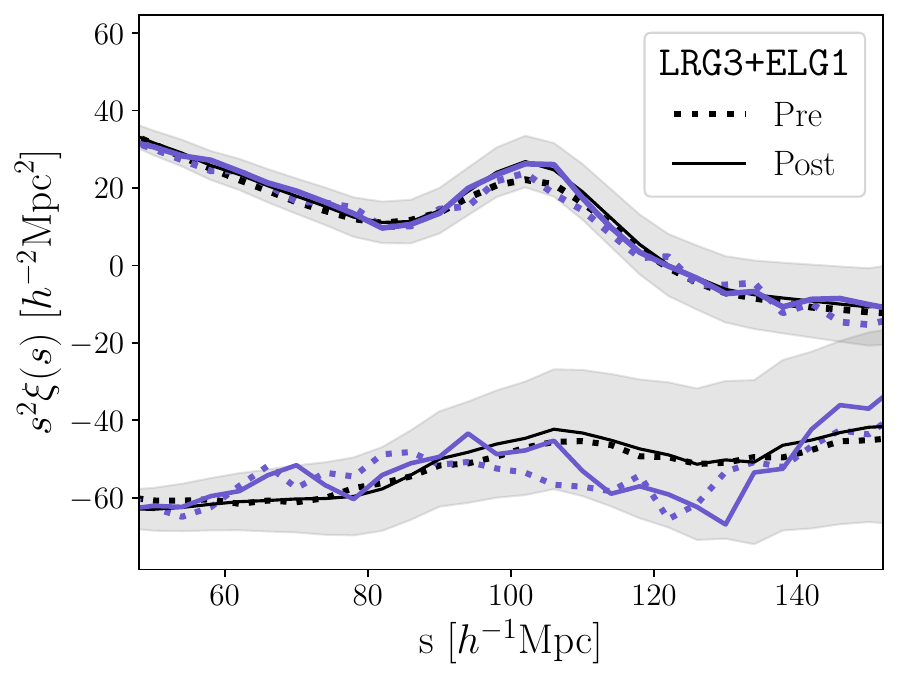}
\includegraphics[width=0.38\linewidth]{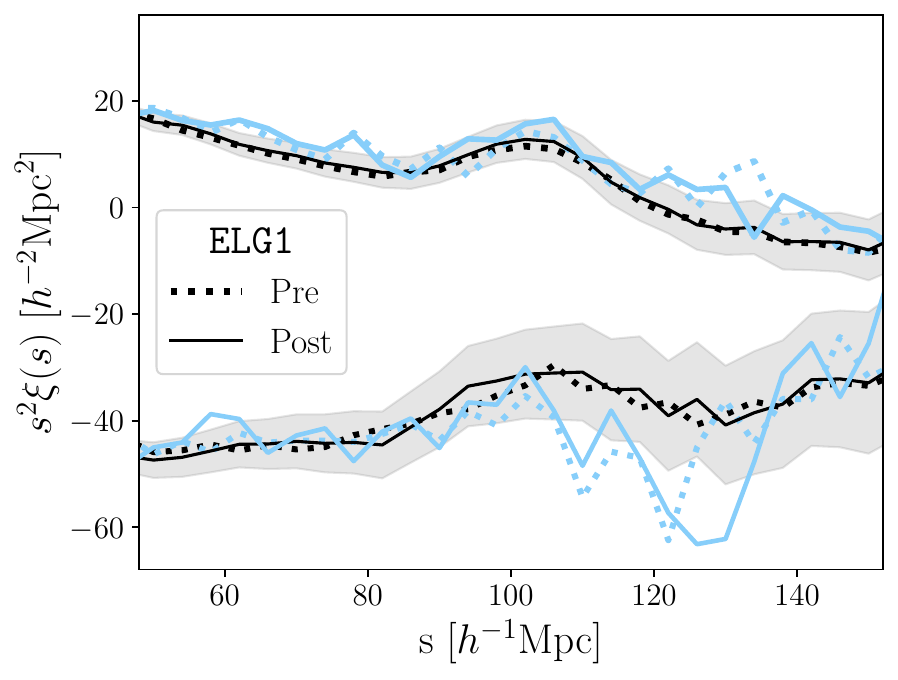}
\includegraphics[width=0.38\linewidth]{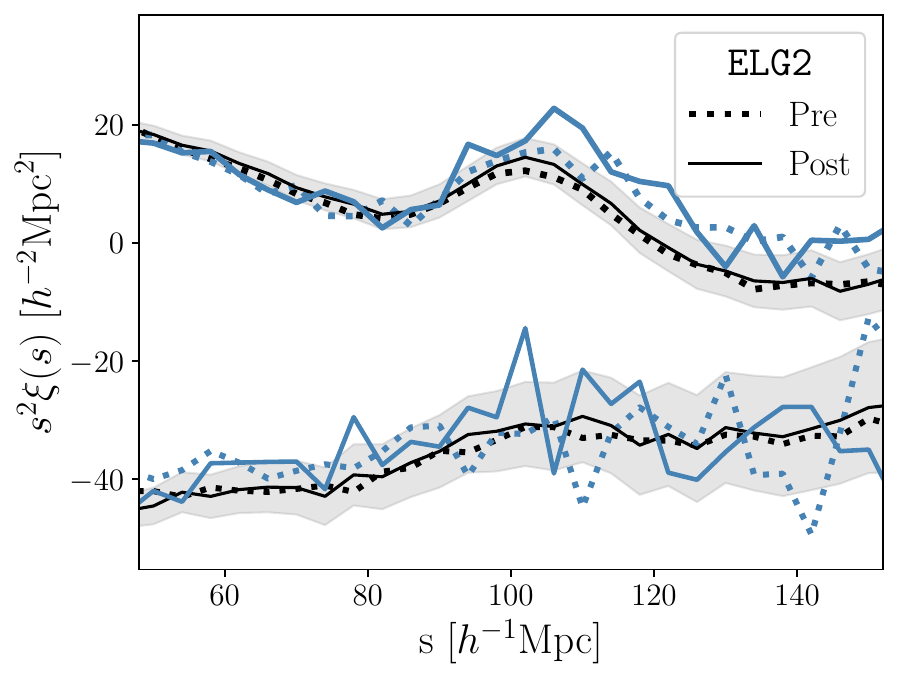}
\includegraphics[width=0.38\linewidth]{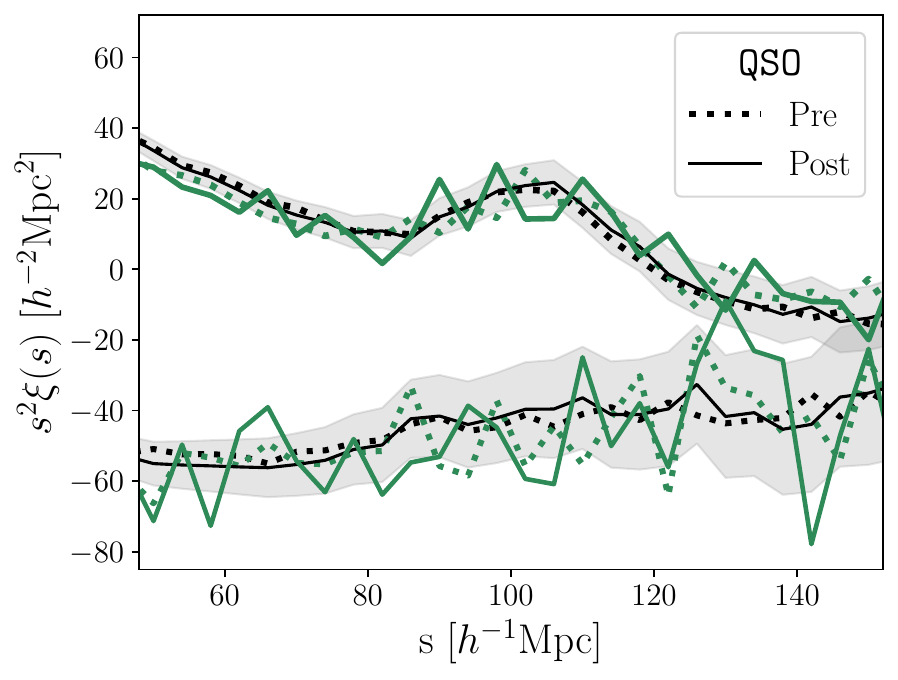}
\caption{Two-point correlation functions of various tracers of the unblinded \desidrone\ (colored lines). We compare the overall clustering amplitude over the BAO fitting range ($48\hMpc<r<152\hMpc$) before (dotted lines) and after reconstruction (solid lines), compared to the corresponding mean of the 25 \abacussecond\ DR1 mocks (black lines). \edited{Gray shading represents the error associated with post-reconstruction \desidrone\ based on \rascalc\ covariance. As a reminder, the data points of $\xi$ are substantially correlated between different separation $r$.}  The upper set of the lines is for the monopole, and the lower set of the lines is for the quadrupole. The plots demonstrate excellent consistency in the overall clustering between the data and the mocks. }
\label{fig:XiY1PrePost}
\end{figure*}

\begin{figure*}
\centering
\includegraphics[width=0.4\linewidth]{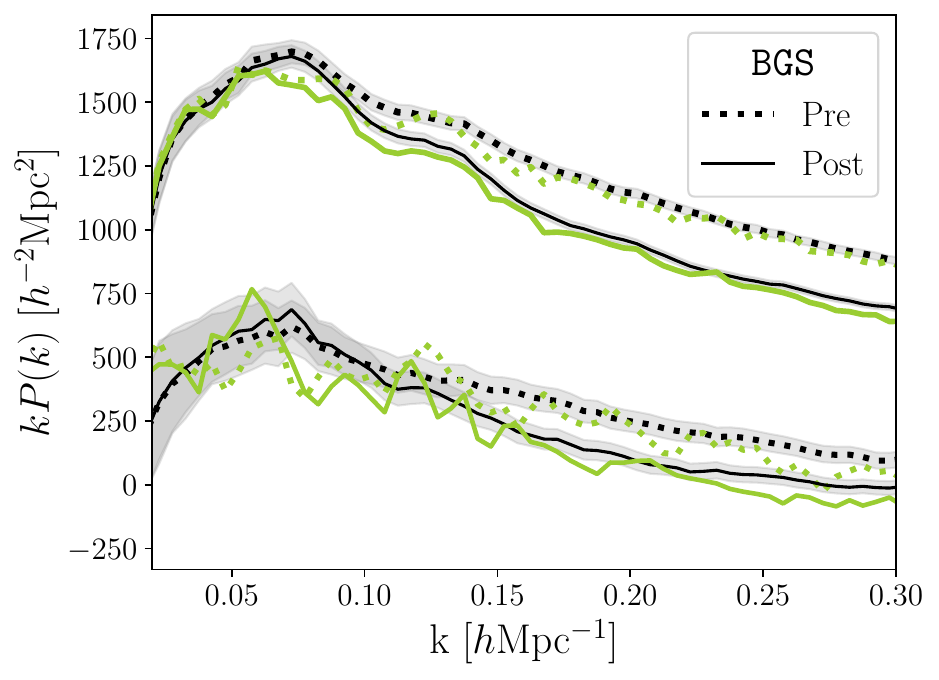}
\includegraphics[width=0.4\linewidth]{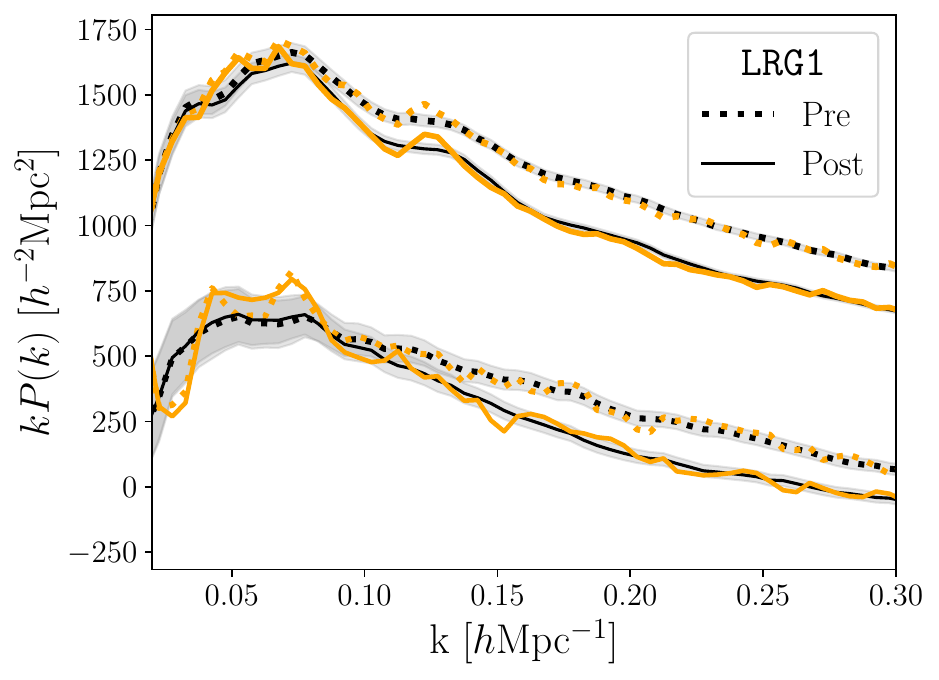}
\includegraphics[width=0.4\linewidth]{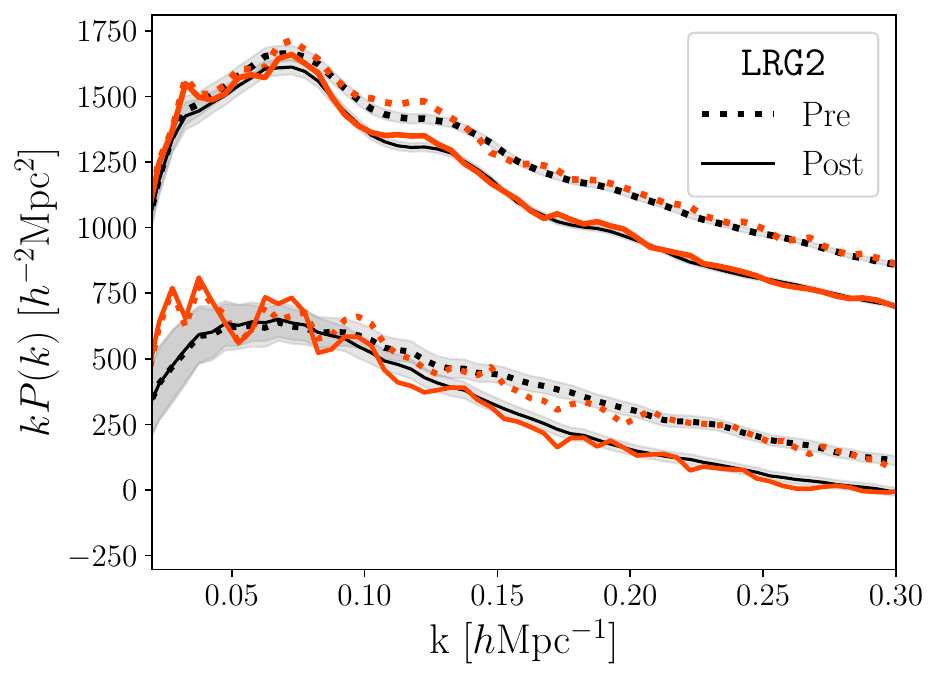}
\includegraphics[width=0.4\linewidth]{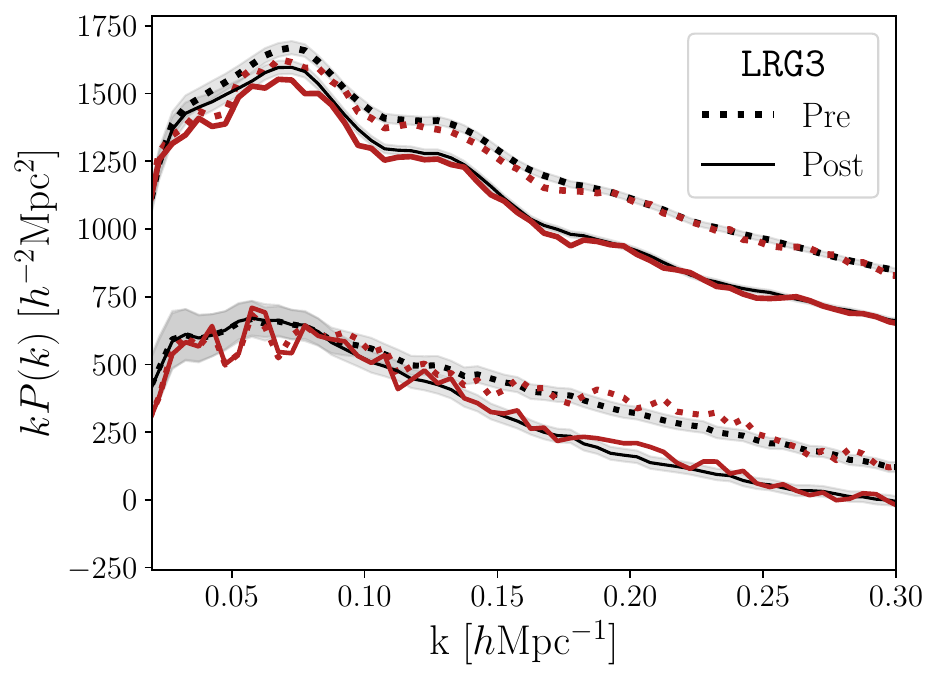}
\includegraphics[width=0.4\linewidth]{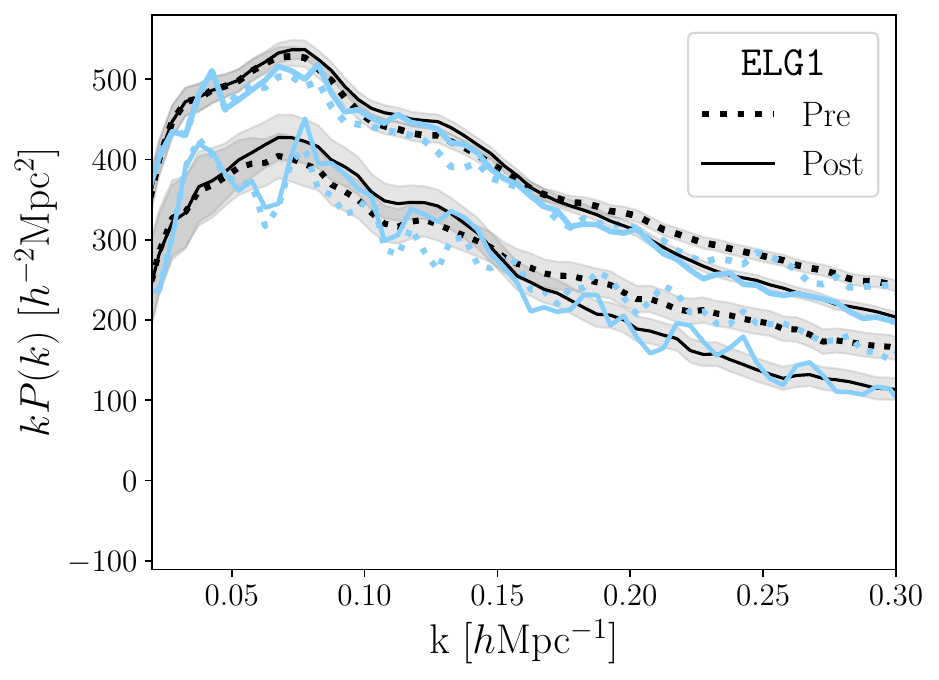}
\includegraphics[width=0.4\linewidth]{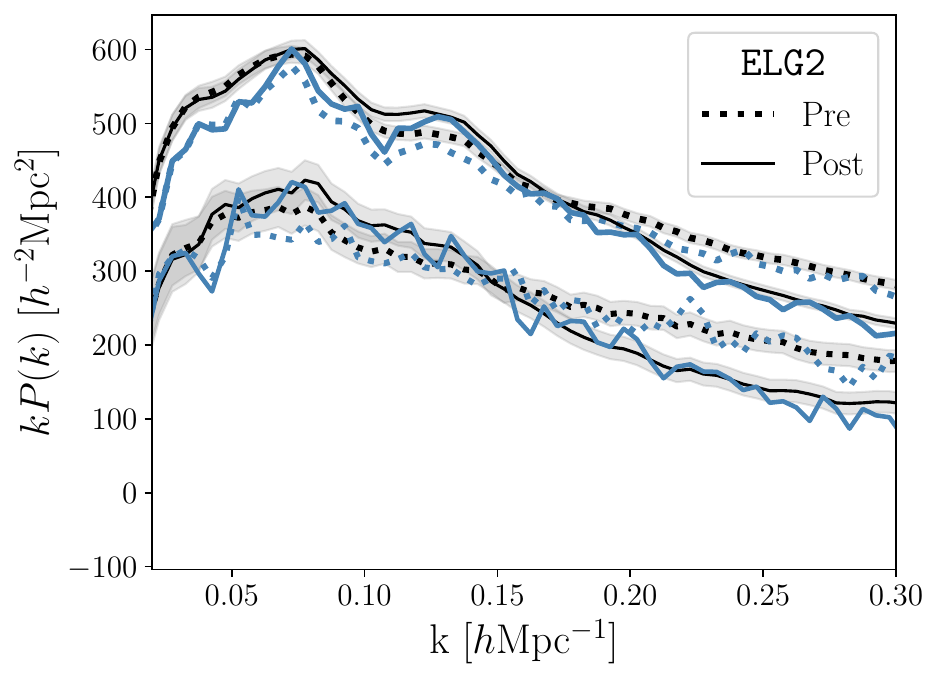}
\includegraphics[width=0.4\linewidth]{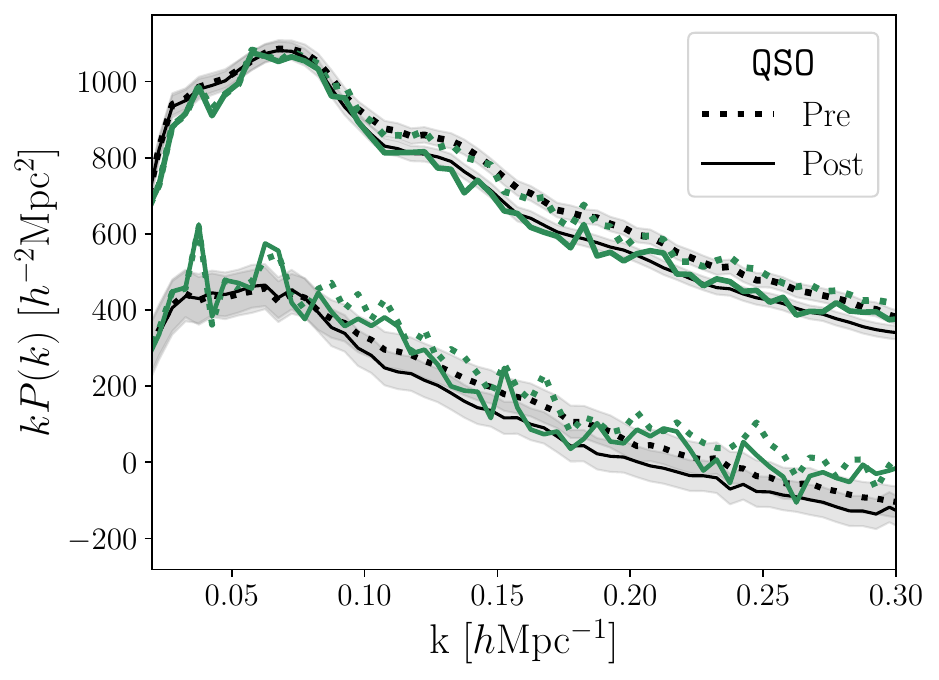}
\caption{Power spectra multipoles of various tracers of the unblinded \desidrone\ (colored lines) in comparison to the \abacussecond\ DR1 mocks (black lines). Similar to \cref{fig:XiY1PrePost}. The gray shading is from the analytic covariance matrices for the \desidrone\ data in \cref{subsec:methods-cov}. Again, the plots demonstrate excellent consistency in the overall clustering between the data and the mocks.}
\label{fig:PkY1PrePost}
\end{figure*}


\begin{figure*}
    \centering
    \includegraphics[width=0.32\textwidth]{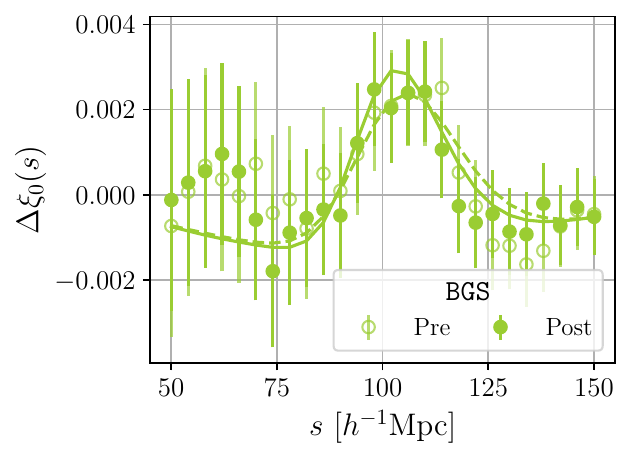}
    \includegraphics[width=0.32\textwidth]{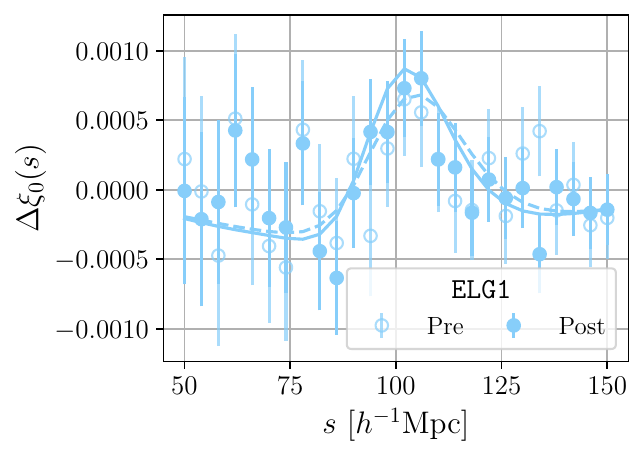}
    \includegraphics[width=0.32\textwidth]{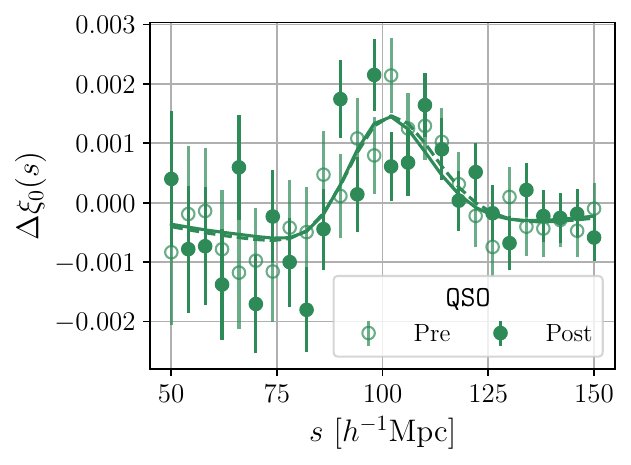}
     \includegraphics[width=0.32\textwidth]{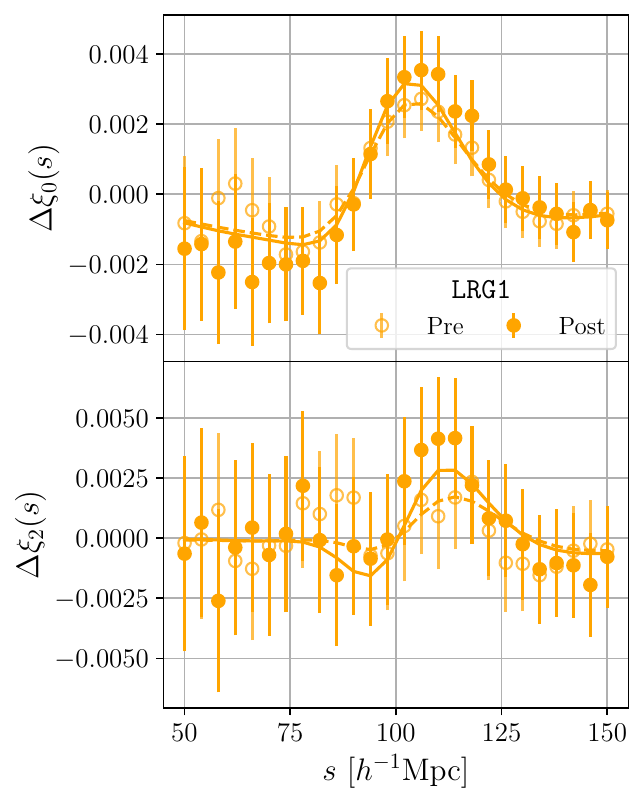}
      \includegraphics[width=0.32\textwidth]{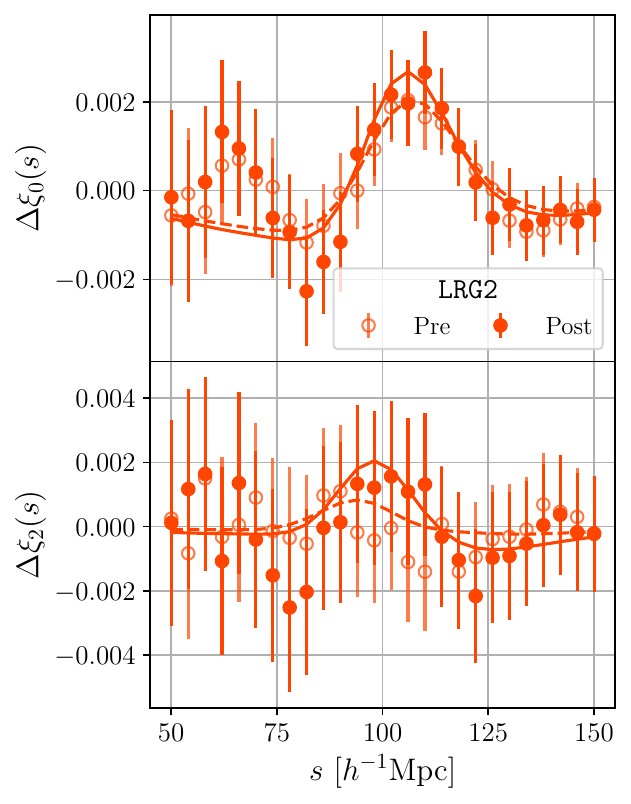}
       \includegraphics[width=0.32\textwidth]{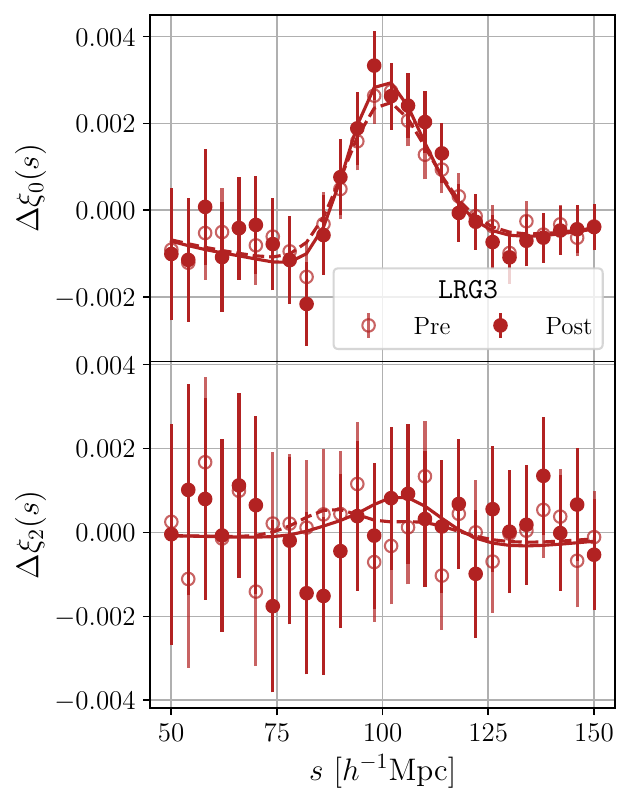}
              \includegraphics[width=0.32\textwidth]{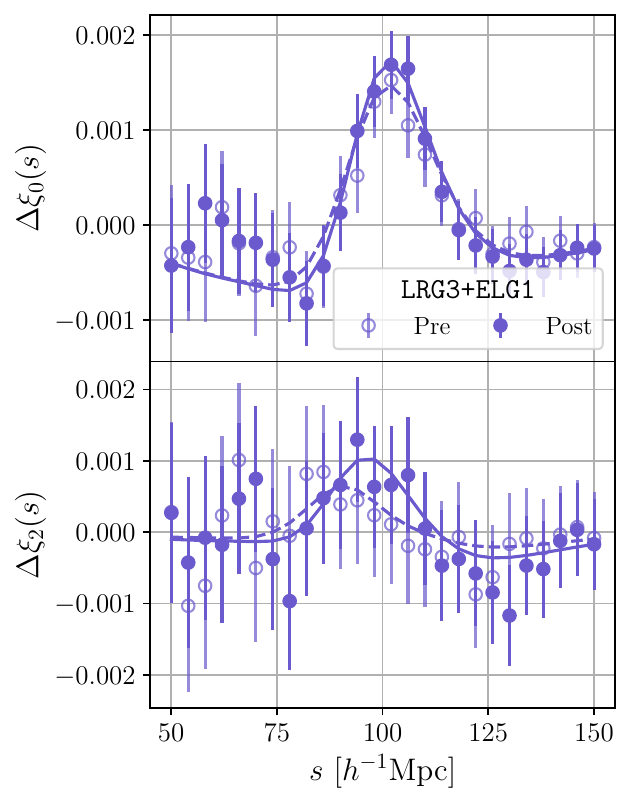}
         \includegraphics[width=0.32\textwidth]{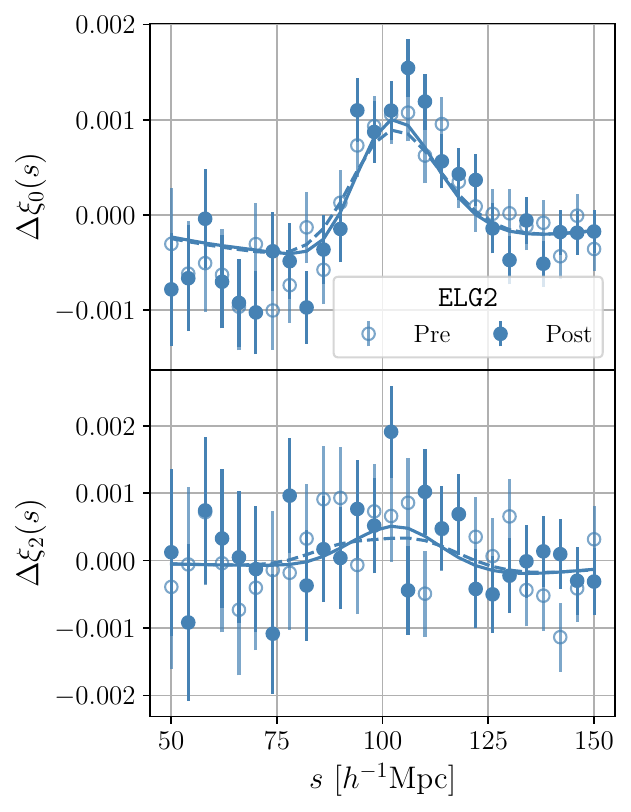}

      \caption{The isolated BAO feature in the correlation function of \desione\ data  before (open circles) and after reconstruction (solid circles). A 1-D BAO fitting is performed for \bgs, \elgo, and \qso, while the rest is fitted for the 2-D BAO scales. The solid and dashsed lines are the best fit BAO models to the unblinded \desidrone\ before (open circles) and  after reconstruction (solid circles), respectively.}
\label{fig:y1xiunblindedwigprepost}
\end{figure*}

\begin{figure*}
    \centering
        \includegraphics[width=0.45\textwidth]{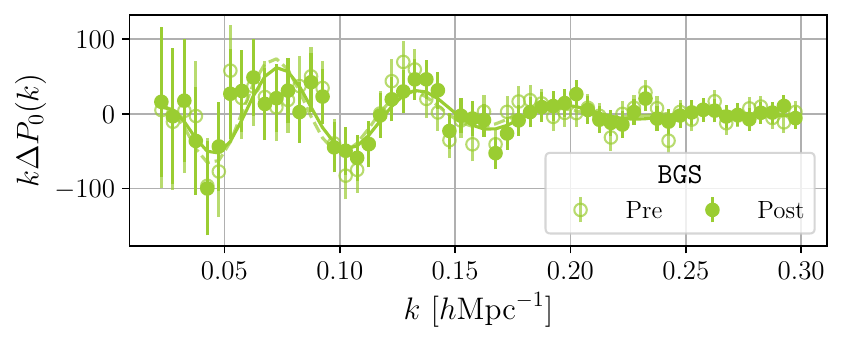}
             \includegraphics[width=0.45\textwidth]{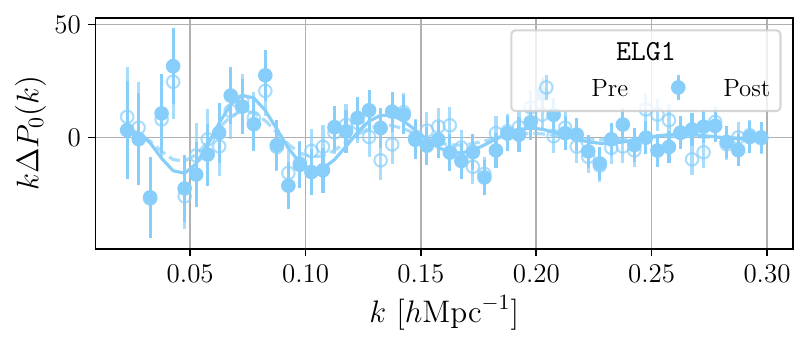}
             \begin{center}
              \includegraphics[width=0.45\textwidth]{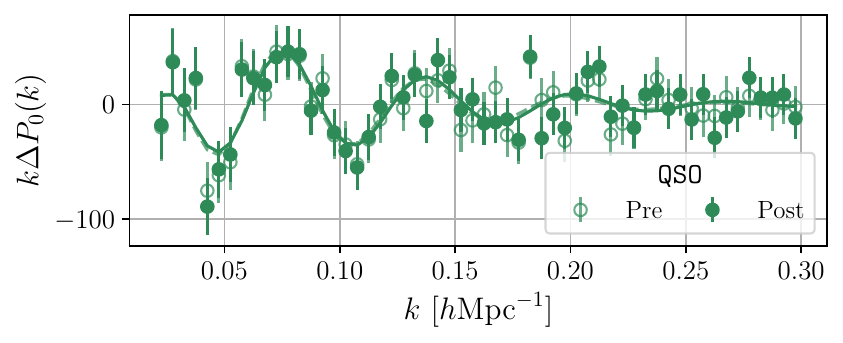}
              \end{center}
     \includegraphics[width=0.45\textwidth]{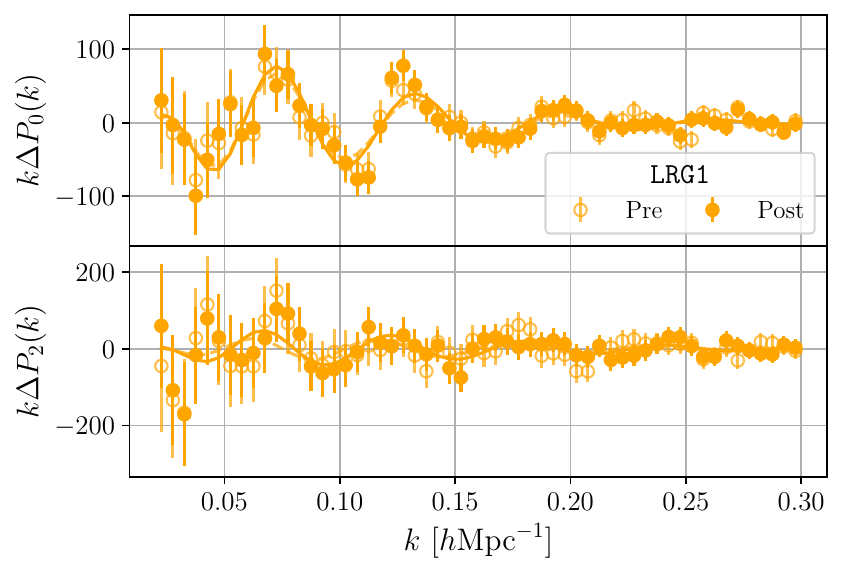}
      \includegraphics[width=0.45\textwidth]{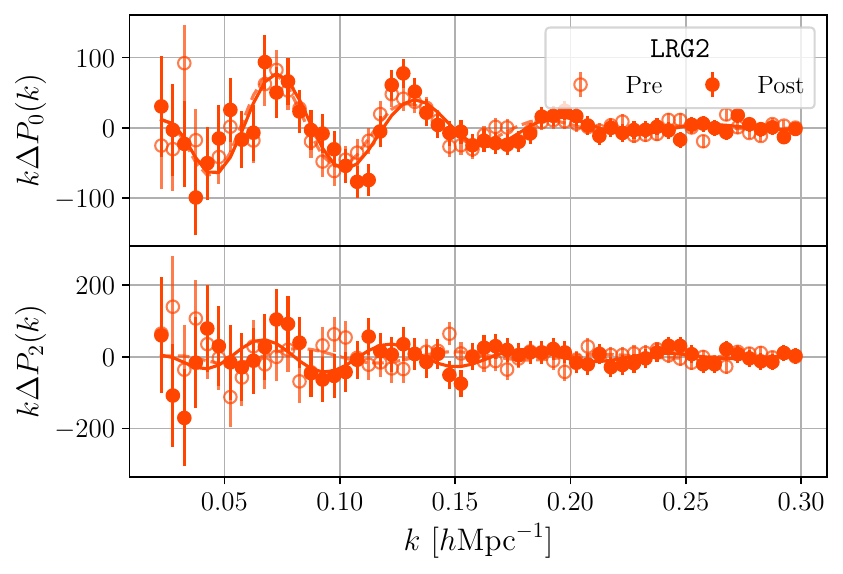}
        \includegraphics[width=0.45\textwidth]{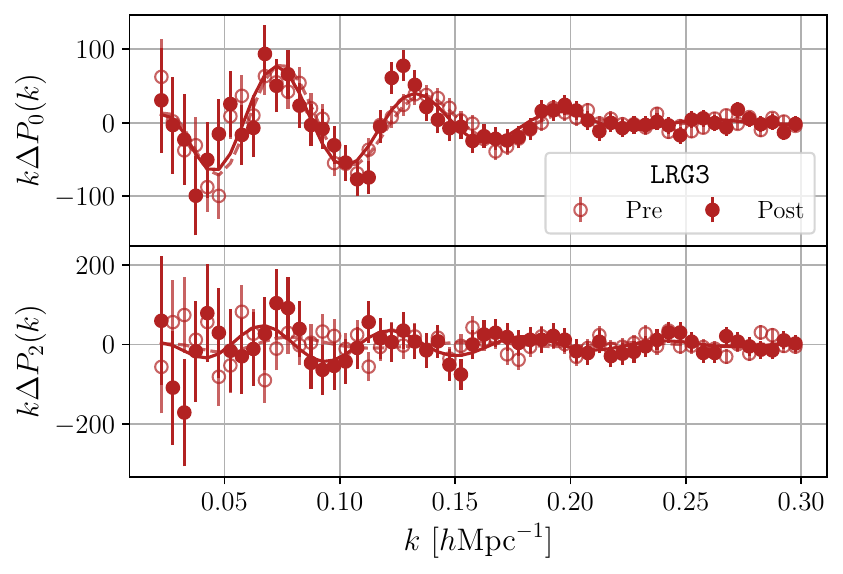}
         \includegraphics[width=0.45\textwidth]{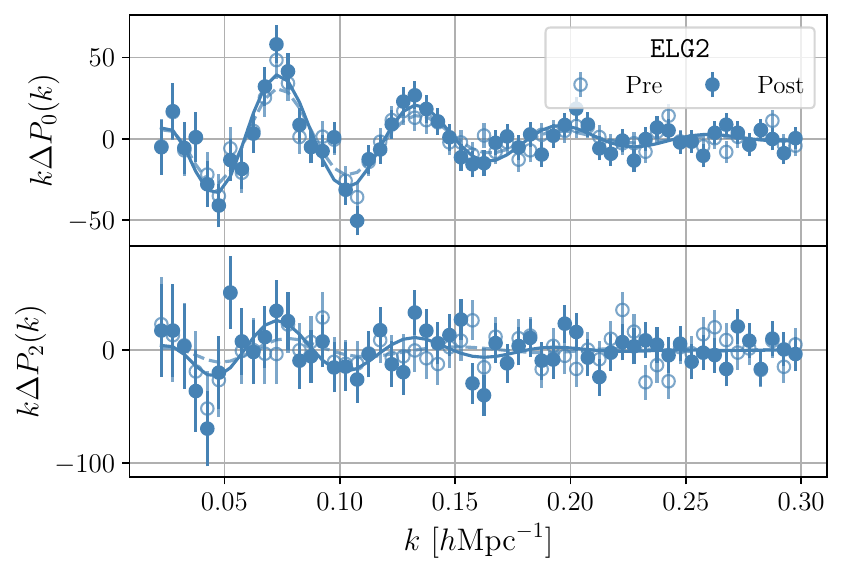}  
      \caption{The isolated BAO feature in the monopole and quadrupole before (open circles) and after reconstruction (solid circles) in the power spectrum. The solid and dashsed lines are the best fit BAO models to the unblinded \desidrone\ before (open circles) and  after reconstruction (solid circles), respectively. The unit in the y axis $h^{-2}{\rm Mpc}^{2}$ is omitted due to the limited space.}
\label{fig:y1pkunblindedwg}
\end{figure*}

\subsection{Reconstructed two-point clustering of \desidrone}

We start with presenting the clustering measurements of the reconstructed catalogs using our default reconstruction method described in \cref{sec:methods}. 
While we refer to \cite{DESI2024.II.KP3} for the full quantitative discussion comparing the mocks and the data, here we note the excellent consistency between the overall clustering amplitude of the \abacussecond\ DR1 mocks and that of the data as a function of scale within the fitting ranges pre and post reconstruction, as shown in \cref{fig:XiY1PrePost,fig:PkY1PrePost}. 


\cref{fig:XiY1PrePost,fig:PkY1PrePost} also compare the pre-reconstruction versus post-reconstruction clustering of \desidrone\ within the fitting ranges. As we adopted the \recsym\ convention, which preserves the linear RSD boost along the line of sight through the displacement of the random particles, the clustering before and after reconstruction is quite consistent on large scales, as expected. On smaller scales we see in \cref{fig:PkY1PrePost} that the post-reconstruction power spectrum is reduced relative to the pre-reconstruction one---this is not unexpected as in general we expect the reconstructed density field to have different nonlinear power due to the coordinate remapping, and there is some analytic evidence for the decrease of power specifically in the case of the post-reconstruction matter power spectrum \cite{Seo2010:0910.5005,White15,Hikage17,Chen19b}. From \cref{fig:PkY1PrePost}, the change in clustering before and after reconstruction of the data (colored solid and dotted lines for monopole and quadrupole, respectively) appears highly consistent with the expectation based on \abacussecond\ DR1 mocks within the $1\sigma$ dispersion expected from the \rascalc\ covariance matrice (the black solid and dashed lines with the gray shade for the $1\sigma$ dispersion).



The BAO feature appears moderately sharpened by reconstruction in the \lrgs\ redshift bins, while the improvement is less obvious for other tracers. One can see that the \abacussecond\ DR1 mocks replicate the observed level of the BAO sharpening in the \desidrone\ data. Hence, we qualitatively find that the reconstruction of the data is performing as expected given the survey configuration and the reconstruction method we chose. In the next section, where the results of the BAO fits are discussed, we make quantitative comparisons on the aforementioned aspects and show that
all tracers had moderate gain from reconstruction at the level consistent with the mocks.  


\subsection{BAO measurements from the \desidrone\ galaxies }\label{sec:baodesione}

In this section, we present the BAO fitting of all galaxy and quasar tracers
using the default fitting method defined in \cref{subsec:methods-fitting}. 
We focus on the fitting results in configuration space for our discussion below, 
although we get consistent results in Fourier space \cref{se:app}.\footnote{Additionally, 
since our analytic and mock-based covariances have not fully converged in our Fourier space analysis, 
we view the configuration space results as more robust.}
All
results except for \elgo\ \edited{and \lrgth} will be directly used for the
\desione cosmology analysis of \cite{DESI2024.VI.KP7A}, \edited{with \elgo\ and
\lrgth\ incorporated into the analysis via \lrgelg.}
Based on the unblinding tests and the decisions presented in \cref{sec:unblinding}, we performed two-dimensional BAO fits ($\alphaiso$ and $\alphaap$) for all tracers other than \bgs, \elgo, and \qso. For the latter, we derive $\alphaiso$ using only the monopole data.


\cref{tab:Y1unblinded}, \cref{tab:Y1unblindedPk}, \cref{fig:y1xiunblindedwigprepost} and \cref{fig:y1pkunblindedwg} summarize the resulting BAO constraints and visualize the isolated BAO features.
\cref{tab:Y1unblinded} shows that the $\chi^2$ values of the 2-D fits to the correlation functions fall between \edited{33.1 for \lrgelg\ and}  58.6 for \elgt, all for 39 degrees of freedom. 
The cases of 1-D fits show $\chi^2=17.3$ for \bgs, $\chi^2=20.4$ for \elgo, and $\chi^2=32.4$ for \qso\ for 19 degrees of freedom. The goodness-of-fit is reasonable; the worst cases, \elgt\ gives an associated p-value of 2.27\% and the \qso\ gives an associated p-value of 2.82\%.
\cref{tab:Y1unblindedPk} in appendix summarises the corresponding fits using the power spectrum multiples and shows a reasonable level of the goodness-of-fit with $P(k)$ as well \edited{(a smallest p-value of 4.35\% for \qso)}. The BAO measurement scales for the two conjugate spaces are very consistent (less than 0.3\% in $\aiso$ and $\alap$ except for \qso). \edited{This level of consistency between the BAO fits between $\xi(r)$ and $P(k)$ agrees with the \abacussecond\ DR1 mocks (\cite{KP4s4-Paillas}).}

\begin{figure*}
\centering
\begin{tabular}{cc}
  \hspace{-0.3cm}\includegraphics[width=0.49\linewidth]{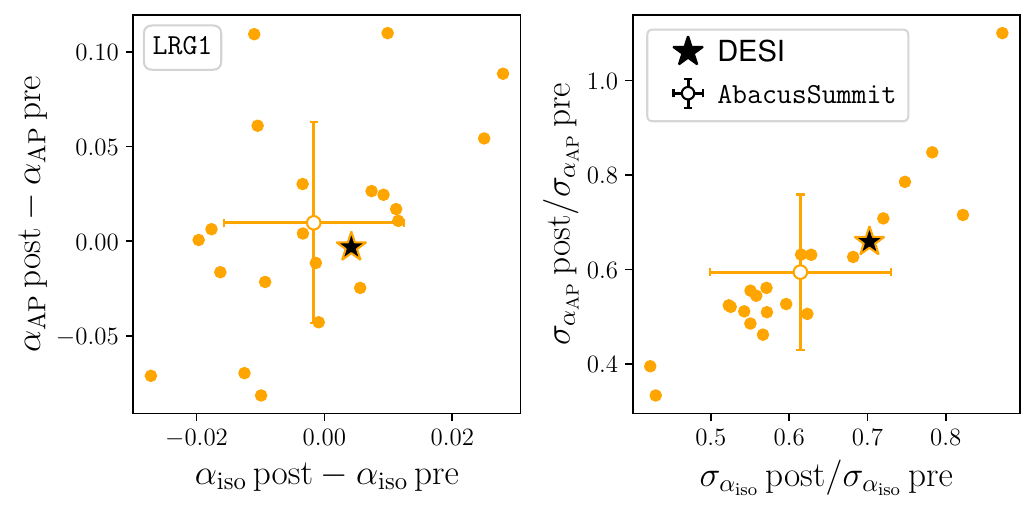}   & \hspace{-0.25cm}\includegraphics[width=0.49\linewidth]{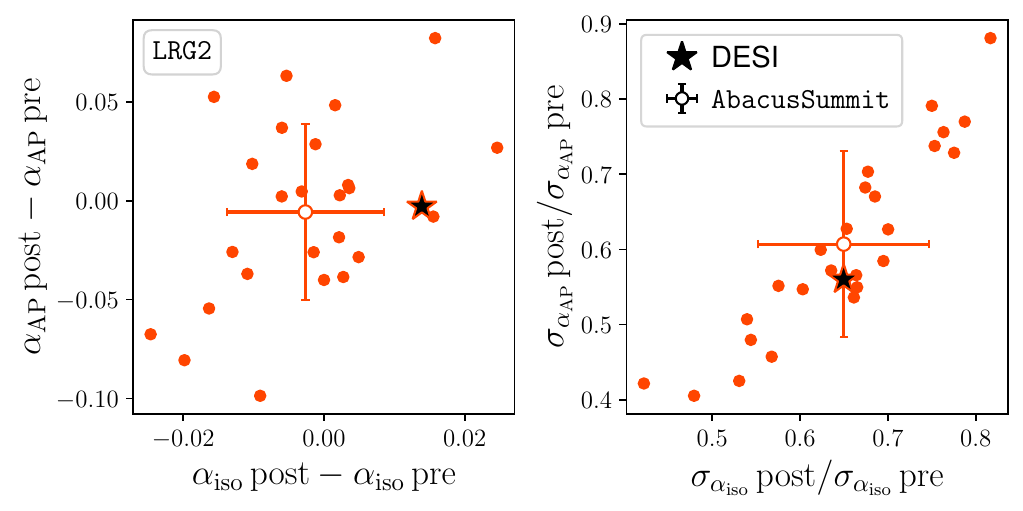} \\
  \hspace{-0.25cm}\includegraphics[width=0.49\linewidth]{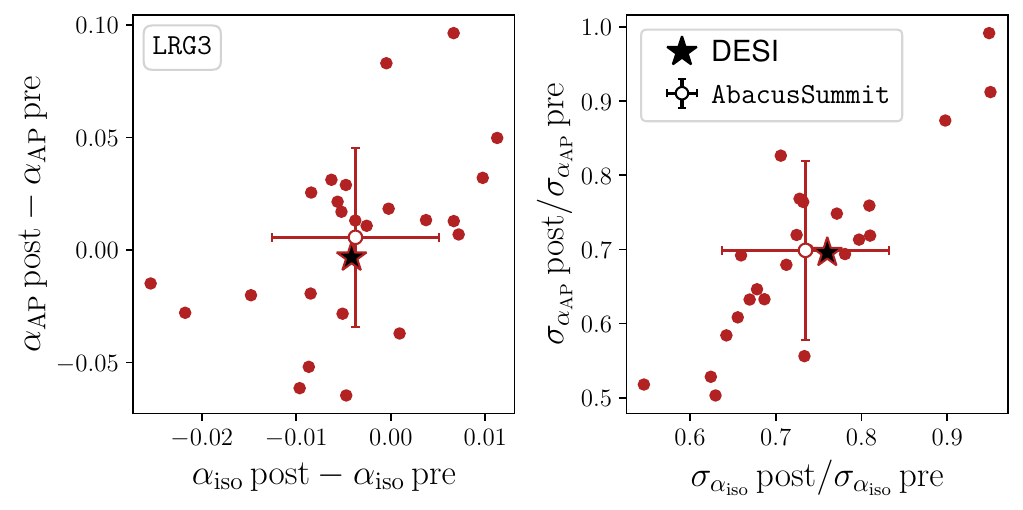} & 
\hspace{-0.25cm}\includegraphics[width=0.49\linewidth]{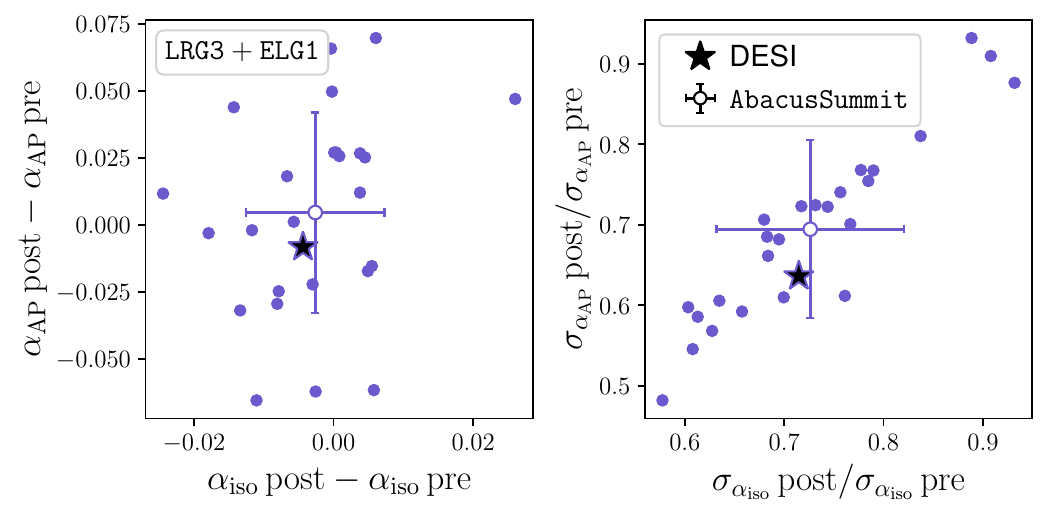} \\
\hspace{-0.25cm}\includegraphics[width=0.49\linewidth]
{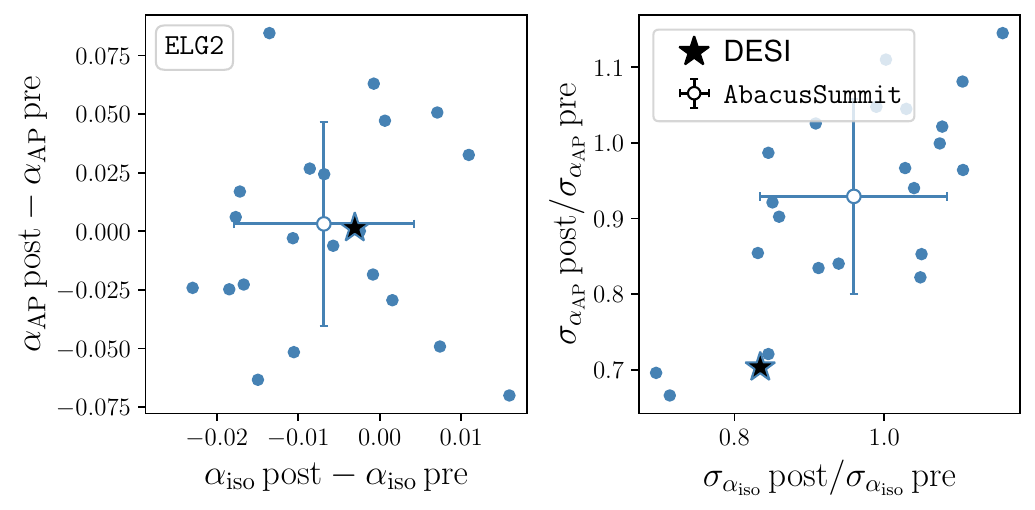}
\end{tabular}
\caption{Pre- and post-reconstruction BAO measurements, comparing \desidrone\ measurements (stars) with the 25 \abacussecond\ DR1 (the open points with the error bars are the means and the standard deviations around the means of the 25 mocks). Due to the discrepancy between the BAO scales imprinted in the mocks and the BAO scales recovered from the data, we focus on the relative comparison between the pre and the post reconstruction. All \desidrone\ fits are consistent with what is expected from the mocks.}\label{fig:scatteralpha}
\end{figure*}

\begin{figure*}
\centering
\begin{tabular}{ccc}
  \includegraphics[width=0.3\textwidth]{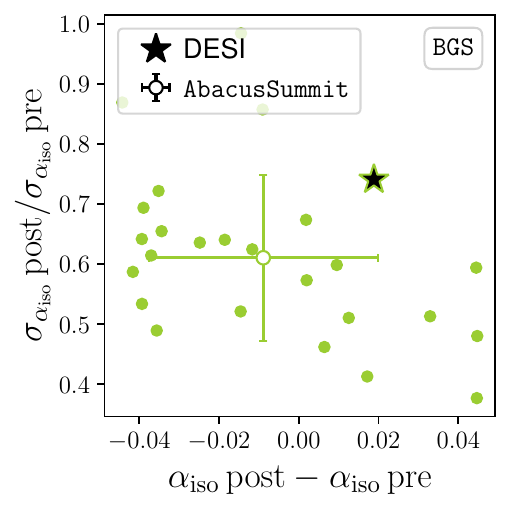}   & \includegraphics[width=0.3\textwidth]{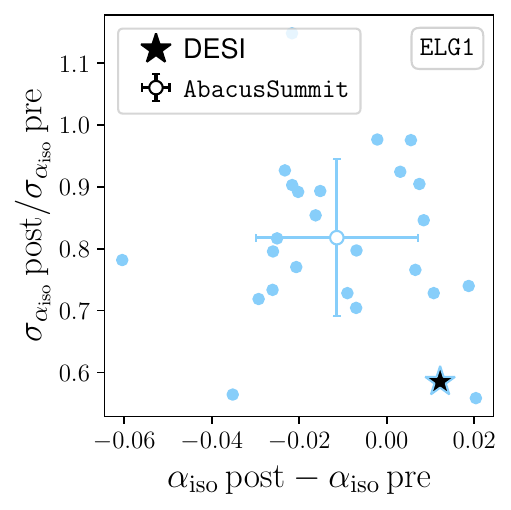} & 
\includegraphics[width=0.3\textwidth]{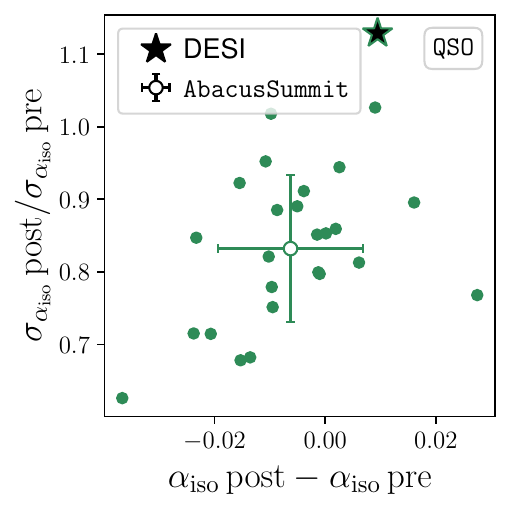} 
\end{tabular}
\caption{Similar to \cref{fig:scatteralpha}, but showing results for \bgs, \elgo, and \qso. The \desidrone\ results are, again, consistent with the range covered by \abacussecond\ DR1, except for \qso; the result of \qso\ shows the reconstruction of \desidrone\ is less efficient than the worst case of the mocks.}\label{fig:scatteralpha1d}
\end{figure*}

\cref{fig:y1xiunblindedwigprepost} displays the BAO features in the correlation function selected for each tracer, compared to the best-fit $\xi$ models for pre-reconstruction (dashed lines) and post-reconstruction data (solid lines). For the 2-D cases, \lrgo\ shows a distinct BAO feature in the post-reconstruction quadrupole measurement. This is consistent with the $2.29\sigma$ offset of $\alphaap -1$ measured in \cref{tab:Y1unblinded}. The more the underlying $\alphaap$ differs from the assumed $\alphaap=1$ of the fiducial cosmology, the greater extent to which the feature in the quadrupole would be shifted from that of the monopole and also amplified, as the BAO feature leaks from the monopole to an extent proportional to $\alphaap-1$.\footnote{From \cite{Padmanabhan08}, for a small change in $(\epsilon+1)^3 \equiv \alphaap$, the monopole and quadrupole change as
\begin{eqnarray}
P_0 &\rightarrow& P_0 - \frac{2\epsilon}{5}\frac{dP_2}{d\ln k}-\frac{6\epsilon}{5}P_2,\\ \label{eq:Padone}
P_2 &\rightarrow & \left(1-\frac{6\epsilon}{7}\right) P_2 - \frac{4\epsilon}{7}\frac{dP_2}{d\ln k}- 2\epsilon\frac{dP_0}{d\ln k}.\label{eq:Padtwo}
\end{eqnarray}}
This behaviour of the BAO feature in the quadrupole is not evident in the mock data, as the mocks consistently assume $\alphaap=1$ when the they were analyzed. That is, the discrepancy between the BAO signal in the mock and the data merely indicates a measurement of $\alphaap \ne 1$ with $2.29\sigma$ significance. A similar quadrupole feature and the consistent $\alphaap$ best fit is derived from the power spectrum fit. 

\begin{figure}
    \centering
    \includegraphics[width=0.6\textwidth]{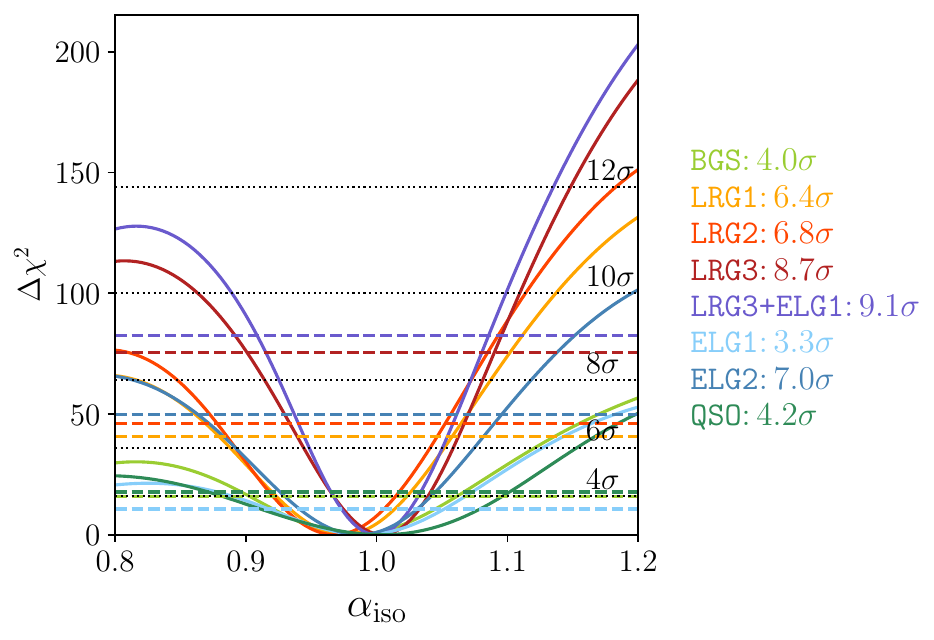}
    \caption{\edited{Detection significance of the BAO features in various \desidrone\ galaxy tracers. Solid lines show $\Delta\chi^2$ of the BAO fits as a function of $\aiso$, i.e., $\chi^2$ offset from the minimum $\chi^2$. The
 dashed lines correspond to $\Delta\chi^2$ of the noBAO fits, i.e., using a model without the BAO wiggles, after subracting the minimum $\chi^2$ of the BAO fit for each tracer. The detection level of the BAO feature for each tracer is shown in the legend. }}
    \label{fig:bao_detection_level}
\end{figure}

The reconstruction applied to the QSO sample is less effective than the other tracers due to its high shot noise, as depicted in \cref{fig:y1xiunblindedwigprepost} and detailed in \cref{tab:Y1unblinded,tab:Y1unblindedPk}. The mock analysis in \cite{KP4s4-Paillas} suggests \edited{$3-8\%$}  gain in the precision after reconstruction, though individual realizations will vary considerably due to high shot noise. Additionally, \cite{KP4s2-Chen} and \cite{SeononlinearBAO} find that on average the reconstruction will unbias the BAO location, even in this regime of high shot noise. 
 Given these findings, in \cref{sec:unblinding}, we opted to use the reconstructed QSO result as our fiducial 
 while recognizing its inefficiency from the unblinding test. This decision ensures consistency across cases rather than relying on the realization-dependent performance. In particular, with the future \desi\ data releases, we anticipate an improved reconstruction performance for the QSO sample, with a reduced fluctuation from realization to realization.

\cref{fig:scatteralpha,fig:scatteralpha1d} compare the pre- and post-reconstruction BAO fits for the data and the 25 \abacussecond\ DR1 mocks. 
Due to the discrepancy between the BAO location assumed in the mocks and its measurement in the data, we focus on the changes in the best fits and precision. 
Except for \qso, the measurements from the data (indicated by stars) lie well within the distribution of the 25 \abacussecond\ DR1 mocks, which is the criteria we set as ``consistent" given the small number of the mocks. For \qso, the blinded data BAO fit was marginally consistent with the upper boundary (the least effective reconstruction limit) reached by the mocks, therefore passing the unblinding test (\cite{KP4s4-Paillas}). However, the reconstruction of the unblinded \desidrone\ is slightly worse than the blinded case and less efficient than in the worst case of the mocks. We believe that such a slight change before and after unblinding is acceptable.

In detail, we observe the change in the best fit $\alphaiso$ with reconstruction is the greatest for \lrgt, giving an 1.6\%   shift \edited{($0.87\sigma$ compared to its pre-reconstruction precision)} in $\xi$.  In terms of the precision, \lrgs\ show the biggest improvement from reconstruction, \edited{a factor of 1.3-1.6 higher precision} in $\sigma(\alphaiso)$ and \edited{a factor of 1.5-1.8 } in $\sigma(\alphaap)$. For other tracers, we find an improvement of \edited{1.4 for \bgs, a factor of 3 for \elgo\footnote{\edited{For \elgo, the pre-reconstruction constraint is overestimated as the posterior distribution deviates from Gaussian distribution due to its low signal-to-noise.}}, and 1.2 ($\aiso$) and 1.4  ($\alap$) for \elgt.} Reconstruction of these tracers is not as effective as \lrgs. \edited{In particular, the reconstruction for \elgt\ is very likely inhibited due to its low completeness and irregular footprint for \desidrone. Again, with the future \desi\ data releases, we anticipate an improved reconstruction performance for the \elgs\ sample.} 
To summarize, the BAO reconstruction fits lead to a substantial gain for the \lrgs, but only a moderate gain for the other tracers.

\cref{fig:bao_detection_level} shows the detection level of the BAO feature. For this plot, one-dimensional BAO fits were performed for all tracers including \lrgs\ and \elgt. Following the standard procedure \cite{Eisenstein05}, the fits with the template without the BAO feature were also made (i.e., setting the second term in \cref{eq:generic_model} to zero, hereafter a `noBAO fit'). The square root of the difference in $\chi^2$ between the pair of the BAO and noBAO fits at the best fit $\aiso$ of the former is quoted as the detection significance.\footnote{\cite{Eisenstein05} presented the detection significance estimation using the same number of free parameters in the BAO fit and the noBAO fit, with the only degree of freedom being the existence of the BAO. In our case, the BAO fit parameters that would have changed the second term in \cref{eq:generic_model} have no impact on the noBAO fit. Therefore, we can effectively consider the BAO and noBAO fits have the same number of free parameters.} The flat likelihood of the noBAO fit is a trivial result of the improved separation between the BAO and noBAO modeling in \cref{eq:generic_model} that the \desione\ analysis adopted, i.e., the noBAO modeling now has naturally no dependence on $\alpha_{\rm iso}$, in contrast to previous configuration space analyses (see, e.g., the right-hand panel of figure 12 in \cite{Anderson14}). \edited{The detection significance, in the increasing order, is 4.0$\sigma$ for \bgs, 4.2$\sigma$ for \qso, 6.4$\sigma$ for \lrgo, 6.8$\sigma$ for \lrgt, 7.0$\sigma$ for \elgt, and 8.7$\sigma$ for \lrgth, with the most significant case being $9.1\sigma$ for  \lrgth+\elgo, closely correlated with the precision.}

\begin{table}
    \centering
    \begin{tabular}{|l|c|c|c|c|c|r|}
    \hline
     Tracer   & Redshift   & Recon   & $\alpha_{\rm iso}$   & $\alpha_{\rm AP}$   &   $r_{\rm off}$ & $\chi^2 / {\rm dof}$   \\
    \hline
     {\tt BGS}       & 0.1--0.4   & Post    & $0.9829 \pm 0.0185$  &                     &                 & 17.3/19                \\
     {\tt LRG1}      & 0.4--0.6   & Post    & $0.9792 \pm 0.0113$  & $0.9148 \pm 0.0372$ &         -0.0558 & 41.8/39                \\
     {\tt LRG2}      & 0.6--0.8   & Post    & $0.9661 \pm 0.0117$  & $1.0459 \pm 0.0432$ &         -0.0514 & 46.5/39                \\
     {\tt LRG3}      & 0.8--1.1   & Post    & $1.0042 \pm 0.0090$  & $1.0027 \pm 0.0299$ &         -0.0597 & 46.2/39                \\
     {\tt LRG3+ELG1} & 0.8--1.1   & Post    & $0.9983 \pm 0.0081$  & $1.0262 \pm 0.0278$ &         -0.1019 & 33.1/39                \\
     {\tt ELG1}      & 0.8--1.1   & Post    & $0.9887 \pm 0.0200$  &                     &                 & 20.4/19                \\
     {\tt ELG2}      & 1.1--1.6   & Post    & $0.9876 \pm 0.0147$  & $0.9904 \pm 0.0466$ &         -0.3031 & 58.6/39                \\
     {\tt QSO}       & 0.8--2.1   & Post    & $1.0015 \pm 0.0256$  &                     &                 & 32.4/19                \\
     \hline
     {\tt BGS}       & 0.1--0.4   & Pre     & $0.9646 \pm 0.0258$  &                     &                 & 14.3/19                \\
     {\tt LRG1}      & 0.4--0.6   & Pre     & $0.9760 \pm 0.0171$  & $0.9258 \pm 0.0590$ &          0.2674 & 25.8/39                \\
     {\tt LRG2}      & 0.6--0.8   & Pre     & $0.9505 \pm 0.0191$  & $1.0332 \pm 0.0769$ &          0.4130 & 31.1/39                \\
     {\tt LRG3}      & 0.8--1.1   & Pre     & $1.0085 \pm 0.0120$  & $1.0034 \pm 0.0446$ &          0.2737 & 48.4/39                \\
     {\tt LRG3+ELG1} & 0.8--1.1   & Pre     & $1.0027 \pm 0.0114$  & $1.0344 \pm 0.0436$ &          0.2569 & 53.0/39                \\
     {\tt ELG1}      & 0.8--1.1   & Pre     & $0.9404 \pm 0.0620$  &                     &                 & 29.2/19                \\
     {\tt ELG2}      & 1.1--1.6   & Pre     & $0.9882 \pm 0.0176$  & $0.9913 \pm 0.0640$ &          0.0484 & 42.7/39                \\
     {\tt QSO}       & 0.8--2.1   & Pre     & $0.9967 \pm 0.0229$  &                     &                 & 8.1/19                 \\
    \hline
    \end{tabular}
    \caption{Mean values and standard deviations from the marginalized posteriors of the BAO scaling parameters from fits to the unblinded \desidrone\ correlation functions in the $\alpha_{\rm iso}$-$\alpha_{\rm AP}$ basis. \edited{These do not include the systematic effect.} $r_{\rm off}=C_{\aiso,\alap}/\sqrt{C_{\aiso,\aiso}C_{\alap,\alap}}$.
    }\label{tab:Y1unblinded}
\end{table}

\begin{table}
    \centering
    \begin{tabular}{|l|c|c|c|c|c|c|}
    \hline
     Tracer   & Redshift   & Recon   & $\alpha_{\parallel}$   & $\alpha_{\perp}$   &   $r_{\rm off}$ & $\chi^2 / {\rm dof}$   \\
    \hline
     {\tt LRG1}      & 0.4--0.6   & Post    & $0.9219 \pm 0.0269$    & $1.0098 \pm 0.0186$ &         -0.4623 & 41.8/39                \\
     {\tt LRG2}      & 0.6--0.8   & Post    & $0.9955 \pm 0.0296$    & $0.9527 \pm 0.0179$ &         -0.4386 & 46.5/39                \\
     {\tt LRG3}      & 0.8--1.1   & Post    & $1.0057 \pm 0.0214$    & $1.0038 \pm 0.0139$ &         -0.4181 & 46.2/39                \\
     {\tt LRG3+ELG1} & 0.8--1.1   & Post    & $1.0148 \pm 0.0193$    & $0.9903 \pm 0.0126$ &         -0.4190 & 33.1/39                \\
     {\tt ELG2}      & 1.1--1.6   & Post    & $0.9797 \pm 0.0297$    & $0.9922 \pm 0.0245$ &         -0.4545 & 58.6/39                \\
     \hline
     {\tt LRG1}      & 0.4--0.6   & Pre     & $0.9242 \pm 0.0485$    & $1.0035 \pm 0.0243$ &         -0.4434 & 25.8/39                \\
     {\tt LRG2}      & 0.6--0.8   & Pre     & $0.9766 \pm 0.0642$    & $0.9402 \pm 0.0248$ &         -0.5015 & 31.1/39                \\
     {\tt LRG3}      & 0.8--1.1   & Pre     & $1.0098 \pm 0.0353$    & $1.0080 \pm 0.0165$ &         -0.4360 & 48.4/39                \\
     {\tt LRG3+ELG1} & 0.8--1.1   & Pre     & $1.0230 \pm 0.0323$    & $0.9930 \pm 0.0151$ &         -0.4549 & 53.6/39                \\
     {\tt ELG2}      & 1.1--1.6   & Pre     & $0.9810 \pm 0.0475$    & $0.9934 \pm 0.0273$ &         -0.4743 & 42.7/39                \\
    \hline
    \end{tabular}
    \caption{Mean values and standard deviations from the marginalized posteriors of the BAO scaling parameters from fits to the unblinded DESI DR1 correlation functions in the $\alpha_\parallel$-$\alpha_\perp$ basis. \edited{These do not include the systematic effect.} $r_{\rm off}=C_{\aperp,\apar}/\sqrt{C_{\aperp,\aperp}C_{\apar,\apar}}$.}
    \label{tab:Y1unblindedperp}
\end{table}

\subsection{BAO constraints projected onto $\alpha_\perp$ and $\alpha_\parallel$}

The parameterisation of the BAO-scale fitting by $\alphaiso$ and $\alphaap$ provides a natural separation of the isotropic and anisotropic dilation of the BAO that can be constrained somewhat independently.  \cref{tab:Y1unblinded} (and \cref{tab:Y1unblindedPk} in Appendix for $P(k)$) displays the cross-correlation coefficients, $r_{\rm off}=C_{\aiso,\alap}/\sqrt{C_{\aiso,\aiso}C_{\alap,\alap}}$.
We find the absolute values of $r_{\rm off}$ are less than $\sim 0.10$ for the three \lrgs\ after reconstruction, although it is larger for \elgt ($\sim 0.3$). The low value of the cross-correlation allows us to incorporate systematic errors ignoring the covariance between the two dilation parameters, i.e., by adding the statistical covariance from \cref{tab:Y1unblinded} with the systematic covariance in \cref{eq:sysisoap}.

We can also decompose the BAO locations into transverse and line-of-sight dilation parameters $\alpha_\perp$ and $\alpha_\parallel$ (\cref{eqn:alpha_defs})
 \edited{that are more directly linked to the angular diameter distance and the Hubble parameter at the given redshift}. 
\cref{tab:Y1unblindedperp} (and \cref{tab:Y1unblindedPkperp} for 
$P(k)$) 
shows the BAO constraints from re-fitting $\xi(r)$ into this parameterization rather than transforming them using \cref{eq:alphatransform}.\footnote{Transforming using \cref{eq:alphatransform}  gives slightly different numbers due to the assumption of two-dimensional Gaussian likelihood} The cross-correlation coefficients in this case, $r_{\rm off}=C_{\aperp,\apar}/\sqrt{C_{\aperp,\aperp}C_{\apar,\apar}}$, are in excellent agreement with the theoretical Fisher matrix prediction when the BAO feature is singled out~\cite{SE2007}. 
The covariance based on \cref{tab:Y1unblindedperp} will be combined with the covariance for systematics (\cref{eq:sysperppar}).

\subsection{Test of the systematics using the unblinded data}

\begin{figure*}
    \centering
    \includegraphics[width=0.95\textwidth]{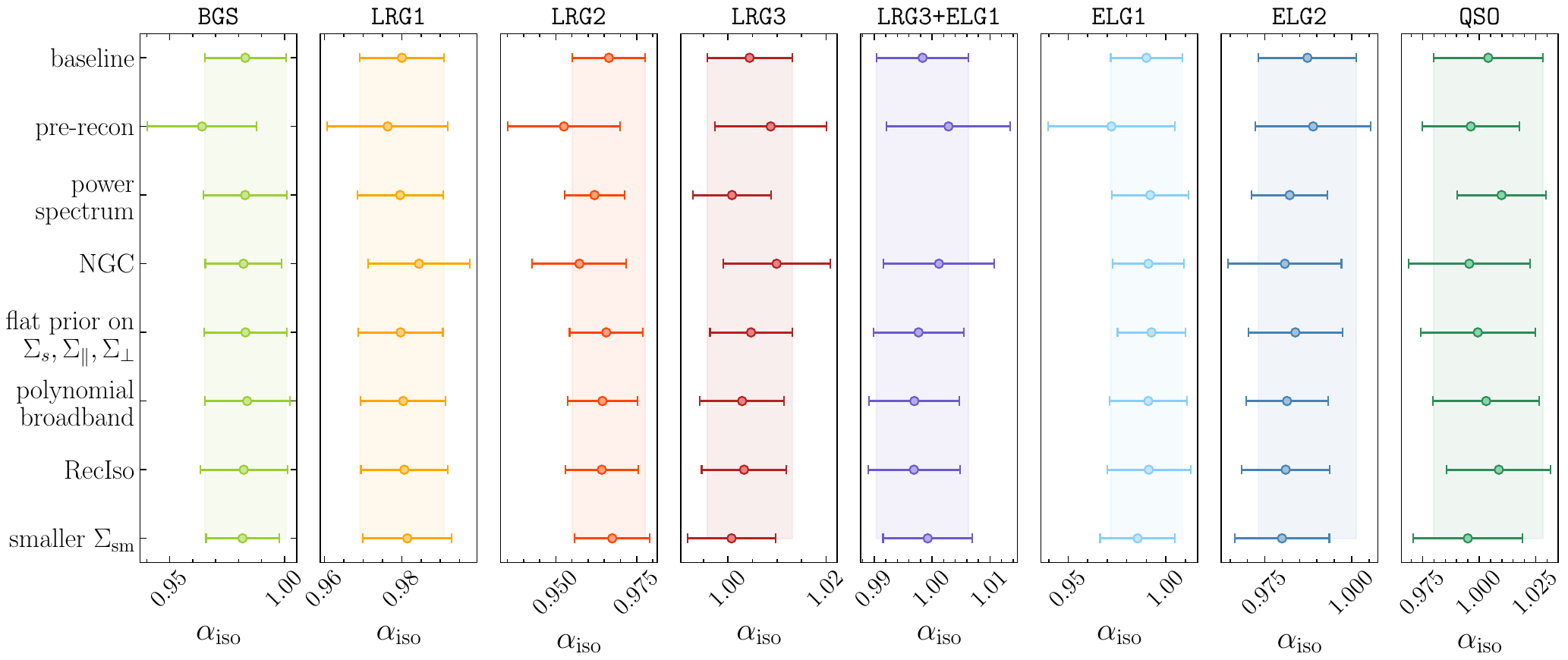}
    \includegraphics[width=0.6\textwidth]{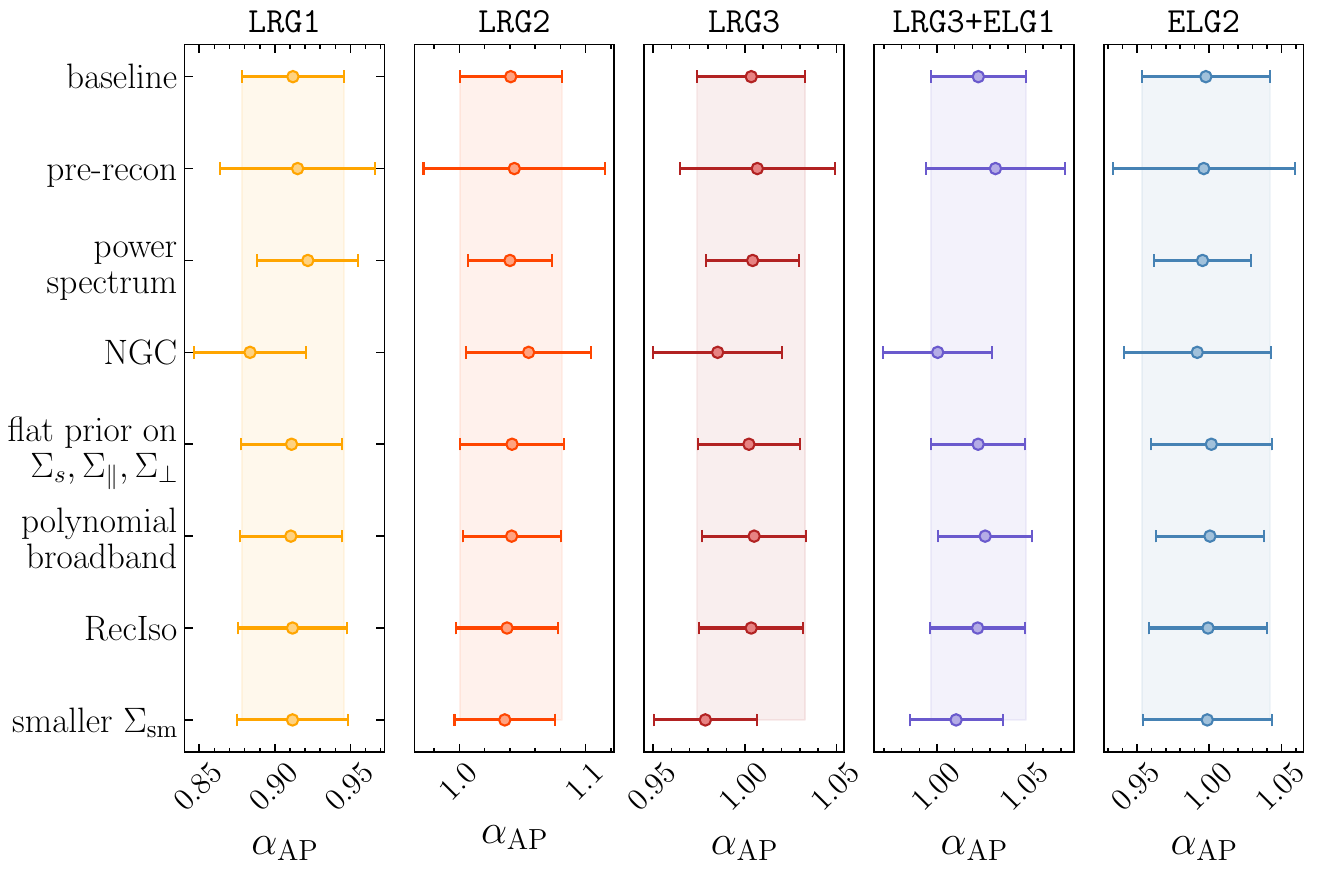}
    \caption{Response of the constraints on the isotropic (top) or anisotropic (bottom) BAO scaling parameters to changes in the data vector or the model assumptions. The baseline configuration adopted for the BAO analysis is shown in the top row, while the remaining rows show single variations around that baseline (see the main text for a description of each case).  The consistency between the $\xi(r)$ and $P(k)$ fits and the consistency between the spline broadband versus polynomial broadband are also very high. We identified the Gaussian prior on the sampling parameters as a more robust choice, but even the suboptimal flat priors return very consistent results. The difference between NGC alone and the default (NGC + SGC) shows mild discrepancies for $\alphaap$ of \lrgo.}\label{fig:y1unblindedwhisker}
\end{figure*}

We replicate the tests conducted on the blinded catalogs with the unblinded catalogs to investigate any indications of systematics. \cref{fig:y1unblindedwhisker} illustrates the BAO measurements when comparably optimal variations are introduced around our fiducial setup. The figure compares the baseline fit (using $\xi(r)$) versus the $P(k)$ fit, a different choice of the damping priors, a different choice of the broadband modeling, and a different choice of the reconstruction convention. These are the choices that we consider equally optimal (e.g., the power spectrum fit) or close to being optimal, compared to the default setup. In addition, we also include the pre-reconstruction results (second row).



The comparison between the baseline  fits  versus the fits using $P(k)$ (third whisker) shows high consistency in the derived BAO scales across all cases, \edited{as discussed in \cref{sec:baodesione}.} 
The BAO precision from the \desione\ $\xi(r)$ fit reasonably agrees with these mock results, especially for $\alphaiso$. In some cases, such as \lrgt\ and \elgt, the  $P(k)$-BAO precision tends to be 20-30\% better than the $\xi(r)$-BAO error. The \abacussecond\ DR1 mock test shows 5-10\% larger error with $\xi$; \lrgt\ lies somewhat beyond the range of the mocks.
Thus, opting for the $\xi-BAO$ fits is deemed a conservative choice.

In the fourth whisker, we display the constraints obtained by only fitting data from the NGC, which differs from the baseline case where both galactic caps (NGC+SGC) are combined at the clustering level. In most cases, we only see a small degradation in the parameter constraints compared to the baseline method. The largest deviations appear in $\alphaap$ of \lrgo\ and \lrgelg, but the comparison with mocks shows that this level of deviation can occur for \desidrone\ \cite{KP4s4-Paillas}.

The fifth whisker in \cref{fig:y1unblindedwhisker} shows the effect of using a flat, non-informative prior on the BAO non-linear damping parameters. Based on \cite{KP4s2-Chen}, we identified the Gaussian prior on the damping parameters as a more robust choice than a flat prior, but even the sub-optimal, flat priors return very consistent results.

Previous BAO analyses from BOSS \cite{Alam17} and eBOSS \cite{SDSS-DR16-cosmology} adopted a different parameterization for the broadband component of the power spectrum and correlation function, using polynomials with varying degrees of freedom and functional forms. The comparison between our default spline broadband analysis and the polynomial-based broadband method (described in \cref{subsubsec:polymodel}) demonstrates excellent consistency (sixth whisker). We note that we have performed this comparison in $\xi(r)$ and have only modified the parameters that capture the broadband component in the modeling, while still integrating all the other theoretical improvements in our analysis.

The second to last whisker shows the constraints using the \reciso\ convention for reconstruction. This convention, which was the default in previous SDSS BAO analyses \cite{Alam17, Tamone2020:2007.09009, Bautista2021} supresses higher-order multipoles by removing linear RSD on large scales. To perform this test, we used the (e)BOSS polynomial broadband parametrization for consistency with previous analysis. Overall, we find excellent agreement in the BAO constraints compared to our baseline case using the \recsym\ convention,  in agreement with what was found with mock galaxy catalogs \cite{KP4s2-Chen} and the blinded DESI data \cite{KP4s4-Paillas}.

Finally, in the last row, we show results using a smaller smoothing scale when reconstructing the density field. This corresponds to $10 \hinvmpc$ for \bgs, \lrgs, and \elgs, and $20 \hinvmpc$ for \qso. We find that the constraints on $\aiso$ are only slightly affected by this choice of scale. For $\alphaap$, the largest shift is of $1\sigma$ for \lrgth. In \cite{KP4s4-Paillas}, tests on the full ensemble of \abacussecond\ DR1 show that no statistically significant shifts are observed when comparing these choices of smoothing scales with the baseline. However, variations in individual mocks are expected due to noise fluctuations and are consistent with what is observed in the data.

In summary, the systematic tests conducted directly using unblinded \desidrone\ data demonstrate high consistency across variations of the fiducial setups and data selections (e.g., NGC versus NGC+SGC). Hence, this test bolsters the robustness of our measurements. \edited{As a caveat, these consistency tests may not identify systematics that shift all the results together.}








\subsection{BAO measurements from the combined tracers}
\label{subsec:results-overlapping}
There are several redshift ranges in which more than one DESI target class substantially overlaps. Over $0.8<z<1.1$, \lrgs, \elgs, and \qso\ overlap, and over $1.1<z<1.6$, \elgs, and \qso\ overlap. In particular, over $0.8<z<1.1$, \lrgth\ and \elgo\ are expected to have comparable densities for {\tt DESI} Y5 and there is a compensatory crossing of the two populations within this redshift range. As both tracers probe the same volume, the two BAO constraints would be highly covariant on large scales, providing a test for consistency in the tracer-independent parameter, such as post-reconstructed BAO locations. Although we have estimated the tracer-dependent systematics on BAO in \cref{subsec:sys-hod} within a reasonable (or limited, depending on a view) range of the HODs for each tracer, these overlapping tracers provide an utmost test of the tracer dependence.

When combining the two BAO constraints we must account for this cross-covariance, either using a set of large mocks in which both tracers are present or with an analytical method. Another approach is to construct a combined catalog. A BAO constraint from such a catalog seamlessly combines the information from two BAO measurements from auto-clustering statistics and one from the cross-clustering between \elgo\ and \lrgth\ (hereafter `\lrgxelg'). The combined catalog, which we defined as \lrgelg\ in \cref{sec:catalog}, may have an additional gain: due to the higher number density, the reconstruction can be more effective. \cite{KP4s5-Valcin} tested the optimal construction of \desione\ \lrgth\ and \elgo.  The work includes the test of systematics using auto-clustering measurements and the cross-clustering between the two tracers. 

For the unblinded \desidrone\ \lrgelg, we find 11\% precision-improvement in $\aiso$ and 14\% improvement in $\alap$ relative to \lrgth, as shown in \cref{fig:y1pkunblindedwg} and \cref{tab:Y1unblinded}. This is highly consistent with the $\sim 10\%$ improvement predicted by  \cite{KP4s5-Valcin} based on the mocks.

We find an offset of 0.4\% in $\aiso$ between \lrgth\ and \lrgxelg\ (\cite{KP4s5-Valcin}) and 0.6\% between \lrgth\ versus \lrgelg\ (\cref{tab:Y1unblinded}). In $\alap$, we find an offset of 0.4\% between \lrgth\ and \lrgxelg\ (\cite{KP4s5-Valcin}) and 1.9\% between \lrgelg\ (\cref{tab:Y1unblinded}). These differences are well within the range covered by the mocks. Therefore, we detect no tracer-dependent bias among the \desidrone\ BAO measurements from \lrgth, \elgo, \lrgxelg, and \lrgelg. 
With Y3 and Y5, as the completeness of the \elgs\ increases, we expect a greater improvement from the combined tracer analysis in terms of reconstruction as well as the systematic test. Note that we perform only the $\xi(r)$ fit to the combined tracer, as we have only the \textsc{RascalC} covariance available for the combined tracer.

\subsection{Comparison to previous analyses}\label{sec:comparison}



\begin{figure}
    \centering
    \includegraphics[width=0.9\textwidth]{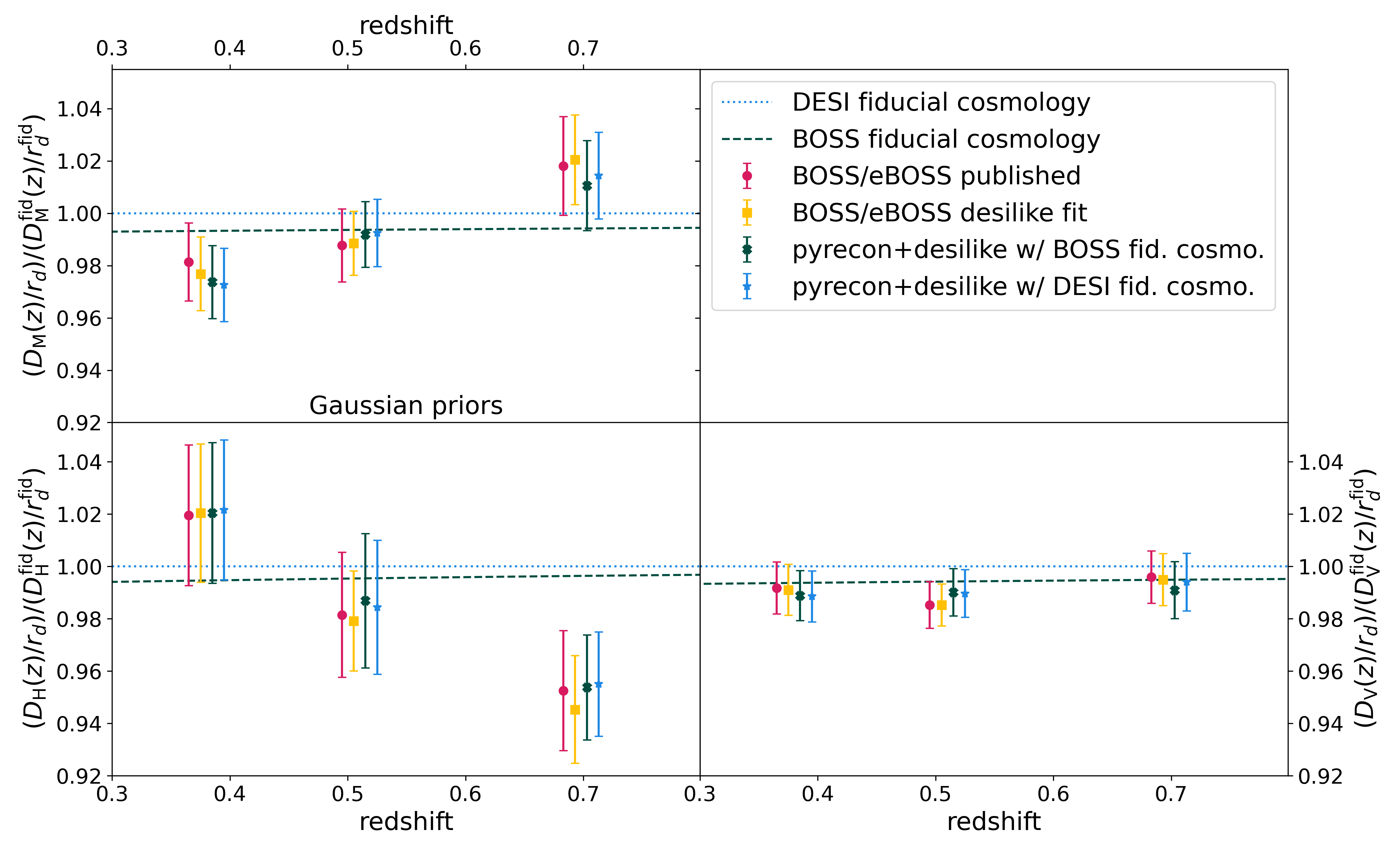}
    \caption{A comparison of the impact of different pipelines on the derived distances, 
    analyzing the published BOSS and eBOSS LRG data. 
    There are three groups of points, one for each of the redshift bins, while the different panels show 
    results for $D_V/\rd$, $D_M/\rd$, and $D_H/\rd$.
    In each group, the leftmost point is the published BOSS/eBOSS result. The second point from the left 
    uses the published correlation functions and covariance matrices, but refits these using the methodology presented in this 
    paper, while the last two points show the results of reprocessing the BOSS and eBOSS catalogs and randoms through the full \desi\ pipeline
    for different fiducial cosmologies.
    To make the comparison more direct (and to leave the covariance matrices unchanged), we use the {\bf RecIso} reconstruction scheme here.
    The fiducial cosmology used to normalize the distance scales on the $y$-axis is the DESI fiducial cosmology 
    used in this paper.} 
    \label{fig:BOSS_compare}
\end{figure}

\begin{table}
    \centering
    \small
    \begin{tabular}{|l|c|c|c|c|c|}
    \hline
     Tracer          &   $z$ & $D_{\rm M}/r_{\rm d}$   & $D_{\rm H}/r_{\rm d}$      & $r_{\rm off}$   \\
    \hline
SDSS published & & & & \\ \hline
BOSS LOWZ+CMASS & 0.2 - 0.5 & $10.23 \pm 0.17$ & $25.00 \pm 0.76$ & -0.39 \\ 
BOSS LOWZ+CMASS & 0.4 - 0.6 & $13.36 \pm 0.21$ & $22.33 \pm 0.58$ &  -0.39 \\ 
eBOSS LRG & 0.6 - 1.0 & $17.86 \pm 0.33$ & $19.33 \pm 0.53$ & -0.33 \\ \hline
SDSS published/desilike refit & & & &   \\ \hline
BOSS LOWZ+CMASS & 0.2 - 0.5 &  $10.18 \pm 0.15$ & $25.1 \pm 0.65$ & -0.425 \\ 
BOSS LOWZ+CMASS & 0.4 - 0.6 & $13.35 \pm 0.17$ & $22.27 \pm 0.43$ &  -0.404 \\ 
eBOSS LRG & 0.6 - 1.0 & $17.9 \pm 0.3$ & $19.16 \pm 0.42$ & -0.416 \\ \hline
DESI pipeline & & & &   \\ \hline
BOSS LOWZ+CMASS & 0.2 - 0.5 & $10.15 \pm 0.15$ & $25.10 \pm 0.66$ &  -0.45 \\ 
BOSS LOWZ+CMASS & 0.4 - 0.6 & $13.39 \pm 0.17 $& $22.44 \pm 0.58$ & -0.43 \\ 
eBOSS LRG & 0.6 - 1.0 & $17.72 \pm 0.30$ & $19.33 \pm 0.41$ &  -0.40 \\ 
    \hline
    \end{tabular}
    \caption{\edited{Comparison between the SDSS published results and the
    reanalysis using the DESI pipeline. Among several options we tested in
    \cref{sec:comparison}, we present the DESI reanalysis with {\tt IFFT} {\bf
    RecIso} under the BOSS fiducial cosmology, to best match the clustering
    amplitude assumed in the published covariance matrices. We did not add the SDSS systematic errors in the middle group. We also did not add the
    DESI systematic errors from \cref{eq:sysperppar} to the bottom group, but the difference is negligible.
    }}
    \label{tab:SDSSvsDESI}
\end{table}

Given the changes described above in our pipeline, we revisit the SDSS (BOSS and eBOSS) data previously analyzed to 
quantify the impact of these changes on previously published results \cite{Alam21}. We do so in 
two stages. First, we refit the published correlation functions (using the published covariance matrices) to 
measure the distance scale. Second, we rerun the reconstruction pipeline using our new convention and then re-fit the updated 
correlation functions. \cref{fig:BOSS_compare} shows the results of these comparisons for the three 
redshift bins used in \edited{the SDSS (\cite{Alam21}).}  

The red point of each group in \cref{fig:BOSS_compare} (first from the left) is the published BOSS/eBOSS result for $\alpha$, 
while the yellow point (second from the left)
uses the published correlation functions and covariance matrices, but refits these using 
the baseline methodology described in this paper. For the fits performed here, we use damping parameters 
with Gaussian priors to match this analysis, while the published analyses used a fixed damping scale.
The differences between these points are due to two primary factors. The 
first are changes in how the BAO features are damped. When computing the theoretical correlation functions 
in the previous analyses, the power spectrum was damped by an additional smoothing of $1\hinvmpc$ to 
accelerate the convergence of the Hankel transform and reduce ringing. This effectively changes the damping 
scales. Furthermore, previous analyses applied the small-scale streaming velocities to both the BAO and the 
broadband shape, while our current analysis only applies it to the broadband shape (\cref{subsec:sdssvsdesifitting}). 
Additionally, the current analysis has a new separation of the BAO and broadband shape.
The top and middle group of Table~\ref{tab:SDSSvsDESI} shows the net change from all of these effects.

We also explore the effect of different reconstruction
conventions using the full DESI pipeline and setup.
We reprocess the BOSS and eBOSS
catalogs and randoms using \textsc{pyrecon} and \textsc{pycorr} under
\textsc{cosmodesi} and fit using the fiducial setup of
\cref{subsec:methods-fitting} implemented in \textsc{desilike}. The two rightmost points in each group of \cref{fig:BOSS_compare} show the results of this reprocessing,
and the bottom group of Table~\ref{tab:SDSSvsDESI} shows their impact.
When comparing these numbers, we emphasize that the exact details of how the data sets are assigned to grids and padded will also affect 
the derived displacement fields. 
These different implementations should be thought of as 
yielding different data sets, even though they are derived from the same initial sets of galaxy positions.
Given that the remaining differences are well within the statistical precision of these samples, we therefore 
conclude that our new analysis does yield consistent results with published results. The companion cosmological 
interpretation paper \cite{DESI2024.VI.KP7A} will present a detailed consistency check at the level of the underlying cosmological 
parameters.

\section{The final distance measurements and the Hubble Diagram}
\label{sec:interpretation}

\begin{figure*}[h]
 \centering
 \begin{tabular}{cc}
    \includegraphics[width=0.4\textwidth]{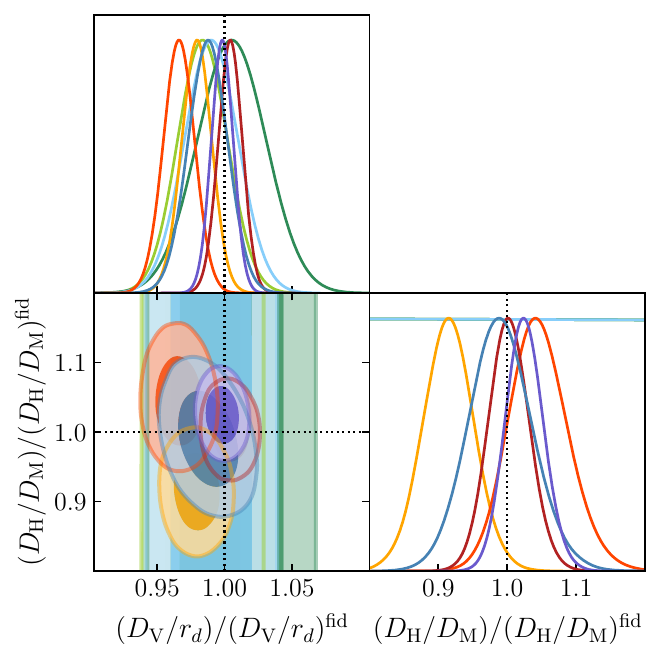}  & \includegraphics[width=0.4\textwidth]{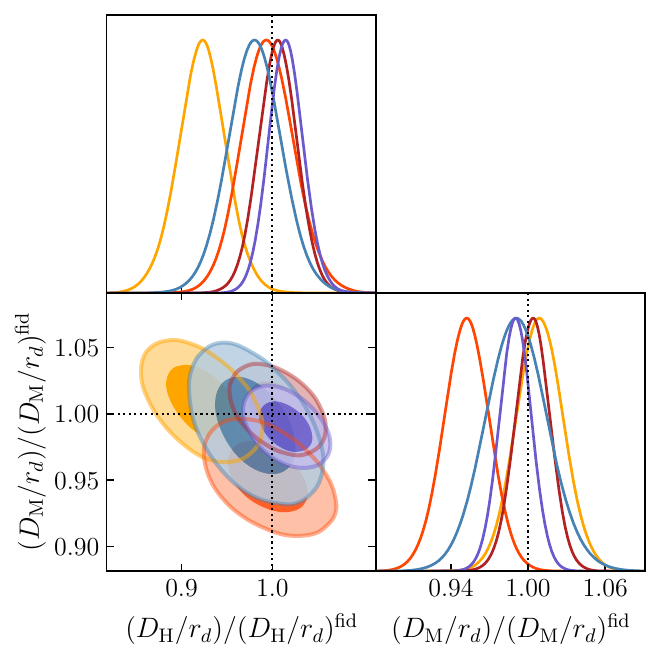} \\
    \multicolumn{2}{c}{\includegraphics[width=0.8\textwidth]{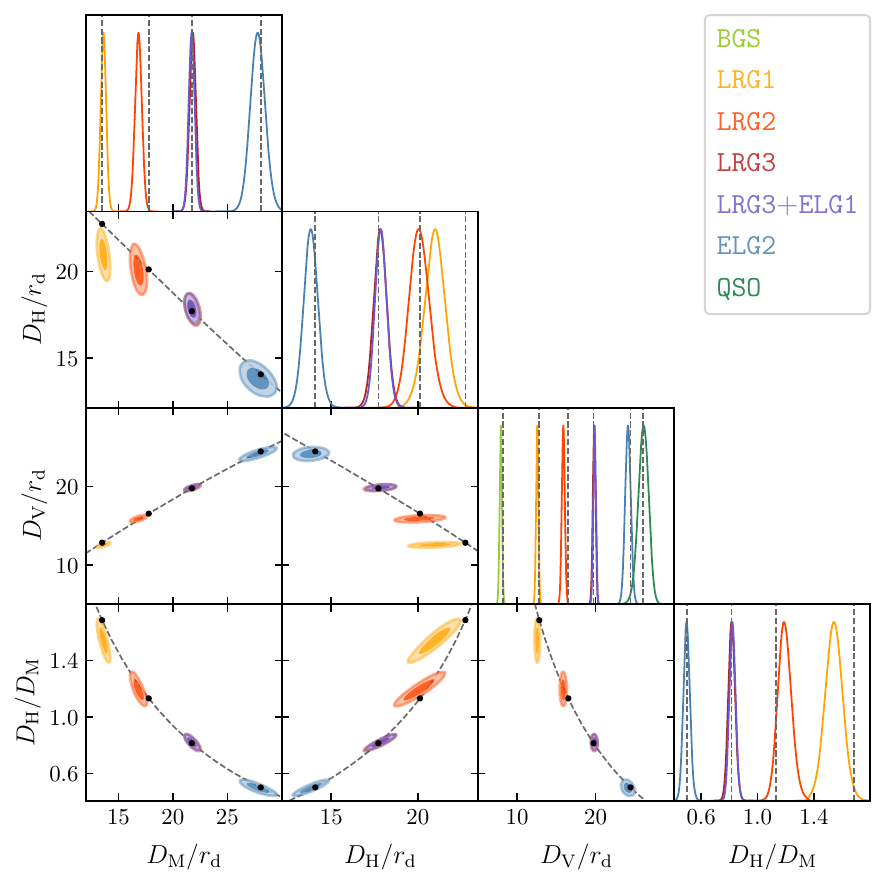}  }
 \end{tabular}
 \caption{The final BAO-only distance posteriors, shown as measurements of $\DMoverrd$, $\DHoverrd$, $\DVoverrd$, and $\DHoverDM$, compared against the fiducial \planck\ distances. In the bottom panel, the prediction at the effective redshift of each sample is shown by the black dots along the dashed lines: $\zeff=0.51$ (\lrgo), 0.71 (\lrgt), 0.93 (\lrgth), and 1.32 (\elgt). The uncertainty intervals on the plot do not take account of the systematic errors but doing so would make little visible difference.}\label{fig:hubblecornerplot}
\end{figure*}

\begin{figure}[h]
 \centering
        \includegraphics[width=0.8\textwidth]{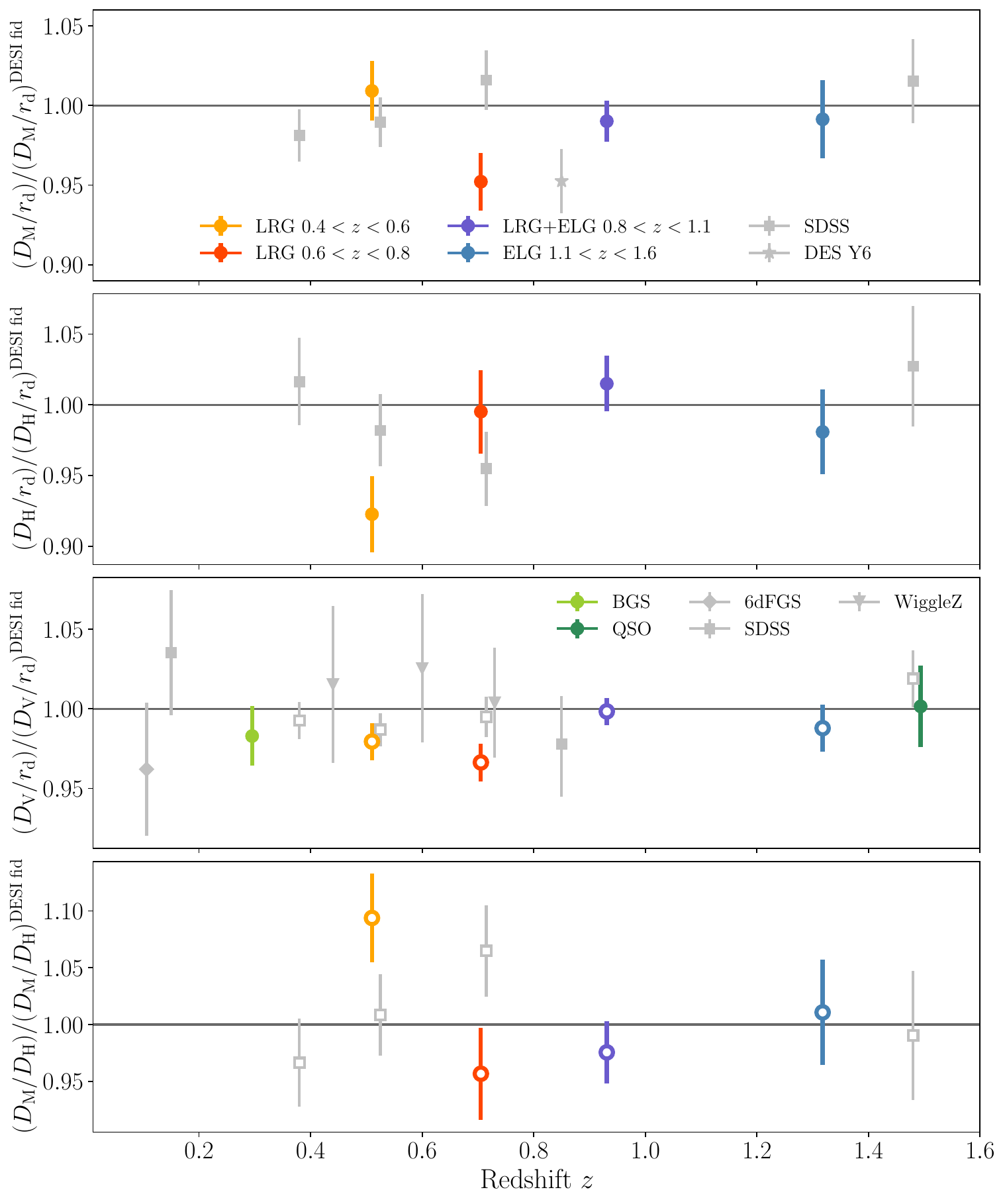}
         \caption{Hubble diagram of the BAO distance scales measured from the unblinded galaxy and quasar data, compared to those from earlier BAO measurements by the 6 degree Field Galaxy Survey (6dFGS, \cite{SixdFBAO}), WiggleZ \cite{Kazin2014:1401.0358}, the Sloan Digital Sky Survey (SDSS, \cite{Alam21}), and the Dark Energy Survey (DES Y6, \cite{desbao}), as labelled. From top to bottom, the panels show $\DMoverrd$, $\DHoverrd$, $\DVoverrd$ and $D_\mathrm{M}/D_\mathrm{H}$, all relative to the respective quantities evaluated in the DESI fiducial cosmology described in \cref{sec:intro}. For 6dFGS, WiggleZ and some redshift bins of SDSS and DESI, only $\DVoverrd$ measurements were possible due to the low signal-to-noise ratio, so these points are only shown in the third panel. For the DESI and SDSS redshift bins where both $\DMoverrd$ and $\DHoverrd$ were measured, results for $\DVoverrd$ and $D_\mathrm{M}/D_\mathrm{H}$ in the third and fourth panels are displayed with open markers to indicate the repetition of information in the top two panels in a different parametrisation. Note that a slight offset has been applied to the effective redshifts of the SDSS results at $z_{\rm eff}=0.51$ and $0.70$ to avoid overlap and ensure visibility in this figure.}\label{fig:hubble}
\end{figure}

In this section, we present our final distance measurements, construct the Hubble diagram, compare with the previous BAO measurements, and discuss any implications.
We adopt the configuration space measurements as our fiducial results.

\begin{table}
    \centering
    \small
    \begin{tabular}{|l|c|c|c|c|c|c|c|c|}
    \hline
     Tracer          &   $z_{\rm eff}$ & $D_{\rm M}/r_{\rm d}$   & $D_{\rm H}/r_{\rm d}$   & $D_{\rm V}/r_{\rm d}$   & $D_{\rm H}/D_{\rm M}$   & $r_{\rm off}$   \\
    \hline
     {\tt BGS}       &            0.30 & ---                     & ---                     & $7.93 \pm 0.15$       & ---                     & ---             \\
     {\tt LRG1}      &            0.51 & $13.62 \pm 0.25$      & $20.98 \pm 0.61$      & $12.56 \pm 0.15$      & $1.542 \pm 0.063$       & $-0.445$        \\
     {\tt LRG2}      &            0.71 & $16.85 \pm 0.32$      & $20.08 \pm 0.60$      & $15.90 \pm 0.20$      & $1.193 \pm 0.049$       & $-0.420$        \\
     {\tt LRG3}      &            0.92 & $21.81 \pm 0.31$      & $17.83 \pm 0.38$      & $19.82 \pm 0.18$      & $0.818 \pm 0.025$       & $-0.393$        \\
     {\tt LRG3+ELG1} &            0.93 & $21.71 \pm 0.28$      & $17.88 \pm 0.35$      & $19.86 \pm 0.17$      & $0.824 \pm 0.022$       & $-0.389$        \\
     {\tt ELG1}      &            0.95 & ---                     & ---                     & $20.01 \pm 0.41$      & ---                     & ---             \\
     {\tt ELG2}      &            1.32 & $27.79 \pm 0.69$      & $13.82 \pm 0.42$      & $24.13 \pm 0.36$      & $0.498 \pm 0.023$       & $-0.444$        \\
     {\tt QSO}       &            1.49 & ---                     & ---                     & $26.07 \pm 0.67$      & ---                     & ---             \\
    \hline
    \end{tabular}
    \caption{Final summary of the derived BAO distance scales from fits to the post-reconstruction correlation function multipoles for each DESI target sample, including the systematics from \cref{eq:sysisoap,eq:sysperppar}. We display results in terms of the mean values and standard deviations from the marginalized posteriors of each parameter. $r_{\rm off}$ is the correlation coefficient between $D_{\rm M}/r_{\rm d}$ and $D_{\rm H}/r_{\rm d}$. \edited{The results in this table are used for the cosmology analysis in \cite{DESI2024.VI.KP7A}.}}
    \label{tab:finaldistance}
\end{table}

\subsection{The final distance measurements with systematics}
\label{subsec:results-finaldistance}

In \cref{tab:finaldistance}, we combine the best-fit BAO measurements, based on
\cref{tab:Y1unblinded,tab:Y1unblindedperp}, with the total systematic error
budget from \cref{eq:sysisoap,eq:sysperppar}, and present the final distance posteriors to be used for cosmology
inference in \cite{DESI2024.VI.KP7A}. \edited{With the knowledge of the fiducial distance-to-redshift relation and the sound horizon scale used in the analysis, we can convert the BAO measurements to the distance observables: comoving angular diameter distance $D_M(z)$,} 
\begin{equation}
 \frac{D_M(z)}{\rd} \equiv \frac{D_A(z)[1+z]}{\rd} = \aperp 
 \frac{D^{\rm fid}_M(z)}{r^{\rm fid}_d},
\end{equation}
the Hubble distance $D_H(z)$,
\begin{equation}
\frac{D_H(z)}{\rd} \equiv \frac{c}{H(z)\rd}
= \alpha_\parallel \frac{D^{\rm fid}_H(z)}{r^{\rm fid}_d},
\end{equation}
and the spherically-averaged distance $D_V(z)$,
\begin{equation}
\frac{D_V(z)}{\rd} \equiv \frac{[zD^2_M(z) D_H(z)]^{1/3}}{\rd} = \aiso  \frac{D^{\rm fid}_V(z)}{r^{\rm fid}_d}.
\end{equation}

\cref{fig:hubblecornerplot} graphically presents our distance constraints of all tracers relative to the fiducial
\planck\ distances. 
The plot shows offsets from the \planck\
predictions for $0.4<z<0.6$ ($\lrgo\ (\zeff=0.51)$) and $0.6<z<0.8$. The deviation at  $\lrgo\ (\zeff=0.51)$ is
mainly for $D_H/\rd$. This means that the line-of-sight BAO scale appears bigger than the transverse BAO scale, when the observed space is mapped to the physical space using the metric of the \planck\ cosmology.

On the other hand, the deviation at $\lrgt\ (\zeff=0.71)$ is \edited{mainly in $D_V/\rd$ in
$D_M/\rd$}. That is, the transverse BAO scale at this redshift appears much larger than the prediction of the \planck, giving a smaller $\DVoverrd$ and $\DMoverrd$  at this redshift.
The cosmological implication of these deviations is extensively tested
in \cite{DESI2024.VI.KP7A}. 
At $z>0.8$, our BAO measurements are consistent with the expansion history predicted by the \planck. 

We then construct a Hubble diagram using \desione\ BAO-only distance measurements and  also with previous spectroscopic BAO measurements overlaid (\cref{fig:hubble}). \edited{Again, we defer the thorough investigation of the cosmological implications of these measurements in combination with the DESI DR1 Ly-$\alpha$ BAO measurement (\cite{DESI2024.IV.KP6}) to \cite{DESI2024.VI.KP7A}.}




\subsection{Remarks on the discrepancy between SDSS and DESI over $0.6<z<0.8$}
\cref{fig:hubble} shows noticeable discrepancy between the \desione\ measurement at
$\zeff=0.71$ (\lrgt) and SDSS {LRG} (eBOSS DR16) at $\zeff=0.7$ ($0.6<z<0.85$). \edited{SDSS {LRG} measured $D_M(0.70)/\rd=17.86\pm 0.33$ and $D_H(0.80)/\rd = 19.33\pm 0.53$, with $r_{\rm off}=-0.32$ \cite{Bautista2021}}, while this paper reports $D_M(0.71)/\rd=16.85\pm 0.32$, $D_H(0.71)/\rd=20.08 \pm 0.60$ with $r_{\rm off} = -0.420$ for \lrgt. 

An
approximate estimate of the cross-correlation between the two tracers in the
power spectrum space is 0.21. Assuming that this directly propagates to the
cross-correlation of the BAO measurements, the discrepancy is close to $\sim$3$\sigma$ for $D_M/\rd$ and $D_V/\rd$ when we account for the difference between $\zeff=0.7$  and  $\zeff=0.71$ using the \planck\ cosmology.\footnote{${D_M(0.71)/D_M(0.7)}]^{\rm fid} = 1.011$.}

An insight gleaned from \cref{sec:comparison} and numerous robustness tests conducted in this paper is that an estimator derived from a particular dataset undergoing different pipelines can be regarded as two distinct yet unbiased estimators of the truth, provided each is individually demonstrated to be unbiased. Hence, the cross-correlation coefficient could be much lower than the theoretical estimate, suggesting that the observed 3$\sigma$ discrepancy between SDSS and \desi\ is likely an upper limit. However, the lower limit, determined assuming no correlation, is still 2.7$\sigma$. 
\edited{The \desi\ reanalysis in \cref{tab:SDSSvsDESI} yields a slightly lower BAO measurement compared to the published SDSS measurements, resulting in significances of 2.8$\sigma$ and 2.5$\sigma$, assuming cross-correlations of 0.21 and 0, respectively.}


While this \edited{2.5$\sigma$--3}$\sigma$ discrepancy warrants attention, we have not identified any sources of non-statistical discrepancy from both sets of the data. We reiterate the significant value of a blinded analysis of \desione, conducted and unblinded without being affected by prior knowledge of any discrepancies compared to previous measurements. 
We anticipate gaining a deeper understanding of its nature through the analysis of \desi\ Y3 and Y5 data.

\section{Conclusion}
\label{sec:conclusion}

In this paper, we presented the first BAO scale measurements\footnote{Previously, we reported the first detection of the \desidrone\ BAO \cite{BAO.EDR.Moon.2023} using the first two months of the DESI data without presenting the BAO scales.} from the DESI galaxies and quasars since the start of its data collection in 2021. The measurements from \desidrone\ include over 5.7 million galaxy and quasar redshifts over the redshift range of $0.1<z<2.1$, with the total effective survey volume of $\sim 18\Gpc^3$.\footnote{Computed using \qso\ only over $1.1<z<1.6$.} \edited{This marks the largest volume and the most redshifts used for any spectroscopic BAO measurements.} 
We summarize the final distance constraints for six redshift bins, detailed in \cref{tab:finaldistance}, as: 

\begin{itemize}
\item Distance precision (i.e., $\DVoverrd$) ranges of 1.9\% (\bgs\ at $\zeff =0.3$), 1.2\% (\lrgo\ at $\zeff=0.51$), 1.2\% (\lrgt\ at $\zeff=0.71$), 0.8\% (\lrgelg\ at $\zeff \sim 0.95$), 1.5\% (\elgt\ at $\zeff=1.32$), and  2.6\% (\qso\ at $\zeff=1.49$). The aggregate precision of all measurements is 0.52\% on the BAO scale.

\item The anisotropic distortion of the BAO (i.e., $\DHoverDM$) is measured with the precision of 4.1\% (\lrgo), 4.1\% (\lrgt), 2.7\% (\lrgelg), and 4.6\% (\qso).
\end{itemize}

Our measurement marks the highest precision on the cosmological distances for $z>0.8$, compared to the previous survey, and also collectively marks the highest aggregate precision. \edited{Comparing this with the SDSS BAO measurements (\cite{Alam21,Neveux2020}), which report an aggregate precision of 0.64\% on $D_V/\rd$ for all galaxy and quasar tracers combined, the aggregate precision of \desidrone\ in this paper returns 0.52\%.} The detection levels range from the highest 9.1$\sigma$ for \lrgelg\ to the lowest 4.0$\sigma$ for \bgs.
We produced the first high-significance BAO detection using galaxies $z>1$, i.e., \elgt\ at the significance of 7.0$\sigma$. Proving the robust nature of the BAO for \elgs, despite its substantial imaging systematics, is an important result for upcoming Stage-V BAO surveys.

Alongside the state-of-the-art dataset, the \desidrone\ BAO analysis incorporated several significant novel elements.
Our analysis is one of the first to rigorously implement blinding in the BAO analysis (c.f. \cite{desbao}) and the first catalog-level blinded BAO analysis (\cref{subsec:blinding}). This approach was adopted to mitigate confirmation bias. The true clustering was unveiled only once the blinded catalogs successfully passed a series of unblinding tests (refer to \cref{sec:unblinding}).
Anticipating the unprecedented precision of \desi\ BAO beyond \desione, our analysis incorporated physically motivated enhancements to the BAO fitting and reconstruction methods from previous surveys. This included the adoption of a new reconstruction method (\cref{subsec:methods-recon}) and improved treatment for separating BAO from broadband signals (\cref{subsec:methods-fitting}). We introduced a spline-based broadband model to reduce parameterization dependence on data signal-to-noise. We constructed and analyzed a combined tracer catalog for \lrgs\ and \elgs\ spanning $0.8 < z < 1.1$, merging information from both tracers and facilitating an additional systematic test (refer to \cref{subsec:results-overlapping}). This approach yielded an improvement of approximately 10\%, with even greater improvements anticipated for future DESI analyses.

We employed a unified BAO analysis pipeline and methodology across all galaxy and quasar tracers, as well as between configuration space and Fourier space, marking a first in BAO surveys. The algorithms utilized in this study are incorporated into a publicly accessible repository developed by the collaboration,\footnote{\url{https://github.com/cosmodesi}} with some currently available and others planned for future release.
We meticulously coordinated a systematic study across all tracers and components. Our investigation covered theoretical modeling effects (\cref{sec:model_fitting}), tracer-specific systematics (\cref{subsec:sys-hod}), assumptions regarding fiducial cosmology (\cref{subsec:sys-fiducialcosmo}), observational systemtaics (\cref{subsec:obssys}), reconstruction algorithm choices (\cref{subsec:methods-recon}), and covariance matrix choice (\cref{subsec:methods-cov}). Summary of these findings is provided in \cref{tab:finalsystematics}, yielding conservative systematics estimates of 0.245\% for the isotropic BAO scale and 0.300\% for the anisotropic distortion of the BAO feature (i.e., the Alcock-Paczynski effect). Further details are available in supporting papers listed in \cref{tab:supportingpapers}.

\edited{To quantify the impact of the changes we made in the DESI pipeline, we revisited the SDSS BAO measurements using the DESI pipelines. We conclude that the two measurement pipelines yield sufficiently similar results for these datasets (refer to \cref{sec:comparison}), although we anticipate that the new method will be  advantageous for the higher-precision dataset expected in the future.}

From the Hubble diagram constructed from our results, we found that our BAO measurements indicate systematically larger observed BAO scales than the prediction of \planck\ at $z<0.8$, and therefore lower distances. 
For $\zeff=0.51$, this discrepancy is sourced from the line-of-sight BAO: it 
 appears bigger than the transverse BAO scale, when the observed space is mapped to the physical space using the metric of the \planck\ cosmology. 
For $\zeff=0.71$, the discrepancy is sourced from the overall size of the BAO (and more transverse than the line-of-sight). The transverse BAO scale at this redshift appears much larger than the prediction of the \planck, implying a smaller $\DVoverrd$ and $\DMoverrd$  at this redshift. \edited{Accordingly, we found a large discrepancy between the DESI BAO measurement at $\zeff=0.71$ and the SDSS measurement at $\zeff=0.7$. The level of discrepancy is $\sim$3$\sigma$ on $\DMoverrd$.}
At $z>0.8$, our BAO measurements are consistent with that of the \planck. 
\edited{We emphasize the value of our blinded analysis in achieving the \desidrone\ BAO measurements without prior knowledge of comparisons with previous measurements. An extensive investigation on the cosmological implication of these discrepancies from \planck, in combination with the BAO measurement from the Lyman-$\alpha$ forest (\cite{DESI2024.IV.KP6})}, is presented in the \desidrone\ cosmology  paper, \cite{DESI2024.VI.KP7A}.

Assuming the same sample
definitions (which are conservative for \desidrone), the final DESI dataset will be more than a factor of 3 in terms of the effective volume (4-5 for \elgs, $\sim 3$ for \lrgs, and 2-3 for \bgs\ and \qso) compared to \desidrone, using more than 20 million unique redshifts. 
Through the analysis of the DESI Y3 and Y5 data, we anticipate to gain a deeper understanding of the nature of the intriguing measurements we observed.

%



\section{Data Availability}

The data used in this analysis will be made public along the Data Release 1 (details in https://data.desi.lbl.gov/doc/releases/).

\acknowledgments

This material is based upon work supported by the U.S. Department of Energy (DOE), Office of Science, Office of High-Energy Physics, under Contract No. DE–AC02–05CH11231, and by the National Energy Research Scientific Computing Center, a DOE Office of Science User Facility under the same contract. Additional support for DESI was provided by the U.S. National Science Foundation (NSF), Division of Astronomical Sciences under Contract No. AST-0950945 to the NSF’s National Optical-Infrared Astronomy Research Laboratory; the Science and Technology Facilities Council of the United Kingdom; the Gordon and Betty Moore Foundation; the Heising-Simons Foundation; the French Alternative Energies and Atomic Energy Commission (CEA); the National Council of Humanities, Science and Technology of Mexico (CONAHCYT); the Ministry of Science and Innovation of Spain (MICINN), and by the DESI Member Institutions: \url{https://www.desi.lbl.gov/collaborating-institutions}. Any opinions, findings, and conclusions or recommendations expressed in this material are those of the author(s) and do not necessarily reflect the views of the U. S. National Science Foundation, the U. S. Department of Energy, or any of the listed funding agencies.

The authors are honored to be permitted to conduct scientific research on Iolkam Du’ag (Kitt Peak), a mountain with particular significance to the Tohono O’odham Nation.

\bibliographystyle{JHEP}
\bibliography{DESI2024,references,theorysys}

\providecommand{\href}[2]{#2}\begingroup\raggedright\begin{thebibliography}{100}

\bibitem{BAOreview2013}
D.H.~{Weinberg}, M.J.~{Mortonson}, D.J.~{Eisenstein}, C.~{Hirata}, A.G.~{Riess}
  and E.~{Rozo}, \emph{{Observational probes of cosmic acceleration}},
  \href{https://doi.org/10.1016/j.physrep.2013.05.001}{\emph{\physrep}
  {\bfseries 530} (2013) 87} [\href{https://arxiv.org/abs/1201.2434}{{\ttfamily
  1201.2434}}].

\bibitem{CosmologyIntertwined}
E.~{Abdalla}, G.F.~{Abell{\'a}n}, A.~{Aboubrahim}, A.~{Agnello},
  {\"O}.~{Akarsu}, Y.~{Akrami} et~al., \emph{{Cosmology intertwined: A review
  of the particle physics, astrophysics, and cosmology associated with the
  cosmological tensions and anomalies}},
  \href{https://doi.org/10.1016/j.jheap.2022.04.002}{\emph{Journal of High
  Energy Astrophysics} {\bfseries 34} (2022) 49}
  [\href{https://arxiv.org/abs/2203.06142}{{\ttfamily 2203.06142}}].

\bibitem{SHOES-2022}
A.G.~{Riess}, W.~{Yuan}, L.M.~{Macri}, D.~{Scolnic}, D.~{Brout}, S.~{Casertano}
  et~al., \emph{{A Comprehensive Measurement of the Local Value of the Hubble
  Constant with 1 km s$^{-1}$ Mpc$^{-1}$ Uncertainty from the Hubble Space
  Telescope and the SH0ES Team}},
  \href{https://doi.org/10.3847/2041-8213/ac5c5b}{\emph{\apjl} {\bfseries 934}
  (2022) L7} [\href{https://arxiv.org/abs/2112.04510}{{\ttfamily 2112.04510}}].

\bibitem{SDSS-DR16-cosmology}
{eBOSS Collaboration}, S.~{Alam}, M.~{Aubert}, S.~{Avila}, C.~{Balland},
  J.E.~{Bautista} et~al., \emph{{Completed SDSS-IV extended Baryon Oscillation
  Spectroscopic Survey: Cosmological implications from two decades of
  spectroscopic surveys at the Apache Point Observatory}},
  \href{https://doi.org/10.1103/PhysRevD.103.083533}{\emph{\prd} {\bfseries
  103} (2021) 083533} [\href{https://arxiv.org/abs/2007.08991}{{\ttfamily
  2007.08991}}].

\bibitem{2005NewAR..49..360E}
D.J.~{Eisenstein}, \emph{{Dark energy and cosmic sound [review article]}},
  \href{https://doi.org/10.1016/j.newar.2005.08.005}{\emph{\nar} {\bfseries 49}
  (2005) 360}.

\bibitem{2010deot.book..246B}
B.~{Bassett} and R.~{Hlozek}, \emph{{Baryon acoustic oscillations}},  in
  \emph{Dark Energy: Observational and Theoretical Approaches},
  P.~{Ruiz-Lapuente}, ed., p.~246 (2010),
  \href{https://doi.org/10.48550/arXiv.0910.5224}{DOI}.

\bibitem{KP4s2-Chen}
S.-F.~{Chen}, C.~{Howlett}, M.~{White}, P.~{McDonald}, A.J.~{Ross}, H.-J.~{Seo}
  et~al., \emph{{Baryon Acoustic Oscillation Theory and Modelling Systematics
  for the DESI 2024 results}},
  \href{https://doi.org/10.48550/arXiv.2402.14070}{\emph{arXiv e-prints} (2024)
  arXiv:2402.14070} [\href{https://arxiv.org/abs/2402.14070}{{\ttfamily
  2402.14070}}].

\bibitem{Peebles70}
P.J.E.~{Peebles} and J.T.~{Yu}, \emph{{Primeval Adiabatic Perturbation in an
  Expanding Universe}}, \href{https://doi.org/10.1086/150713}{\emph{\apj}
  {\bfseries 162} (1970) 815}.

\bibitem{Sunyaev70}
R.A.~{Sunyaev} and Y.B.~{Zeldovich}, \emph{{Small-Scale Fluctuations of Relic
  Radiation}}, \href{https://doi.org/10.1007/BF00653471}{\emph{\apss}
  {\bfseries 7} (1970) 3}.

\bibitem{Alcock1979}
C.~{Alcock} and B.~{Paczynski}, \emph{{An evolution free test for non-zero
  cosmological constant}}, \href{https://doi.org/10.1038/281358a0}{\emph{\nat}
  {\bfseries 281} (1979) 358}.

\bibitem{Padmanabhan08}
N.~{Padmanabhan} and M.~{White}, \emph{{Constraining anisotropic baryon
  oscillations}}, \href{https://doi.org/10.1103/PhysRevD.77.123540}{\emph{\prd}
  {\bfseries 77} (2008) 123540}
  [\href{https://arxiv.org/abs/0804.0799}{{\ttfamily 0804.0799}}].

\bibitem{Eisenstein07}
D.J.~{Eisenstein}, H.-J.~{Seo}, E.~{Sirko} and D.N.~{Spergel}, \emph{{Improving
  Cosmological Distance Measurements by Reconstruction of the Baryon Acoustic
  Peak}}, \href{https://doi.org/10.1086/518712}{\emph{\apj} {\bfseries 664}
  (2007) 675} [\href{https://arxiv.org/abs/astro-ph/0604362}{{\ttfamily
  astro-ph/0604362}}].

\bibitem{Seo08}
H.-J.~{Seo}, E.R.~{Siegel}, D.J.~{Eisenstein} and M.~{White}, \emph{{Nonlinear
  Structure Formation and the Acoustic Scale}},
  \href{https://doi.org/10.1086/589921}{\emph{\apj} {\bfseries 686} (2008) 13}
  [\href{https://arxiv.org/abs/0805.0117}{{\ttfamily 0805.0117}}].

\bibitem{Padmanabhan09a}
N.~{Padmanabhan}, M.~{White} and J.D.~{Cohn}, \emph{{Reconstructing baryon
  oscillations: A Lagrangian theory perspective}},
  \href{https://doi.org/10.1103/PhysRevD.79.063523}{\emph{\prd} {\bfseries 79}
  (2009) 063523} [\href{https://arxiv.org/abs/0812.2905}{{\ttfamily
  0812.2905}}].

\bibitem{Seo2010}
H.-J.~{Seo}, J.~{Eckel}, D.J.~{Eisenstein}, K.~{Mehta}, M.~{Metchnik},
  N.~{Padmanabhan} et~al., \emph{{High-precision Predictions for the Acoustic
  Scale in the Nonlinear Regime}},
  \href{https://doi.org/10.1088/0004-637X/720/2/1650}{\emph{\apj} {\bfseries
  720} (2010) 1650} [\href{https://arxiv.org/abs/0910.5005}{{\ttfamily
  0910.5005}}].

\bibitem{Noh09}
Y.~{Noh}, M.~{White} and N.~{Padmanabhan}, \emph{{Reconstructing baryon
  oscillations}}, \href{https://doi.org/10.1103/PhysRevD.80.123501}{\emph{\prd}
  {\bfseries 80} (2009) 123501}
  [\href{https://arxiv.org/abs/0909.1802}{{\ttfamily 0909.1802}}].

\bibitem{White15}
M.~{White}, \emph{{Reconstruction within the Zeldovich approximation}},
  \href{https://doi.org/10.1093/mnras/stv842}{\emph{\mnras} {\bfseries 450}
  (2015) 3822} [\href{https://arxiv.org/abs/1504.03677}{{\ttfamily
  1504.03677}}].

\bibitem{Hikage20}
C.~{Hikage}, K.~{Koyama} and R.~{Takahashi}, \emph{{Perturbation theory for the
  redshift-space matter power spectra after reconstruction}},
  \href{https://doi.org/10.1103/PhysRevD.101.043510}{\emph{\prd} {\bfseries
  101} (2020) 043510} [\href{https://arxiv.org/abs/1911.06461}{{\ttfamily
  1911.06461}}].

\bibitem{Anderson12}
L.~{Anderson}, E.~{Aubourg}, S.~{Bailey}, D.~{Bizyaev}, M.~{Blanton},
  A.S.~{Bolton} et~al., \emph{{The clustering of galaxies in the SDSS-III
  Baryon Oscillation Spectroscopic Survey: baryon acoustic oscillations in the
  Data Release 9 spectroscopic galaxy sample}},
  \href{https://doi.org/10.1111/j.1365-2966.2012.22066.x}{\emph{\mnras}
  {\bfseries 427} (2012) 3435}
  [\href{https://arxiv.org/abs/1203.6594}{{\ttfamily 1203.6594}}].

\bibitem{Ross17}
A.J.~{Ross}, F.~{Beutler}, C.-H.~{Chuang}, M.~{Pellejero-Ibanez}, H.-J.~{Seo},
  M.~{Vargas-Maga{\~n}a} et~al., \emph{{The clustering of galaxies in the
  completed SDSS-III Baryon Oscillation Spectroscopic Survey: observational
  systematics and baryon acoustic oscillations in the correlation function}},
  \href{https://doi.org/10.1093/mnras/stw2372}{\emph{\mnras} {\bfseries 464}
  (2017) 1168} [\href{https://arxiv.org/abs/1607.03145}{{\ttfamily
  1607.03145}}].

\bibitem{Ata2018}
M.~{Ata}, F.~{Baumgarten}, J.~{Bautista}, F.~{Beutler}, D.~{Bizyaev},
  M.R.~{Blanton} et~al., \emph{{The clustering of the SDSS-IV extended Baryon
  Oscillation Spectroscopic Survey DR14 quasar sample: first measurement of
  baryon acoustic oscillations between redshift 0.8 and 2.2}},
  \href{https://doi.org/10.1093/mnras/stx2630}{\emph{\mnras} {\bfseries 473}
  (2018) 4773} [\href{https://arxiv.org/abs/1705.06373}{{\ttfamily
  1705.06373}}].

\bibitem{MerzImagingSysBAO}
G.~{Merz}, M.~{Rezaie}, H.-J.~{Seo}, R.~{Neveux}, A.J.~{Ross}, F.~{Beutler}
  et~al., \emph{{The clustering of the SDSS-IV extended Baryon Oscillation
  Spectroscopic Survey quasar sample: testing observational systematics on the
  Baryon Acoustic Oscillation measurement}},
  \href{https://doi.org/10.1093/mnras/stab1887}{\emph{\mnras} {\bfseries 506}
  (2021) 2503} [\href{https://arxiv.org/abs/2105.10463}{{\ttfamily
  2105.10463}}].

\bibitem{KP3s2-Rosado}
{Rosado et al.}{\emph{, in preparation} (2024) }.

\bibitem{KP3s4-Yu}
{Yu et al.}{\emph{, in preparation} (2024) }.

\bibitem{BAO-discovery}
D.J.~{Eisenstein}, I.~{Zehavi}, D.W.~{Hogg}, R.~{Scoccimarro}, M.R.~{Blanton},
  R.C.~{Nichol} et~al., \emph{{Detection of the Baryon Acoustic Peak in the
  Large-Scale Correlation Function of SDSS Luminous Red Galaxies}},
  \href{https://doi.org/10.1086/466512}{\emph{\apj} {\bfseries 633} (2005) 560}
  [\href{https://arxiv.org/abs/astro-ph/0501171}{{\ttfamily
  astro-ph/0501171}}].

\bibitem{BAO2dF}
S.~{Cole}, W.J.~{Percival}, J.A.~{Peacock}, P.~{Norberg}, C.M.~{Baugh},
  C.S.~{Frenk} et~al., \emph{{The 2dF Galaxy Redshift Survey: power-spectrum
  analysis of the final data set and cosmological implications}},
  \href{https://doi.org/10.1111/j.1365-2966.2005.09318.x}{\emph{\mnras}
  {\bfseries 362} (2005) 505}
  [\href{https://arxiv.org/abs/astro-ph/0501174}{{\ttfamily
  astro-ph/0501174}}].

\bibitem{SixdFBAO}
F.~{Beutler}, C.~{Blake}, M.~{Colless}, D.H.~{Jones}, L.~{Staveley-Smith},
  L.~{Campbell} et~al., \emph{{The 6dF Galaxy Survey: baryon acoustic
  oscillations and the local Hubble constant}},
  \href{https://doi.org/10.1111/j.1365-2966.2011.19250.x}{\emph{\mnras}
  {\bfseries 416} (2011) 3017}
  [\href{https://arxiv.org/abs/1106.3366}{{\ttfamily 1106.3366}}].

\bibitem{SDSS-DR12-cosmology}
S.~{Alam}, M.~{Ata}, S.~{Bailey}, F.~{Beutler}, D.~{Bizyaev}, J.A.~{Blazek}
  et~al., \emph{{The clustering of galaxies in the completed SDSS-III Baryon
  Oscillation Spectroscopic Survey: cosmological analysis of the DR12 galaxy
  sample}}, \href{https://doi.org/10.1093/mnras/stx721}{\emph{\mnras}
  {\bfseries 470} (2017) 2617}
  [\href{https://arxiv.org/abs/1607.03155}{{\ttfamily 1607.03155}}].

\bibitem{Wigglez}
C.~{Blake}, S.~{Brough}, M.~{Colless}, C.~{Contreras}, W.~{Couch}, S.~{Croom}
  et~al., \emph{{The WiggleZ Dark Energy Survey: joint measurements of the
  expansion and growth history at z < 1}},
  \href{https://doi.org/10.1111/j.1365-2966.2012.21473.x}{\emph{\mnras}
  {\bfseries 425} (2012) 405}
  [\href{https://arxiv.org/abs/1204.3674}{{\ttfamily 1204.3674}}].

\bibitem{LyABAO}
H.~{du Mas des Bourboux}, J.~{Rich}, A.~{Font-Ribera}, V.~{de Sainte Agathe},
  J.~{Farr}, T.~{Etourneau} et~al., \emph{{The Completed SDSS-IV Extended
  Baryon Oscillation Spectroscopic Survey: Baryon Acoustic Oscillations with
  Ly{\ensuremath{\alpha}} Forests}},
  \href{https://doi.org/10.3847/1538-4357/abb085}{\emph{\apj} {\bfseries 901}
  (2020) 153} [\href{https://arxiv.org/abs/2007.08995}{{\ttfamily
  2007.08995}}].

\bibitem{Planck2018}
{Planck Collaboration}, N.~{Aghanim}, Y.~{Akrami}, M.~{Ashdown}, J.~{Aumont},
  C.~{Baccigalupi} et~al., \emph{{Planck 2018 results. VI. Cosmological
  parameters}}, \href{https://doi.org/10.1051/0004-6361/201833910}{\emph{\aap}
  {\bfseries 641} (2020) A6}
  [\href{https://arxiv.org/abs/1807.06209}{{\ttfamily 1807.06209}}].

\bibitem{Addison13}
G.E.~{Addison}, G.~{Hinshaw} and M.~{Halpern}, \emph{{Cosmological constraints
  from baryon acoustic oscillations and clustering of large-scale structure}},
  \href{https://doi.org/10.1093/mnras/stt1687}{\emph{\mnras} {\bfseries 436}
  (2013) 1674} [\href{https://arxiv.org/abs/1304.6984}{{\ttfamily 1304.6984}}].

\bibitem{DESICollaboration2016a}
{DESI Collaboration}, A.~{Aghamousa}, J.~{Aguilar}, S.~{Ahlen}, S.~{Alam},
  L.E.~{Allen} et~al., \emph{{The DESI Experiment Part I: Science,Targeting,
  and Survey Design}}, {\emph{arXiv e-prints} (2016) arXiv:1611.00036}
  [\href{https://arxiv.org/abs/1611.00036}{{\ttfamily 1611.00036}}].

\bibitem{DESI2016b.Instr}
{DESI Collaboration}, A.~{Aghamousa}, J.~{Aguilar}, S.~{Ahlen}, S.~{Alam},
  L.E.~{Allen} et~al., \emph{{The DESI Experiment Part II: Instrument Design}},
  {\emph{arXiv e-prints} (2016) arXiv:1611.00037}
  [\href{https://arxiv.org/abs/1611.00037}{{\ttfamily 1611.00037}}].

\bibitem{DESI2022.KP1.Instr}
{DESI Collaboration}, B.~{Abareshi}, J.~{Aguilar}, S.~{Ahlen}, S.~{Alam},
  D.M.~{Alexander} et~al., \emph{{Overview of the Instrumentation for the Dark
  Energy Spectroscopic Instrument}},
  \href{https://doi.org/10.3847/1538-3881/ac882b}{\emph{\aj} {\bfseries 164}
  (2022) 207} [\href{https://arxiv.org/abs/2205.10939}{{\ttfamily
  2205.10939}}].

\bibitem{DESItarget}
A.D.~Myers et~al., \emph{{The Target Selection Pipeline for the Dark Energy
  Spectroscopic Instrument}},
  \href{https://arxiv.org/abs/2208.08518}{{\ttfamily 2208.08518}}.

\bibitem{DESIspec}
J.~Guy et~al., \emph{{The Spectroscopic Data Processing Pipeline for the Dark
  Energy Spectroscopic Instrument}},
  \href{https://doi.org/10.3847/1538-3881/acb212}{\emph{Astron. J.} {\bfseries
  165} (2023) 144} [\href{https://arxiv.org/abs/2209.14482}{{\ttfamily
  2209.14482}}].

\bibitem{DESI2023a.KP1.SV}
{DESI Collaboration}, A.G.~{Adame}, J.~{Aguilar}, S.~{Ahlen}, S.~{Alam},
  G.~{Aldering} et~al., \emph{{Validation of the Scientific Program for the
  Dark Energy Spectroscopic Instrument}},
  \href{https://doi.org/10.3847/1538-3881/ad0b08}{\emph{\aj} {\bfseries 167}
  (2024) 62} [\href{https://arxiv.org/abs/2306.06307}{{\ttfamily 2306.06307}}].

\bibitem{DESI2023b.KP1.EDR}
{DESI Collaboration}, A.G.~{Adame}, J.~{Aguilar}, S.~{Ahlen}, S.~{Alam},
  G.~{Aldering} et~al., \emph{{The Early Data Release of the Dark Energy
  Spectroscopic Instrument}},
  \href{https://doi.org/10.48550/arXiv.2306.06308}{\emph{arXiv e-prints} (2023)
  arXiv:2306.06308} [\href{https://arxiv.org/abs/2306.06308}{{\ttfamily
  2306.06308}}].

\bibitem{BAO.EDR.Moon.2023}
J.~{Moon}, D.~{Valcin}, M.~{Rashkovetskyi}, C.~{Saulder}, J.N.~{Aguilar},
  S.~{Ahlen} et~al., \emph{{First detection of the BAO signal from early DESI
  data}}, \href{https://doi.org/10.1093/mnras/stad2618}{\emph{\mnras}
  {\bfseries 525} (2023) 5406}
  [\href{https://arxiv.org/abs/2304.08427}{{\ttfamily 2304.08427}}].

\bibitem{LyaBAO.EDR.Gordon.2023}
C.~{Gordon}, A.~{Cuceu}, J.~{Chaves-Montero}, A.~{Font-Ribera},
  A.X.~{Gonz{\'a}lez-Morales}, J.~{Aguilar} et~al., \emph{{3D correlations in
  the Lyman-{\ensuremath{\alpha}} forest from early DESI data}},
  \href{https://doi.org/10.1088/1475-7516/2023/11/045}{\emph{\jcap} {\bfseries
  2023} (2023) 045} [\href{https://arxiv.org/abs/2308.10950}{{\ttfamily
  2308.10950}}].

\bibitem{DESI2024.IV.KP6}
{DESI Collaboration}, \emph{{DESI 2024 IV: Baryon Acoustic Oscillations from
  the Lyman Alpha Forest}}, {\emph{in preparation} (2024) }.

\bibitem{DESI2024.VI.KP7A}
{DESI Collaboration}, \emph{{DESI 2024 VI: Cosmological Constraints from the
  Measurements of Baryon Acoustic Oscillations}}, {\emph{in preparation} (2024)
  }.

\bibitem{2023MNRAS.524.3894R}
M.~{Rashkovetskyi}, D.J.~{Eisenstein}, J.N.~{Aguilar}, D.~{Brooks},
  T.~{Claybaugh}, S.~{Cole} et~al., \emph{{Validation of semi-analytical,
  semi-empirical covariance matrices for two-point correlation function for
  early DESI data}},
  \href{https://doi.org/10.1093/mnras/stad2078}{\emph{\mnras} {\bfseries 524}
  (2023) 3894} [\href{https://arxiv.org/abs/2306.06320}{{\ttfamily
  2306.06320}}].

\bibitem{KP4s3-Chen}
{Chen et al.}{\emph{, in preparation} (2024) }.

\bibitem{KP4s4-Paillas}
{Paillas et al.}{\emph{, in preparation} (2024) }.

\bibitem{KP4s5-Valcin}
{Valcin et al.}{\emph{, in preparation} (2024) }.

\bibitem{KP4s6-Forero-Sanchez}
{Forero-Sanchez et al.}{\emph{, in preparation} (2024) }.

\bibitem{KP4s7-Rashkovetskyi}
{Rashkovetskyi et al.}{\emph{, in preparation} (2024) }.

\bibitem{KP4s8-Alves}
{Alves et al.}{\emph{, in preparation} (2024) }.

\bibitem{KP4s9-Perez-Fernandez}
{Perez-Fernandez et al.}{\emph{, in preparation} (2024) }.

\bibitem{KP4s10-Mena-Fernandez}
{Mena-Fernandez et al.}{\emph{, in preparation} (2024) }.

\bibitem{KP4s11-Garcia-Quintero}
{Garcia-Quintero et al.}{\emph{, in preparation} (2024) }.

\bibitem{KP3s9-Andrade}
{Andrade et al.}{\emph{, in preparation} (2024) }.

\bibitem{AbacusSummit}
N.A.~{Maksimova}, L.H.~{Garrison}, D.J.~{Eisenstein}, B.~{Hadzhiyska},
  S.~{Bose} and T.P.~{Satterthwaite}, \emph{{ABACUSSUMMIT: a massive set of
  high-accuracy, high-resolution N-body simulations}},
  \href{https://doi.org/10.1093/mnras/stab2484}{\emph{\mnras} {\bfseries 508}
  (2021) 4017} [\href{https://arxiv.org/abs/2110.11398}{{\ttfamily
  2110.11398}}].

\bibitem{DESI2024.I.DR1}
{DESI Collaboration}, \emph{{DESI 2024 I: Data Release 1 of the Dark Energy
  Spectroscopic Instrument}}, {\emph{in preparation} (2024) }.

\bibitem{DESI_instrument}
B.~{Abareshi}, J.~{Aguilar}, S.~{Ahlen}, S.~{Alam}, D.M.~{Alexander},
  R.~{Alfarsy} et~al., \emph{{Overview of the Instrumentation for the Dark
  Energy Spectroscopic Instrument}},
  \href{https://doi.org/10.3847/1538-3881/ac882b}{\emph{\aj} {\bfseries 164}
  (2022) 207} [\href{https://arxiv.org/abs/2205.10939}{{\ttfamily
  2205.10939}}].

\bibitem{DESIsv}
{DESI Collaboration}, A.G.~{Adame}, J.~{Aguilar}, S.~{Ahlen}, S.~{Alam},
  G.~{Aldering} et~al., \emph{{Validation of the Scientific Program for the
  Dark Energy Spectroscopic Instrument}},
  \href{https://doi.org/10.48550/arXiv.2306.06307}{\emph{arXiv e-prints} (2023)
  arXiv:2306.06307} [\href{https://arxiv.org/abs/2306.06307}{{\ttfamily
  2306.06307}}].

\bibitem{DESIfocalplane}
J.H.~Silber, P.~Fagrelius, K.~Fanning, M.~Schubnell, J.N.~Aguilar, S.~Ahlen
  et~al., \emph{The robotic multiobject focal plane system of the dark energy
  spectroscopic instrument ({DESI})},
  \href{https://doi.org/10.3847/1538-3881/ac9ab1}{\emph{The Astronomical
  Journal} {\bfseries 165} (2022) 9}.

\bibitem{DESIfa}
{Raichoor et al. in}, \emph{{Dark Energy Spectroscopic Instrument Fiber
  Assignment}},  \href{https://arxiv.org/abs/xxxx.xxxxx}{{\ttfamily
  xxxx.xxxxx}}.

\bibitem{DESIsops}
{Schlafly et al. in}, \emph{{Dark Energy Spectroscopic Instrument Survey
  Operations}},  \href{https://arxiv.org/abs/xxxx.xxxxx}{{\ttfamily
  xxxx.xxxxx}}.

\bibitem{DESIbgstarget}
C.~Hahn et~al., \emph{{DESI Bright Galaxy Survey: Final Target Selection,
  Design, and Validation}},  \href{https://arxiv.org/abs/2208.08512}{{\ttfamily
  2208.08512}}.

\bibitem{DESIlrgtarget}
R.~Zhou et~al., \emph{{Target Selection and Validation of DESI Luminous Red
  Galaxies}},  \href{https://arxiv.org/abs/2208.08515}{{\ttfamily 2208.08515}}.

\bibitem{DESIqsotarget}
E.~Chaussidon et~al., \emph{{Target Selection and Validation of DESI Quasars}},
   \href{https://arxiv.org/abs/2208.08511}{{\ttfamily 2208.08511}}.

\bibitem{DESIelgtarget}
A.~Raichoor et~al., \emph{{Target Selection and Validation of DESI Emission
  Line Galaxies}},  \href{https://arxiv.org/abs/2208.08513}{{\ttfamily
  2208.08513}}.

\bibitem{desilss}
A.J.~Ross et~al., \emph{{Construction of Large-scale Structure Catalogs for
  Data from the Dark Energy Spectroscopic Instrument}}, {\emph{in prep.} (2024)
  } [\href{https://arxiv.org/abs/xxxx.xxxxx}{{\ttfamily xxxx.xxxxx}}].

\bibitem{DESI2024.II.KP3}
{DESI Collaboration}, \emph{{DESI 2024 II: Two Point Clustering Measurements
  and Validation}}, {\emph{in preparation} (2024) }.

\bibitem{LS.Overview.Dey.2019}
A.~{Dey}, D.J.~{Schlegel}, D.~{Lang}, R.~{Blum}, K.~{Burleigh}, X.~{Fan}
  et~al., \emph{{Overview of the DESI Legacy Imaging Surveys}},
  \href{https://doi.org/10.3847/1538-3881/ab089d}{\emph{\aj} {\bfseries 157}
  (2019) 168} [\href{https://arxiv.org/abs/1804.08657}{{\ttfamily
  1804.08657}}].

\bibitem{KP3s1-Zhou}
{Zhou et al.}{\emph{, in preparation} (2024) }.

\bibitem{KP3s10-Chaussidon}
{Chaussidon et al.}{\emph{, in preparation} (2024) }.

\bibitem{KP3s3-Krolewski}
{Krolewski et al.}{\emph{, in preparation} (2024) }.

\bibitem{KP3s7-Lasker}
{Lasker et al.}{\emph{, in preparation} (2024) }.

\bibitem{KP3s6-Bianchi}
{Bianchi et al.}{\emph{, in preparation} (2024) }.

\bibitem{KP3s5-Pinon}
{Pinon et al.}{\emph{, in preparation} (2024) }.

\bibitem{fkp}
H.A.~Feldman, N.~Kaiser and J.A.~Peacock, \emph{{Power spectrum analysis of
  three-dimensional redshift surveys}},
  \href{https://doi.org/10.1086/174036}{\emph{Astrophys. J.} {\bfseries 426}
  (1994) 23} [\href{https://arxiv.org/abs/astro-ph/9304022}{{\ttfamily
  astro-ph/9304022}}].

\bibitem{FontRibera2014}
A.~Font-Ribera, P.~McDonald, N.~Mostek, B.A.~Reid, H.-J.~Seo and A.~Slosar,
  \emph{{DESI and other dark energy experiments in the era of neutrino mass
  measurements}},
  \href{https://doi.org/10.1088/1475-7516/2014/05/023}{\emph{JCAP} {\bfseries
  05} (2014) 023} [\href{https://arxiv.org/abs/1308.4164}{{\ttfamily
  1308.4164}}].

\bibitem{BGS.TS.Hahn.2023}
C.~{Hahn}, M.J.~{Wilson}, O.~{Ruiz-Macias}, S.~{Cole}, D.H.~{Weinberg},
  J.~{Moustakas} et~al., \emph{{The DESI Bright Galaxy Survey: Final Target
  Selection, Design, and Validation}},
  \href{https://doi.org/10.3847/1538-3881/accff8}{\emph{\aj} {\bfseries 165}
  (2023) 253} [\href{https://arxiv.org/abs/2208.08512}{{\ttfamily
  2208.08512}}].

\bibitem{fastspecfit}
J.~Moustakas et~al., \emph{{fastspecfit }}, {\emph{in prep.} (2024) }
  [\href{https://arxiv.org/abs/xxxx.xxxxx}{{\ttfamily xxxx.xxxxx}}].

\bibitem{fastspecfit_code}
J.~{Moustakas}, D.~{Scholte}, B.~{Dey} and A.~{Khederlarian}, ``{FastSpecFit:
  Fast spectral synthesis and emission-line fitting of DESI spectra}.''
  Astrophysics Source Code Library, record ascl:2308.005, Aug., 2023.

\bibitem{EDR_HOD_LRGQSO2023}
S.~{Yuan}, H.~{Zhang}, A.J.~{Ross}, J.~{Donald-McCann}, B.~{Hadzhiyska},
  R.H.~{Wechsler} et~al., \emph{{The DESI One-Percent Survey: Exploring the
  Halo Occupation Distribution of Luminous Red Galaxies and Quasi-Stellar
  Objects with AbacusSummit}},
  \href{https://doi.org/10.48550/arXiv.2306.06314}{\emph{arXiv e-prints} (2023)
  arXiv:2306.06314} [\href{https://arxiv.org/abs/2306.06314}{{\ttfamily
  2306.06314}}].

\bibitem{brieden2020}
S.~{Brieden}, H.~{Gil-Mar{\'\i}n}, L.~{Verde} and J.L.~{Bernal}, \emph{{Blind
  Observers of the Sky}},
  \href{https://doi.org/10.1088/1475-7516/2020/09/052}{\emph{\jcap} {\bfseries
  2020} (2020) 052} [\href{https://arxiv.org/abs/2006.10857}{{\ttfamily
  2006.10857}}].

\bibitem{Hadzhiyska23}
B.~{Hadzhiyska}, M.J.~{White}, X.~{Chen}, L.H.~{Garrison}, J.~{DeRose},
  N.~{Padmanabhan} et~al., \emph{{Mitigating the noise of DESI mocks using
  analytic control variates}},
  \href{https://doi.org/10.21105/astro.2308.12343}{\emph{The Open Journal of
  Astrophysics} {\bfseries 6} (2023) 38}
  [\href{https://arxiv.org/abs/2308.12343}{{\ttfamily 2308.12343}}].

\bibitem{abacusnbody}
L.H.~{Garrison}, D.J.~{Eisenstein} and P.A.~{Pinto}, \emph{{A high-fidelity
  realization of the Euclid code comparison N-body simulation with ABACUS}},
  \href{https://doi.org/10.1093/mnras/stz634}{\emph{\mnras} {\bfseries 485}
  (2019) 3370} [\href{https://arxiv.org/abs/1810.02916}{{\ttfamily
  1810.02916}}].

\bibitem{compaso}
B.~{Hadzhiyska}, D.~{Eisenstein}, S.~{Bose}, L.H.~{Garrison} and
  N.~{Maksimova}, \emph{{COMPASO: A new halo finder for competitive assignment
  to spherical overdensities}},
  \href{https://doi.org/10.1093/mnras/stab2980}{\emph{\mnras} {\bfseries 509}
  (2022) 501} [\href{https://arxiv.org/abs/2110.11408}{{\ttfamily
  2110.11408}}].

\bibitem{2021Bose}
S.~{Bose}, D.J.~{Eisenstein}, B.~{Hadzhiyska}, L.H.~{Garrison} and S.~{Yuan},
  \emph{{Constructing high-fidelity halo merger trees in ABACUSSUMMIT}},
  \href{https://doi.org/10.1093/mnras/stac555}{\emph{\mnras} {\bfseries 512}
  (2022) 837} [\href{https://arxiv.org/abs/2110.11409}{{\ttfamily
  2110.11409}}].

\bibitem{abacushod}
S.~{Yuan}, L.H.~{Garrison}, B.~{Hadzhiyska}, S.~{Bose} and D.J.~{Eisenstein},
  \emph{{ABACUSHOD: A highly efficient extended multi-tracer HOD framework and
  its application to BOSS and eBOSS data}},
  \href{https://doi.org/10.1093/mnras/stab3355}{\emph{\mnras} {\bfseries 510}
  (2021) 3301} [\href{https://arxiv.org/abs/2110.11412}{{\ttfamily
  2110.11412}}].

\bibitem{2022AbacusHOD}
S.~{Yuan}, L.H.~{Garrison}, B.~{Hadzhiyska}, S.~{Bose} and D.J.~{Eisenstein},
  \emph{{ABACUSHOD: a highly efficient extended multitracer HOD framework and
  its application to BOSS and eBOSS data}},
  \href{https://doi.org/10.1093/mnras/stab3355}{\emph{\mnras} {\bfseries 510}
  (2022) 3301} [\href{https://arxiv.org/abs/2110.11412}{{\ttfamily
  2110.11412}}].

\bibitem{AlamMulti}
S.~{Alam}, J.A.~{Peacock}, K.~{Kraljic}, A.J.~{Ross} and J.~{Comparat},
  \emph{{Multitracer extension of the halo model: probing quenching and
  conformity in eBOSS}},
  \href{https://doi.org/10.1093/mnras/staa1956}{\emph{\mnras} {\bfseries 497}
  (2020) 581} [\href{https://arxiv.org/abs/1910.05095}{{\ttfamily
  1910.05095}}].

\bibitem{EDR_HOD_ELG2023}
A.~{Rocher}, V.~{Ruhlmann-Kleider}, E.~{Burtin}, S.~{Yuan}, A.~{de Mattia},
  A.J.~{Ross} et~al., \emph{{The DESI One-Percent survey: exploring the Halo
  Occupation Distribution of Emission Line Galaxies with ABACUSSUMMIT
  simulations}},
  \href{https://doi.org/10.1088/1475-7516/2023/10/016}{\emph{\jcap} {\bfseries
  2023} (2023) 016} [\href{https://arxiv.org/abs/2306.06319}{{\ttfamily
  2306.06319}}].

\bibitem{EDR_BGS_ABACUS}
A.~{Smith}, C.~{Grove}, S.~{Cole}, P.~{Norberg}, P.~{Zarrouk}, S.~{Yuan}
  et~al., \emph{{Generating mock galaxy catalogues for flux-limited samples
  like the DESI Bright Galaxy Survey}},
  \href{https://doi.org/10.48550/arXiv.2312.08792}{\emph{arXiv e-prints} (2023)
  arXiv:2312.08792} [\href{https://arxiv.org/abs/2312.08792}{{\ttfamily
  2312.08792}}].

\bibitem{Chuang:2014vfa}
C.-H.~Chuang, F.-S.~Kitaura, F.~Prada, C.~Zhao and G.~Yepes, \emph{{EZmocks:
  extending the Zel'dovich approximation to generate mock galaxy catalogues
  with accurate clustering statistics}},
  \href{https://doi.org/10.1093/mnras/stu2301}{\emph{\mnras} {\bfseries 446}
  (2015) 2621} [\href{https://arxiv.org/abs/1409.1124}{{\ttfamily 1409.1124}}].

\bibitem{Zhao2021}
C.~{Zhao}, C.-H.~{Chuang}, J.~{Bautista}, A.~{de Mattia}, A.~{Raichoor},
  A.J.~{Ross} et~al., \emph{{The completed SDSS-IV extended Baryon Oscillation
  Spectroscopic Survey: 1000 multi-tracer mock catalogues with redshift
  evolution and systematics for galaxies and quasars of the final data
  release}}, \href{https://doi.org/10.1093/mnras/stab510}{\emph{\mnras}
  {\bfseries 503} (2021) 1149}
  [\href{https://arxiv.org/abs/2007.08997}{{\ttfamily 2007.08997}}].

\bibitem{Zarrouk:2020hha}
P.~{Zarrouk}, M.~{Rezaie}, A.~{Raichoor}, A.J.~{Ross}, S.~{Alam}, R.~{Blum}
  et~al., \emph{{Baryon acoustic oscillations in the projected
  cross-correlation function between the eBOSS DR16 quasars and photometric
  galaxies from the DESI Legacy Imaging Surveys}},
  \href{https://doi.org/10.1093/mnras/stab298}{\emph{\mnras} {\bfseries 503}
  (2021) 2562} [\href{https://arxiv.org/abs/2009.02308}{{\ttfamily
  2009.02308}}].

\bibitem{Zeldovich:1969sb}
Y.~Zel'dovich, \emph{{Gravitational instability: An Approximate theory for
  large density perturbations}}, {\emph{\aa} {\bfseries 5} (1970) 84}.

\bibitem{KP3s8-Zhao}
{Zhao et al.}{\emph{, in preparation} (2024) }.

\bibitem{KP3}
{DESI Collaboration}, \emph{{The Dark Energy Survey Instrument Year 1 Results:
  2-Point Clustering Measurements and their Validation}}, {\emph{in
  preparation} (2024) }.

\bibitem{deMattia19IC}
A.~de~Mattia and V.~Ruhlmann-Kleider, \emph{{Integral constraints in
  spectroscopic surveys}},
  \href{https://doi.org/10.1088/1475-7516/2019/08/036}{\emph{JCAP} {\bfseries
  08} (2019) 036} [\href{https://arxiv.org/abs/1904.08851}{{\ttfamily
  1904.08851}}].

\bibitem{Landy1993}
S.D.~{Landy} and A.S.~{Szalay}, \emph{{Bias and Variance of Angular Correlation
  Functions}}, \href{https://doi.org/10.1086/172900}{\emph{\apj} {\bfseries
  412} (1993) 64}.

\bibitem{Padmanabhan12}
N.~{Padmanabhan}, X.~{Xu}, D.J.~{Eisenstein}, R.~{Scalzo}, A.J.~{Cuesta},
  K.T.~{Mehta} et~al., \emph{{A 2 per cent distance to z = 0.35 by
  reconstructing baryon acoustic oscillations - I. Methods and application to
  the Sloan Digital Sky Survey}},
  \href{https://doi.org/10.1111/j.1365-2966.2012.21888.x}{\emph{\mnras}
  {\bfseries 427} (2012) 2132}
  [\href{https://arxiv.org/abs/1202.0090}{{\ttfamily 1202.0090}}].

\bibitem{yamamoto2006}
K.~Yamamoto, M.~Nakamichi, A.~Kamino, B.A.~Bassett and H.~Nishioka, \emph{A
  {Measurement} of the {Quadrupole} {Power} {Spectrum} in the {Clustering} of
  the {2dF} {QSO} {Survey}},
  \href{https://doi.org/10.1093/pasj/58.1.93}{\emph{Publications of the
  Astronomical Society of Japan} {\bfseries 58} (2006) 93}.

\bibitem{Hand2017:1712.05834v1}
N.~{Hand}, Y.~{Feng}, F.~{Beutler}, Y.~{Li}, C.~{Modi}, U.~{Seljak} et~al.,
  \emph{{nbodykit: An Open-source, Massively Parallel Toolkit for Large-scale
  Structure}}, \href{https://doi.org/10.3847/1538-3881/aadae0}{\emph{\aj}
  {\bfseries 156} (2018) 160}
  [\href{https://arxiv.org/abs/1712.05834}{{\ttfamily 1712.05834}}].

\bibitem{Tojeiro2014:1401.1768}
R.~{Tojeiro}, A.J.~{Ross}, A.~{Burden}, L.~{Samushia}, M.~{Manera},
  W.J.~{Percival} et~al., \emph{{The clustering of galaxies in the SDSS-III
  Baryon Oscillation Spectroscopic Survey: galaxy clustering measurements in
  the low-redshift sample of Data Release 11}},
  \href{https://doi.org/10.1093/mnras/stu371}{\emph{\mnras} {\bfseries 440}
  (2014) 2222} [\href{https://arxiv.org/abs/1401.1768}{{\ttfamily 1401.1768}}].

\bibitem{Kazin2014:1401.0358}
E.A.~{Kazin}, J.~{Koda}, C.~{Blake}, N.~{Padmanabhan}, S.~{Brough},
  M.~{Colless} et~al., \emph{{The WiggleZ Dark Energy Survey: improved distance
  measurements to z = 1 with reconstruction of the baryonic acoustic feature}},
  \href{https://doi.org/10.1093/mnras/stu778}{\emph{\mnras} {\bfseries 441}
  (2014) 3524} [\href{https://arxiv.org/abs/1401.0358}{{\ttfamily 1401.0358}}].

\bibitem{Alam17}
S.~{Alam}, M.~{Ata}, S.~{Bailey}, F.~{Beutler}, D.~{Bizyaev}, J.A.~{Blazek}
  et~al., \emph{{The clustering of galaxies in the completed SDSS-III Baryon
  Oscillation Spectroscopic Survey: cosmological analysis of the DR12 galaxy
  sample}}, \href{https://doi.org/10.1093/mnras/stx721}{\emph{\mnras}
  {\bfseries 470} (2017) 2617}
  [\href{https://arxiv.org/abs/1607.03155}{{\ttfamily 1607.03155}}].

\bibitem{Gil-Marin2020}
H.~{Gil-Mar{\'\i}n}, J.E.~{Bautista}, R.~{Paviot}, M.~{Vargas-Maga{\~n}a},
  S.~{de la Torre}, S.~{Fromenteau} et~al., \emph{{The Completed SDSS-IV
  extended Baryon Oscillation Spectroscopic Survey: measurement of the BAO and
  growth rate of structure of the luminous red galaxy sample from the
  anisotropic power spectrum between redshifts 0.6 and 1.0}},
  \href{https://doi.org/10.1093/mnras/staa2455}{\emph{\mnras} {\bfseries 498}
  (2020) 2492} [\href{https://arxiv.org/abs/2007.08994}{{\ttfamily
  2007.08994}}].

\bibitem{Seo2010:0910.5005}
H.-J.~{Seo}, J.~{Eckel}, D.J.~{Eisenstein}, K.~{Mehta}, M.~{Metchnik},
  N.~{Padmanabhan} et~al., \emph{{High-precision Predictions for the Acoustic
  Scale in the Nonlinear Regime}},
  \href{https://doi.org/10.1088/0004-637X/720/2/1650}{\emph{\apj} {\bfseries
  720} (2010) 1650} [\href{https://arxiv.org/abs/0910.5005}{{\ttfamily
  0910.5005}}].

\bibitem{Schmittfull2017:1704.06634}
M.~{Schmittfull}, T.~{Baldauf} and M.~{Zaldarriaga}, \emph{{Iterative initial
  condition reconstruction}},
  \href{https://doi.org/10.1103/PhysRevD.96.023505}{\emph{\prd} {\bfseries 96}
  (2017) 023505} [\href{https://arxiv.org/abs/1704.06634}{{\ttfamily
  1704.06634}}].

\bibitem{Hada2018:1804.04738}
R.~{Hada} and D.J.~{Eisenstein}, \emph{{An iterative reconstruction of
  cosmological initial density fields}},
  \href{https://doi.org/10.1093/mnras/sty1203}{\emph{\mnras} {\bfseries 478}
  (2018) 1866} [\href{https://arxiv.org/abs/1804.04738}{{\ttfamily
  1804.04738}}].

\bibitem{SeoBAOmodel2016}
H.-J.~{Seo}, F.~{Beutler}, A.J.~{Ross} and S.~{Saito}, \emph{{Modeling the
  reconstructed BAO in Fourier space}},
  \href{https://doi.org/10.1093/mnras/stw1138}{\emph{\mnras} {\bfseries 460}
  (2016) 2453} [\href{https://arxiv.org/abs/1511.00663}{{\ttfamily
  1511.00663}}].

\bibitem{Burden2015:1504.02591v2}
A.~{Burden}, W.J.~{Percival} and C.~{Howlett}, \emph{{Reconstruction in Fourier
  space}}, \href{https://doi.org/10.1093/mnras/stv1581}{\emph{\mnras}
  {\bfseries 453} (2015) 456}
  [\href{https://arxiv.org/abs/1504.02591}{{\ttfamily 1504.02591}}].

\bibitem{Bautista2021}
J.E.~{Bautista}, R.~{Paviot}, M.~{Vargas Maga{\~n}a}, S.~{de la Torre},
  S.~{Fromenteau}, H.~{Gil-Mar{\'\i}n} et~al., \emph{{The completed SDSS-IV
  extended Baryon Oscillation Spectroscopic Survey: measurement of the BAO and
  growth rate of structure of the luminous red galaxy sample from the
  anisotropic correlation function between redshifts 0.6 and 1}},
  \href{https://doi.org/10.1093/mnras/staa2800}{\emph{\mnras} {\bfseries 500}
  (2021) 736} [\href{https://arxiv.org/abs/2007.08993}{{\ttfamily
  2007.08993}}].

\bibitem{Anderson14}
L.~{Anderson}, {\'E}.~{Aubourg}, S.~{Bailey}, F.~{Beutler}, V.~{Bhardwaj},
  M.~{Blanton} et~al., \emph{{The clustering of galaxies in the SDSS-III Baryon
  Oscillation Spectroscopic Survey: baryon acoustic oscillations in the Data
  Releases 10 and 11 Galaxy samples}},
  \href{https://doi.org/10.1093/mnras/stu523}{\emph{\mnras} {\bfseries 441}
  (2014) 24} [\href{https://arxiv.org/abs/1312.4877}{{\ttfamily 1312.4877}}].

\bibitem{VargasMagana2014}
M.~{Vargas-Maga{\~n}a}, S.~{Ho}, X.~{Xu}, A.G.~{S{\'a}nchez}, R.~{O'Connell},
  D.J.~{Eisenstein} et~al., \emph{{The clustering of Galaxies in the SDSS-III
  Baryon Oscillation Spectroscopic Survey: potential systematics in fitting of
  baryon acoustic feature}},
  \href{https://doi.org/10.1093/mnras/stu1681}{\emph{\mnras} {\bfseries 445}
  (2014) 2}.

\bibitem{VargasMagana2016:1610.03506v2}
M.~{Vargas-Maga{\~n}a}, S.~{Ho}, A.J.~{Cuesta}, R.~{O'Connell}, A.J.~{Ross},
  D.J.~{Eisenstein} et~al., \emph{{The clustering of galaxies in the completed
  SDSS-III Baryon Oscillation Spectroscopic Survey: theoretical systematics and
  Baryon Acoustic Oscillations in the galaxy correlation function}},
  \href{https://doi.org/10.1093/mnras/sty571}{\emph{\mnras} {\bfseries 477}
  (2018) 1153} [\href{https://arxiv.org/abs/1610.03506}{{\ttfamily
  1610.03506}}].

\bibitem{Beutler17}
F.~{Beutler}, H.-J.~{Seo}, A.J.~{Ross}, P.~{McDonald}, S.~{Saito},
  A.S.~{Bolton} et~al., \emph{{The clustering of galaxies in the completed
  SDSS-III Baryon Oscillation Spectroscopic Survey: baryon acoustic
  oscillations in the Fourier space}},
  \href{https://doi.org/10.1093/mnras/stw2373}{\emph{\mnras} {\bfseries 464}
  (2017) 3409} [\href{https://arxiv.org/abs/1607.03149}{{\ttfamily
  1607.03149}}].

\bibitem{Brieden2022}
S.~{Brieden}, H.~{Gil-Mar{\'\i}n} and L.~{Verde}, \emph{{Model-agnostic
  interpretation of 10 billion years of cosmic evolution traced by BOSS and
  eBOSS data}},
  \href{https://doi.org/10.1088/1475-7516/2022/08/024}{\emph{\jcap} {\bfseries
  2022} (2022) 024} [\href{https://arxiv.org/abs/2204.11868}{{\ttfamily
  2204.11868}}].

\bibitem{Seo2016}
H.-J.~{Seo}, F.~{Beutler}, A.J.~{Ross} and S.~{Saito}, \emph{{Modeling the
  reconstructed BAO in Fourier space}},
  \href{https://doi.org/10.1093/mnras/stw1138}{\emph{\mnras} {\bfseries 460}
  (2016) 2453} [\href{https://arxiv.org/abs/1511.00663}{{\ttfamily
  1511.00663}}].

\bibitem{Jackson72}
J.C.~{Jackson}, \emph{{A critique of Rees's theory of primordial gravitational
  radiation}}, \href{https://doi.org/10.1093/mnras/156.1.1P}{\emph{\mnras}
  {\bfseries 156} (1972) 1P} [\href{https://arxiv.org/abs/0810.3908}{{\ttfamily
  0810.3908}}].

\bibitem{Park94}
C.~{Park}, M.S.~{Vogeley}, M.J.~{Geller} and J.P.~{Huchra}, \emph{{Power
  Spectrum, Correlation Function, and Tests for Luminosity Bias in the CfA
  Redshift Survey}}, \href{https://doi.org/10.1086/174508}{\emph{\apj}
  {\bfseries 431} (1994) 569}.

\bibitem{Kaiser1987}
N.~Kaiser, \emph{{Clustering in real space and in redshift space}}, {\emph{Mon.
  Not. Roy. Astron. Soc.} {\bfseries 227} (1987) 1}.

\bibitem{Sugiyama24}
N.~{Sugiyama}, \emph{{Developing a Theoretical Model for the Resummation of
  Infrared Effects in the Post-Reconstruction Power Spectrum}},
  \href{https://doi.org/10.48550/arXiv.2402.06142}{\emph{arXiv e-prints} (2024)
  arXiv:2402.06142} [\href{https://arxiv.org/abs/2402.06142}{{\ttfamily
  2402.06142}}].

\bibitem{Chaniotis2004}
A.K.~{Chaniotis} and D.~{Poulikakos}, \emph{{High order interpolation and
  differentiation using B-splines}},
  \href{https://doi.org/10.1016/j.jcp.2003.11.026}{\emph{Journal of
  Computational Physics} {\bfseries 197} (2004) 253}.

\bibitem{Sefusatti2016}
E.~{Sefusatti}, M.~{Crocce}, R.~{Scoccimarro} and H.M.P.~{Couchman},
  \emph{{Accurate estimators of correlation functions in Fourier space}},
  \href{https://doi.org/10.1093/mnras/stw1229}{\emph{\mnras} {\bfseries 460}
  (2016) 3624} [\href{https://arxiv.org/abs/1512.07295}{{\ttfamily
  1512.07295}}].

\bibitem{Ross2013}
A.J.~{Ross}, W.J.~{Percival}, A.~{Carnero}, G.-b.~{Zhao}, M.~{Manera},
  A.~{Raccanelli} et~al., \emph{{The clustering of galaxies in the SDSS-III DR9
  Baryon Oscillation Spectroscopic Survey: constraints on primordial
  non-Gaussianity}}, \href{https://doi.org/10.1093/mnras/sts094}{\emph{\mnras}
  {\bfseries 428} (2013) 1116}
  [\href{https://arxiv.org/abs/1208.1491}{{\ttfamily 1208.1491}}].

\bibitem{Beutler21}
F.~{Beutler} and P.~{McDonald}, \emph{{Unified galaxy power spectrum
  measurements from 6dFGS, BOSS, and eBOSS}},
  \href{https://doi.org/10.1088/1475-7516/2021/11/031}{\emph{\jcap} {\bfseries
  2021} (2021) 031} [\href{https://arxiv.org/abs/2106.06324}{{\ttfamily
  2106.06324}}].

\bibitem{jax2018github}
J.~Bradbury, R.~Frostig, P.~Hawkins, M.J.~Johnson, C.~Leary, D.~Maclaurin
  et~al., \emph{{JAX}: composable transformations of {P}ython+{N}um{P}y
  programs},  2018.

\bibitem{emcee}
D.~{Foreman-Mackey}, D.W.~{Hogg}, D.~{Lang} and J.~{Goodman}, \emph{emcee: The
  mcmc hammer}, \href{https://doi.org/10.1086/670067}{\emph{PASP} {\bfseries
  125} (2013) 306} [\href{https://arxiv.org/abs/1202.3665}{{\ttfamily
  1202.3665}}].

\bibitem{minuit}
F.~{James} and M.~{Roos}, \emph{{Minuit - a system for function minimization
  and analysis of the parameter errors and correlations}},
  \href{https://doi.org/10.1016/0010-4655(75)90039-9}{\emph{Computer Physics
  Communications} {\bfseries 10} (1975) 343}.

\bibitem{Hinton2020}
S.R.~{Hinton}, C.~{Howlett} and T.M.~{Davis}, \emph{{BARRY and the BAO model
  comparison}}, \href{https://doi.org/10.1093/mnras/staa361}{\emph{\mnras}
  {\bfseries 493} (2020) 4078}
  [\href{https://arxiv.org/abs/1912.01175}{{\ttfamily 1912.01175}}].

\bibitem{rascal}
R.~{O'Connell}, D.~{Eisenstein}, M.~{Vargas}, S.~{Ho} and N.~{Padmanabhan},
  \emph{{Large covariance matrices: smooth models from the two-point
  correlation function}},
  \href{https://doi.org/10.1093/mnras/stw1821}{\emph{\mnras} {\bfseries 462}
  (2016) 2681} [\href{https://arxiv.org/abs/1510.01740}{{\ttfamily
  1510.01740}}].

\bibitem{rascal-jackknife}
R.~{O'Connell} and D.J.~{Eisenstein}, \emph{{Large covariance matrices:
  accurate models without mocks}},
  \href{https://doi.org/10.1093/mnras/stz1359}{\emph{\mnras} {\bfseries 487}
  (2019) 2701} [\href{https://arxiv.org/abs/1808.05978}{{\ttfamily
  1808.05978}}].

\bibitem{RascalC}
O.H.E.~{Philcox}, D.J.~{Eisenstein}, R.~{O'Connell} and A.~{Wiegand},
  \emph{{RASCALC: a jackknife approach to estimating single- and multitracer
  galaxy covariance matrices}},
  \href{https://doi.org/10.1093/mnras/stz3218}{\emph{\mnras} {\bfseries 491}
  (2020) 3290} [\href{https://arxiv.org/abs/1904.11070}{{\ttfamily
  1904.11070}}].

\bibitem{RascalC-legendre-3}
O.H.E.~{Philcox} and D.J.~{Eisenstein}, \emph{{Estimating covariance matrices
  for two- and three-point correlation function moments in Arbitrary Survey
  Geometries}}, \href{https://doi.org/10.1093/mnras/stz2896}{\emph{\mnras}
  {\bfseries 490} (2019) 5931}
  [\href{https://arxiv.org/abs/1910.04764}{{\ttfamily 1910.04764}}].

\bibitem{RascalC-DA02}
M.~{Rashkovetskyi}, D.J.~{Eisenstein}, J.N.~{Aguilar}, D.~{Brooks},
  T.~{Claybaugh}, S.~{Cole} et~al., \emph{{Validation of semi-analytical,
  semi-empirical covariance matrices for two-point correlation function for
  early DESI data}},
  \href{https://doi.org/10.1093/mnras/stad2078}{\emph{\mnras} {\bfseries 524}
  (2023) 3894} [\href{https://arxiv.org/abs/2306.06320}{{\ttfamily
  2306.06320}}].

\bibitem{Wadekar2019}
D.~Wadekar and R.~Scoccimarro, \emph{{Galaxy power spectrum multipoles
  covariance in perturbation theory}},
  \href{https://doi.org/10.1103/PhysRevD.102.123517}{\emph{Phys. Rev. D}
  {\bfseries 102} (2020) 123517}
  [\href{https://arxiv.org/abs/1910.02914}{{\ttfamily 1910.02914}}].

\bibitem{ESW2007}
D.J.~{Eisenstein}, H.-J.~{Seo} and M.~{White}, \emph{{On the Robustness of the
  Acoustic Scale in the Low-Redshift Clustering of Matter}},
  \href{https://doi.org/10.1086/518755}{\emph{\apj} {\bfseries 664} (2007) 660}
  [\href{https://arxiv.org/abs/astro-ph/0604361}{{\ttfamily
  astro-ph/0604361}}].

\bibitem{Crocce2008}
M.~{Crocce} and R.~{Scoccimarro}, \emph{{Nonlinear evolution of baryon acoustic
  oscillations}}, \href{https://doi.org/10.1103/PhysRevD.77.023533}{\emph{\prd}
  {\bfseries 77} (2008) 023533}
  [\href{https://arxiv.org/abs/0704.2783}{{\ttfamily 0704.2783}}].

\bibitem{SeononlinearBAO}
H.-J.~{Seo}, E.R.~{Siegel}, D.J.~{Eisenstein} and M.~{White}, \emph{{Nonlinear
  Structure Formation and the Acoustic Scale}},
  \href{https://doi.org/10.1086/589921}{\emph{\apj} {\bfseries 686} (2008) 13}
  [\href{https://arxiv.org/abs/0805.0117}{{\ttfamily 0805.0117}}].

\bibitem{Padmanabhan09b}
N.~{Padmanabhan} and M.~{White}, \emph{{Calibrating the baryon oscillation
  ruler for matter and halos}},
  \href{https://doi.org/10.1103/PhysRevD.80.063508}{\emph{\prd} {\bfseries 80}
  (2009) 063508} [\href{https://arxiv.org/abs/0906.1198}{{\ttfamily
  0906.1198}}].

\bibitem{Sherwin12}
B.D.~{Sherwin} and M.~{Zaldarriaga}, \emph{{Shift of the baryon acoustic
  oscillation scale: A simple physical picture}},
  \href{https://doi.org/10.1103/PhysRevD.85.103523}{\emph{\prd} {\bfseries 85}
  (2012) 103523} [\href{https://arxiv.org/abs/1202.3998}{{\ttfamily
  1202.3998}}].

\bibitem{Rossi:2020wxx}
G.~Rossi et~al., \emph{{The completed SDSS-IV extended Baryon Oscillation
  Spectroscopic Survey: N-body mock challenge for galaxy clustering
  measurements}}, \href{https://doi.org/10.1093/mnras/staa3955}{\emph{Mon. Not.
  Roy. Astron. Soc.} {\bfseries 505} (2021) 377}
  [\href{https://arxiv.org/abs/2007.09002}{{\ttfamily 2007.09002}}].

\bibitem{Ross2017}
{\scshape BOSS} collaboration, \emph{{The clustering of galaxies in the
  completed SDSS-III Baryon Oscillation Spectroscopic Survey: Observational
  systematics and baryon acoustic oscillations in the correlation function}},
  \href{https://doi.org/10.1093/mnras/stw2372}{\emph{Mon. Not. Roy. Astron.
  Soc.} {\bfseries 464} (2017) 1168}
  [\href{https://arxiv.org/abs/1607.03145}{{\ttfamily 1607.03145}}].

\bibitem{EDR_SHAM_UNIT2024}
J.~{Yu}, C.~{Zhao}, V.~{Gonzalez-Perez}, C.-H.~{Chuang}, A.~{Brodzeller},
  A.~{de Mattia} et~al., \emph{{The DESI One-Percent Survey: exploring a
  generalized SHAM for multiple tracers with the UNIT simulation}},
  \href{https://doi.org/10.1093/mnras/stad3559}{\emph{\mnras} {\bfseries 527}
  (2024) 6950} [\href{https://arxiv.org/abs/2306.06313}{{\ttfamily
  2306.06313}}].

\bibitem{ELG.TS.Raichoor.2023}
A.~{Raichoor}, J.~{Moustakas}, J.A.~{Newman}, T.~{Karim}, S.~{Ahlen}, S.~{Alam}
  et~al., \emph{{Target Selection and Validation of DESI Emission Line
  Galaxies}}, \href{https://doi.org/10.3847/1538-3881/acb213}{\emph{\aj}
  {\bfseries 165} (2023) 126}
  [\href{https://arxiv.org/abs/2208.08513}{{\ttfamily 2208.08513}}].

\bibitem{Ballinger96}
W.E.~{Ballinger}, J.A.~{Peacock} and A.F.~{Heavens}, \emph{{Measuring the
  cosmological constant with redshift surveys}},
  \href{https://doi.org/10.1093/mnras/282.3.877}{\emph{\mnras} {\bfseries 282}
  (1996) 877} [\href{https://arxiv.org/abs/astro-ph/9605017}{{\ttfamily
  astro-ph/9605017}}].

\bibitem{SeoBAOFisher}
H.-J.~{Seo} and D.J.~{Eisenstein}, \emph{{Probing Dark Energy with Baryonic
  Acoustic Oscillations from Future Large Galaxy Redshift Surveys}},
  \href{https://doi.org/10.1086/379122}{\emph{\apj} {\bfseries 598} (2003) 720}
  [\href{https://arxiv.org/abs/astro-ph/0307460}{{\ttfamily
  astro-ph/0307460}}].

\bibitem{Thepsuriya_2015}
K.~Thepsuriya and A.~Lewis, \emph{Accuracy of cosmological parameters using the
  baryon acoustic scale},
  \href{https://doi.org/10.1088/1475-7516/2015/01/034}{\emph{Journal of
  Cosmology and Astroparticle Physics} {\bfseries 2015} (2015) 034–034}.

\bibitem{Carter2020}
P.~{Carter}, F.~{Beutler}, W.J.~{Percival}, J.~{DeRose}, R.H.~{Wechsler} and
  C.~{Zhao}, \emph{{The impact of the fiducial cosmology assumption on BAO
  distance scale measurements}},
  \href{https://doi.org/10.1093/mnras/staa761}{\emph{\mnras} {\bfseries 494}
  (2020) 2076} [\href{https://arxiv.org/abs/1906.03035}{{\ttfamily
  1906.03035}}].

\bibitem{Bernal20}
J.L.~{Bernal}, T.L.~{Smith}, K.K.~{Boddy} and M.~{Kamionkowski},
  \emph{{Robustness of baryon acoustic oscillation constraints for
  early-Universe modifications of {\ensuremath{\Lambda}} CDM cosmology}},
  \href{https://doi.org/10.1103/PhysRevD.102.123515}{\emph{\prd} {\bfseries
  102} (2020) 123515} [\href{https://arxiv.org/abs/2004.07263}{{\ttfamily
  2004.07263}}].

\bibitem{Pan2023}
J.~Pan, D.~Huterer, F.~Andrade-Oliveira and C.~Avestruz, \emph{Compressed
  baryon acoustic oscillation analysis is robust to modified-gravity models},
  2023.

\bibitem{Sanzwuhl_2024}
S.~Sanz-Wuhl, H.~Gil-Marín, A.J.~Cuesta and L.~Verde, \emph{Bao cosmology in
  non-spatially flat background geometry with application to boss and eboss},
  2024.

\bibitem{Maksimova2021}
N.A.~{Maksimova}, L.H.~{Garrison}, D.J.~{Eisenstein}, B.~{Hadzhiyska},
  S.~{Bose} and T.P.~{Satterthwaite}, \emph{{ABACUSSUMMIT: a massive set of
  high-accuracy, high-resolution N-body simulations}},
  \href{https://doi.org/10.1093/mnras/stab2484}{\emph{\mnras} {\bfseries 508}
  (2021) 4017} [\href{https://arxiv.org/abs/2110.11398}{{\ttfamily
  2110.11398}}].

\bibitem{Baumann_2017}
D.~Baumann, D.~Green and M.~Zaldarriaga, \emph{Phases of new physics in the bao
  spectrum}, \href{https://doi.org/10.1088/1475-7516/2017/11/007}{\emph{Journal
  of Cosmology and Astroparticle Physics} {\bfseries 2017} (2017) 007–007}.

\bibitem{Hikage17}
C.~{Hikage}, K.~{Koyama} and A.~{Heavens}, \emph{{Perturbation theory for BAO
  reconstructed fields: One-loop results in the real-space matter density
  field}}, \href{https://doi.org/10.1103/PhysRevD.96.043513}{\emph{\prd}
  {\bfseries 96} (2017) 043513}
  [\href{https://arxiv.org/abs/1703.07878}{{\ttfamily 1703.07878}}].

\bibitem{Chen19b}
S.-F.~{Chen}, Z.~{Vlah} and M.~{White}, \emph{{The reconstructed power spectrum
  in the Zeldovich approximation}},
  \href{https://doi.org/10.1088/1475-7516/2019/09/017}{\emph{\jcap} {\bfseries
  2019} (2019) 017} [\href{https://arxiv.org/abs/1907.00043}{{\ttfamily
  1907.00043}}].

\bibitem{Eisenstein05}
D.J.~{Eisenstein}, I.~{Zehavi}, D.W.~{Hogg}, R.~{Scoccimarro}, M.R.~{Blanton},
  R.C.~{Nichol} et~al., \emph{{Detection of the Baryon Acoustic Peak in the
  Large-Scale Correlation Function of SDSS Luminous Red Galaxies}},
  \href{https://doi.org/10.1086/466512}{\emph{\apj} {\bfseries 633} (2005) 560}
  [\href{https://arxiv.org/abs/astro-ph/0501171}{{\ttfamily
  astro-ph/0501171}}].

\bibitem{SE2007}
H.-J.~{Seo} and D.J.~{Eisenstein}, \emph{{Improved Forecasts for the Baryon
  Acoustic Oscillations and Cosmological Distance Scale}},
  \href{https://doi.org/10.1086/519549}{\emph{\apj} {\bfseries 665} (2007) 14}
  [\href{https://arxiv.org/abs/astro-ph/0701079}{{\ttfamily
  astro-ph/0701079}}].

\bibitem{Tamone2020:2007.09009}
A.~{Tamone}, A.~{Raichoor}, C.~{Zhao}, A.~{de Mattia}, C.~{Gorgoni},
  E.~{Burtin} et~al., \emph{{The completed SDSS-IV extended baryon oscillation
  spectroscopic survey: growth rate of structure measurement from anisotropic
  clustering analysis in configuration space between redshift 0.6 and 1.1 for
  the emission-line galaxy sample}},
  \href{https://doi.org/10.1093/mnras/staa3050}{\emph{\mnras} {\bfseries 499}
  (2020) 5527} [\href{https://arxiv.org/abs/2007.09009}{{\ttfamily
  2007.09009}}].

\bibitem{Alam21}
S.~{Alam}, M.~{Aubert}, S.~{Avila}, C.~{Balland}, J.E.~{Bautista},
  M.A.~{Bershady} et~al., \emph{{Completed SDSS-IV extended Baryon Oscillation
  Spectroscopic Survey: Cosmological implications from two decades of
  spectroscopic surveys at the Apache Point Observatory}},
  \href{https://doi.org/10.1103/PhysRevD.103.083533}{\emph{\prd} {\bfseries
  103} (2021) 083533} [\href{https://arxiv.org/abs/2007.08991}{{\ttfamily
  2007.08991}}].

\bibitem{desbao}
{DES Collaboration}, T.M.C.~{Abbott}, M.~{Adamow}, M.~{Aguena}, S.~{Allam},
  O.~{Alves} et~al., \emph{{Dark Energy Survey: A 2.1\% measurement of the
  angular Baryonic Acoustic Oscillation scale at redshift $z_{\rm eff}$=0.85
  from the final dataset}},
  \href{https://doi.org/10.48550/arXiv.2402.10696}{\emph{arXiv e-prints} (2024)
  arXiv:2402.10696} [\href{https://arxiv.org/abs/2402.10696}{{\ttfamily
  2402.10696}}].

\bibitem{Neveux2020}
R.~{Neveux}, E.~{Burtin}, A.~{de Mattia}, A.~{Smith}, A.J.~{Ross}, J.~{Hou}
  et~al., \emph{{The completed SDSS-IV extended Baryon Oscillation
  Spectroscopic Survey: BAO and RSD measurements from the anisotropic power
  spectrum of the quasar sample between redshift 0.8 and 2.2}},
  \href{https://doi.org/10.1093/mnras/staa2780}{\emph{\mnras} {\bfseries 499}
  (2020) 210} [\href{https://arxiv.org/abs/2007.08999}{{\ttfamily
  2007.08999}}].

\end{thebibliography}\endgroup

\appendix
\section{BAO measurements in Fourier space}\label{se:app}
\begin{table}
    \centering
    \begin{tabular}{|l|c|c|c|c|c|r|}
    \hline
     Tracer   & Redshift   & Recon   & $\alpha_{\rm iso}$   & $\alpha_{\rm AP}$   &   $r_{\rm off}$ & $\chi^2 / {\rm dof}$   \\
    \hline
     {\tt BGS}  & 0.1--0.4   & Post    & $0.9848 \pm 0.0242$  &                     &                 & 27.4/41                \\
     {\tt LRG1} & 0.4--0.6   & Post    & $0.9799 \pm 0.0118$  & $0.9231 \pm 0.0371$ &         -0.1413 & 85.9/87                \\
     {\tt LRG2} & 0.6--0.8   & Post    & $0.9624 \pm 0.0098$  & $1.0431 \pm 0.0356$ &         -0.0373 & 82.1/87                \\
     {\tt LRG3} & 0.8--1.1   & Post    & $1.0005 \pm 0.0081$  & $1.0057 \pm 0.0261$ &         -0.0991 & 100.8/87               \\
     {\tt ELG1} & 0.8--1.1   & Post    & $0.9913 \pm 0.0249$  &                     &                 & 47.6/41                \\
     {\tt ELG2} & 1.1--1.6   & Post    & $0.9827 \pm 0.0112$  & $0.9986 \pm 0.0352$ &         -0.2217 & 103.8/87               \\
     {\tt QSO}  & 0.8--2.1   & Post    & $1.0094 \pm 0.0212$  &                     &                 & 57.7/41                \\
     \hline
     {\tt BGS}  & 0.1--0.4   & Pre     & $0.9657 \pm 0.0364$  &                     &                 & 50.0/41                \\
     {\tt LRG1} & 0.4--0.6   & Pre     & $0.9774 \pm 0.0179$  & $0.9316 \pm 0.0595$ &          0.2927 & 87.1/87                \\
     {\tt LRG2} & 0.6--0.8   & Pre     & $0.9487 \pm 0.0148$  & $1.0324 \pm 0.0610$ &          0.4315 & 87.5/87                \\
     {\tt LRG3} & 0.8--1.1   & Pre     & $1.0022 \pm 0.0107$  & $1.0132 \pm 0.0387$ &          0.2259 & 108.0/87               \\
     {\tt ELG1} & 0.8--1.1   & Pre     & $0.9819 \pm 0.0619$  &                     &                 & 51.2/41                \\
     {\tt ELG2} & 1.1--1.6   & Pre     & $0.9873 \pm 0.0152$  & $1.0191 \pm 0.0558$ &          0.1395 & 95.6/87                \\
     {\tt QSO}  & 0.8--2.1   & Pre     & $1.0006 \pm 0.0206$  &                     &                 & 49.3/41                \\
    \hline
    \end{tabular}
    \caption{Mean values and standard deviations from the marginalized posteriors of the BAO scaling parameters from fits to the unblinded \desidrone\ power spectra in the $\alpha_{\rm iso}$-$\alpha_{\rm AP}$ basis. \edited{Systematic effects are not included. The $P(k)$ fits do not include \lrgelg, as we do not have a covariance available for the combined tracer in $P(k)$. $r_{\rm off}=C_{\aiso,\alap}/\sqrt{C_{\aiso,\aiso}C_{\alap,\alap}}$.}}\label{tab:Y1unblindedPk}
\end{table}

\begin{table}
    \centering
    \begin{tabular}{|c|c|c|c|c|c|r|}
    \hline
     Tracer   & Redshift   & Recon   & $\alpha_{\parallel}$   & $\alpha_{\perp}$   &   $r_{\rm off}$ & $\chi^2 / {\rm dof}$   \\
    \hline
     \lrgo      & 0.4-0.6    & Post    & $0.928 \pm 0.023$      & $1.007 \pm 0.018$  &          -0.429 & 85.9/87                \\
     \lrgt      & 0.6-0.8    & Post    & $0.987 \pm 0.022$      & $0.949 \pm 0.014$  &          -0.419 & 82.1/87                \\
     \lrgth     & 0.8-1.1    & Post    & $1.004 \pm 0.018$      & $0.999 \pm 0.012$  &          -0.401 & 100.8/87               \\
     \elgt      & 1.1-1.6    & Post    & $0.979 \pm 0.022$      & $0.984 \pm 0.018$  &          -0.416 & 103.8/87               \\
     \hline
     \lrgo      & 0.4-0.6    & Pre     & $0.924 \pm 0.040$      & $1.003 \pm 0.021$  &          -0.399 & 87.1/87                \\
     \lrgt      & 0.6-0.8    & Pre     & $0.975 \pm 0.044$      & $0.937 \pm 0.017$  &          -0.440 & 87.5/87                \\
     \lrgth     & 0.8-1.1    & Pre     & $1.012 \pm 0.029$      & $0.997 \pm 0.014$  &          -0.406 & 108.0/87               \\
     \elgt      & 1.1-1.6    & Pre     & $1.004 \pm 0.041$      & $0.979 \pm 0.020$  &          -0.419 & 95.6/87                \\
    \hline
    \end{tabular}
    \caption{Mean values and standard deviations from the marginalized posteriors of the BAO scaling parameters from fits to the unblinded \desidrone\ power spectra in the $\alpha_\parallel$-$\alpha_\perp$ basis. \edited{Systematic effects are not included. $r_{\rm off}=C_{\aperp,\apar}/\sqrt{C_{\aperp,\aperp}C_{\apar,\apar}}$.}}\label{tab:Y1unblindedPkperp}
\end{table}
\cref{tab:Y1unblindedPk,tab:Y1unblindedPkperp} shows the BAO measurements using the power spectrum measurements \edited{that should be directly compared with the measurements in \cref{tab:Y1unblinded,tab:Y1unblindedperp}. As noted in the main section, our analytical and mock-based covariances have not fully converged in our Fourier space analysis, thus we consider the power spectrum results less robust.
Nevertheless, the power spectrum results serve as a consistency check of our results based on the correlation function as the two are in good agreement, in line with expectations derived from mock tests.}


\section{Author Affiliations}
\label{sec:affiliations}

\noindent \hangindent=.5cm $^{1}${Instituto de F\'{\i}sica Te\'{o}rica (IFT) UAM/CSIC, Universidad Aut\'{o}noma de Madrid, Cantoblanco, E-28049, Madrid, Spain}

\noindent \hangindent=.5cm $^{2}${Lawrence Berkeley National Laboratory, 1 Cyclotron Road, Berkeley, CA 94720, USA}

\noindent \hangindent=.5cm $^{3}${Physics Dept., Boston University, 590 Commonwealth Avenue, Boston, MA 02215, USA}

\noindent \hangindent=.5cm $^{4}${Tata Institute of Fundamental Research, Homi Bhabha Road, Mumbai 400005, India}

\noindent \hangindent=.5cm $^{5}${Centre for Extragalactic Astronomy, Department of Physics, Durham University, South Road, Durham, DH1 3LE, UK}

\noindent \hangindent=.5cm $^{6}${Institute for Computational Cosmology, Department of Physics, Durham University, South Road, Durham DH1 3LE, UK}

\noindent \hangindent=.5cm $^{7}${Department of Physics, University of Michigan, Ann Arbor, MI 48109, USA}

\noindent \hangindent=.5cm $^{8}${Leinweber Center for Theoretical Physics, University of Michigan, 450 Church Street, Ann Arbor, Michigan 48109-1040, USA}

\noindent \hangindent=.5cm $^{9}${IRFU, CEA, Universit\'{e} Paris-Saclay, F-91191 Gif-sur-Yvette, France}

\noindent \hangindent=.5cm $^{10}${Institut de F\'{i}sica d’Altes Energies (IFAE), The Barcelona Institute of Science and Technology, Campus UAB, 08193 Bellaterra Barcelona, Spain}

\noindent \hangindent=.5cm $^{11}${Instituto Avanzado de Cosmolog\'{\i}a A.~C., San Marcos 11 - Atenas 202. Magdalena Contreras, 10720. Ciudad de M\'{e}xico, M\'{e}xico}

\noindent \hangindent=.5cm $^{12}${Instituto de Ciencias F\'{\i}sicas, Universidad Aut\'onoma de M\'exico, Cuernavaca, Morelos, 62210, (M\'exico)}

\noindent \hangindent=.5cm $^{13}${Physics Department, Yale University, P.O. Box 208120, New Haven, CT 06511, USA}

\noindent \hangindent=.5cm $^{14}${Department of Physics and Astronomy, University of California, Irvine, 92697, USA}

\noindent \hangindent=.5cm $^{15}${Department of Physics, Kansas State University, 116 Cardwell Hall, Manhattan, KS 66506, USA}

\noindent \hangindent=.5cm $^{16}${Department of Physics \& Astronomy, University of Rochester, 206 Bausch and Lomb Hall, P.O. Box 270171, Rochester, NY 14627-0171, USA}

\noindent \hangindent=.5cm $^{17}${Institute for Astronomy, University of Edinburgh, Royal Observatory, Blackford Hill, Edinburgh EH9 3HJ, UK}

\noindent \hangindent=.5cm $^{18}${Dipartimento di Fisica ``Aldo Pontremoli'', Universit\`a degli Studi di Milano, Via Celoria 16, I-20133 Milano, Italy}

\noindent \hangindent=.5cm $^{19}${Centre for Astrophysics \& Supercomputing, Swinburne University of Technology, P.O. Box 218, Hawthorn, VIC 3122, Australia}

\noindent \hangindent=.5cm $^{20}${NSF NOIRLab, 950 N. Cherry Ave., Tucson, AZ 85719, USA}

\noindent \hangindent=.5cm $^{21}${Department of Physics and Astronomy, The University of Utah, 115 South 1400 East, Salt Lake City, UT 84112, USA}

\noindent \hangindent=.5cm $^{22}${Department of Physics \& Astronomy, University College London, Gower Street, London, WC1E 6BT, UK}

\noindent \hangindent=.5cm $^{23}${Department of Astronomy and Astrophysics, University of Chicago, 5640 South Ellis Avenue, Chicago, IL 60637, USA}

\noindent \hangindent=.5cm $^{24}${Fermi National Accelerator Laboratory, PO Box 500, Batavia, IL 60510, USA}

\noindent \hangindent=.5cm $^{25}${Korea Astronomy and Space Science Institute, 776, Daedeokdae-ro, Yuseong-gu, Daejeon 34055, Republic of Korea}

\noindent \hangindent=.5cm $^{26}${Institute of Cosmology and Gravitation, University of Portsmouth, Dennis Sciama Building, Portsmouth, PO1 3FX, UK}

\noindent \hangindent=.5cm $^{27}${Departamento de Astrof\'{\i}sica, Universidad de La Laguna (ULL), E-38206, La Laguna, Tenerife, Spain}

\noindent \hangindent=.5cm $^{28}${Instituto de Astrof\'{\i}sica de Canarias, C/ V\'{\i}a L\'{a}ctea, s/n, E-38205 La Laguna, Tenerife, Spain}

\noindent \hangindent=.5cm $^{29}${Department of Physics and Astronomy, University of Sussex, Brighton BN1 9QH, U.K}

\noindent \hangindent=.5cm $^{30}${Departamento de F\'{i}sica, Instituto Nacional de Investigaciones Nucleares, Carreterra M\'{e}xico-Toluca S/N, La Marquesa,  Ocoyoacac, Edo. de M\'{e}xico C.P. 52750,  M\'{e}xico}

\noindent \hangindent=.5cm $^{31}${Institute for Advanced Study, 1 Einstein Drive, Princeton, NJ 08540, USA}

\noindent \hangindent=.5cm $^{32}${Center for Cosmology and AstroParticle Physics, The Ohio State University, 191 West Woodruff Avenue, Columbus, OH 43210, USA}

\noindent \hangindent=.5cm $^{33}${NASA Einstein Fellow}

\noindent \hangindent=.5cm $^{34}${School of Mathematics and Physics, University of Queensland, 4072, Australia}

\noindent \hangindent=.5cm $^{35}${Instituto de F\'{\i}sica, Universidad Nacional Aut\'{o}noma de M\'{e}xico,  Cd. de M\'{e}xico  C.P. 04510,  M\'{e}xico}

\noindent \hangindent=.5cm $^{36}${CIEMAT, Avenida Complutense 40, E-28040 Madrid, Spain}

\noindent \hangindent=.5cm $^{37}${Department of Physics \& Astronomy and Pittsburgh Particle Physics, Astrophysics, and Cosmology Center (PITT PACC), University of Pittsburgh, 3941 O'Hara Street, Pittsburgh, PA 15260, USA}

\noindent \hangindent=.5cm $^{38}${Department of Astronomy, School of Physics and Astronomy, Shanghai Jiao Tong University, Shanghai 200240, China}

\noindent \hangindent=.5cm $^{39}${Space Sciences Laboratory, University of California, Berkeley, 7 Gauss Way, Berkeley, CA  94720, USA}

\noindent \hangindent=.5cm $^{40}${University of California, Berkeley, 110 Sproul Hall \#5800 Berkeley, CA 94720, USA}

\noindent \hangindent=.5cm $^{41}${Universities Space Research Association, NASA Ames Research Centre}

\noindent \hangindent=.5cm $^{42}${Center for Astrophysics $|$ Harvard \& Smithsonian, 60 Garden Street, Cambridge, MA 02138, USA}

\noindent \hangindent=.5cm $^{43}${Department of Physics, The Ohio State University, 191 West Woodruff Avenue, Columbus, OH 43210, USA}

\noindent \hangindent=.5cm $^{44}${The Ohio State University, Columbus, 43210 OH, USA}

\noindent \hangindent=.5cm $^{45}${Kavli Institute for Particle Astrophysics and Cosmology, Stanford University, Menlo Park, CA 94305, USA}

\noindent \hangindent=.5cm $^{46}${SLAC National Accelerator Laboratory, Menlo Park, CA 94305, USA}

\noindent \hangindent=.5cm $^{47}${Instituto de Astrof\'{i}sica de Andaluc\'{i}a (CSIC), Glorieta de la Astronom\'{i}a, s/n, E-18008 Granada, Spain}

\noindent \hangindent=.5cm $^{48}${Ecole Polytechnique F\'{e}d\'{e}rale de Lausanne, CH-1015 Lausanne, Switzerland}

\noindent \hangindent=.5cm $^{49}${Departamento de F\'isica, Universidad de los Andes, Cra. 1 No. 18A-10, Edificio Ip, CP 111711, Bogot\'a, Colombia}

\noindent \hangindent=.5cm $^{50}${Observatorio Astron\'omico, Universidad de los Andes, Cra. 1 No. 18A-10, Edificio H, CP 111711 Bogot\'a, Colombia}

\noindent \hangindent=.5cm $^{51}${Department of Physics, The University of Texas at Dallas, Richardson, TX 75080, USA}

\noindent \hangindent=.5cm $^{52}${Institut d'Estudis Espacials de Catalunya (IEEC), 08034 Barcelona, Spain}

\noindent \hangindent=.5cm $^{53}${Institute of Space Sciences, ICE-CSIC, Campus UAB, Carrer de Can Magrans s/n, 08913 Bellaterra, Barcelona, Spain}

\noindent \hangindent=.5cm $^{54}${Departament de F\'{\i}sica Qu\`{a}ntica i Astrof\'{\i}sica, Universitat de Barcelona, Mart\'{\i} i Franqu\`{e}s 1, E08028 Barcelona, Spain}

\noindent \hangindent=.5cm $^{55}${Institut de Ci\`encies del Cosmos (ICCUB), Universitat de Barcelona (UB), c. Mart\'i i Franqu\`es, 1, 08028 Barcelona, Spain.}

\noindent \hangindent=.5cm $^{56}${Consejo Nacional de Ciencia y Tecnolog\'{\i}a, Av. Insurgentes Sur 1582. Colonia Cr\'{e}dito Constructor, Del. Benito Ju\'{a}rez C.P. 03940, M\'{e}xico D.F. M\'{e}xico}

\noindent \hangindent=.5cm $^{57}${Departamento de F\'{i}sica, Universidad de Guanajuato - DCI, C.P. 37150, Leon, Guanajuato, M\'{e}xico}

\noindent \hangindent=.5cm $^{58}${Centro de Investigaci\'{o}n Avanzada en F\'{\i}sica Fundamental (CIAFF), Facultad de Ciencias, Universidad Aut\'{o}noma de Madrid, ES-28049 Madrid, Spain}

\noindent \hangindent=.5cm $^{59}${Excellence Cluster ORIGINS, Boltzmannstrasse 2, D-85748 Garching, Germany}

\noindent \hangindent=.5cm $^{60}${University Observatory, Faculty of Physics, Ludwig-Maximilians-Universit\"{a}t, Scheinerstr. 1, 81677 M\"{u}nchen, Germany}

\noindent \hangindent=.5cm $^{61}${Department of Astrophysical Sciences, Princeton University, Princeton NJ 08544, USA}

\noindent \hangindent=.5cm $^{62}${Kavli Institute for Cosmology, University of Cambridge, Madingley Road, Cambridge CB3 0HA, UK}

\noindent \hangindent=.5cm $^{63}${Department of Astronomy, The Ohio State University, 4055 McPherson Laboratory, 140 W 18th Avenue, Columbus, OH 43210, USA}

\noindent \hangindent=.5cm $^{64}${Department of Physics, Southern Methodist University, 3215 Daniel Avenue, Dallas, TX 75275, USA}

\noindent \hangindent=.5cm $^{65}${Department of Physics and Astronomy, University of Waterloo, 200 University Ave W, Waterloo, ON N2L 3G1, Canada}

\noindent \hangindent=.5cm $^{66}${Perimeter Institute for Theoretical Physics, 31 Caroline St. North, Waterloo, ON N2L 2Y5, Canada}

\noindent \hangindent=.5cm $^{67}${Waterloo Centre for Astrophysics, University of Waterloo, 200 University Ave W, Waterloo, ON N2L 3G1, Canada}

\noindent \hangindent=.5cm $^{68}${Graduate Institute of Astrophysics and Department of Physics, National Taiwan University, No. 1, Sec. 4, Roosevelt Rd., Taipei 10617, Taiwan}

\noindent \hangindent=.5cm $^{69}${Sorbonne Universit\'{e}, CNRS/IN2P3, Laboratoire de Physique Nucl\'{e}aire et de Hautes Energies (LPNHE), FR-75005 Paris, France}

\noindent \hangindent=.5cm $^{70}${Department of Astronomy and Astrophysics, UCO/Lick Observatory, University of California, 1156 High Street, Santa Cruz, CA 95064, USA}

\noindent \hangindent=.5cm $^{71}${Department of Astronomy and Astrophysics, University of California, Santa Cruz, 1156 High Street, Santa Cruz, CA 95065, USA}

\noindent \hangindent=.5cm $^{72}${Department of Astronomy \& Astrophysics, University of Toronto, Toronto, ON M5S 3H4, Canada}

\noindent \hangindent=.5cm $^{73}${University of Science and Technology, 217 Gajeong-ro, Yuseong-gu, Daejeon 34113, Republic of Korea}

\noindent \hangindent=.5cm $^{74}${Departament de F\'{i}sica, Serra H\'{u}nter, Universitat Aut\`{o}noma de Barcelona, 08193 Bellaterra (Barcelona), Spain}

\noindent \hangindent=.5cm $^{75}${Laboratoire de Physique Subatomique et de Cosmologie, 53 Avenue des Martyrs, 38000 Grenoble, France}

\noindent \hangindent=.5cm $^{76}${Instituci\'{o} Catalana de Recerca i Estudis Avan\c{c}ats, Passeig de Llu\'{\i}s Companys, 23, 08010 Barcelona, Spain}

\noindent \hangindent=.5cm $^{77}${Max Planck Institute for Extraterrestrial Physics, Gie\ss enbachstra\ss e 1, 85748 Garching, Germany}

\noindent \hangindent=.5cm $^{78}${Department of Physics and Astronomy, Siena College, 515 Loudon Road, Loudonville, NY 12211, USA}

\noindent \hangindent=.5cm $^{79}${Department of Physics \& Astronomy, University  of Wyoming, 1000 E. University, Dept.~3905, Laramie, WY 82071, USA}

\noindent \hangindent=.5cm $^{80}${National Astronomical Observatories, Chinese Academy of Sciences, A20 Datun Rd., Chaoyang District, Beijing, 100012, P.R. China}

\noindent \hangindent=.5cm $^{81}${Aix Marseille Univ, CNRS, CNES, LAM, Marseille, France}

\noindent \hangindent=.5cm $^{82}${Ruhr University Bochum, Faculty of Physics and Astronomy, Astronomical Institute (AIRUB), German Centre for Cosmological Lensing, 44780 Bochum, Germany}

\noindent \hangindent=.5cm $^{83}${Departament de F\'isica, EEBE, Universitat Polit\`ecnica de Catalunya, c/Eduard Maristany 10, 08930 Barcelona, Spain}

\noindent \hangindent=.5cm $^{84}${University of California Observatories, 1156 High Street, Sana Cruz, CA 95065, USA}

\noindent \hangindent=.5cm $^{85}${Department of Physics \& Astronomy, Ohio University, Athens, OH 45701, USA}

\noindent \hangindent=.5cm $^{86}${Department of Physics and Astronomy, Sejong University, Seoul, 143-747, Korea}

\noindent \hangindent=.5cm $^{87}${Abastumani Astrophysical Observatory, Tbilisi, GE-0179, Georgia}

\noindent \hangindent=.5cm $^{88}${Faculty of Natural Sciences and Medicine, Ilia State University, 0194 Tbilisi, Georgia}

\noindent \hangindent=.5cm $^{89}${Space Telescope Science Institute, 3700 San Martin Drive, Baltimore, MD 21218, USA}

\noindent \hangindent=.5cm $^{90}${Centre for Advanced Instrumentation, Department of Physics, Durham University, South Road, Durham DH1 3LE, UK}

\noindent \hangindent=.5cm $^{91}${Physics Department, Brookhaven National Laboratory, Upton, NY 11973, USA}

\noindent \hangindent=.5cm $^{92}${Beihang University, Beijing 100191, China}

\noindent \hangindent=.5cm $^{93}${Department of Astronomy, Tsinghua University, 30 Shuangqing Road, Haidian District, Beijing, China, 100190}

\noindent \hangindent=.5cm $^{94}${Physics Department, Stanford University, Stanford, CA 93405, USA}

\noindent \hangindent=.5cm $^{95}${Department of Physics, University of California, Berkeley, 366 LeConte Hall MC 7300, Berkeley, CA 94720-7300, USA}



\end{document}